\documentclass[twocolumn]{aastex63}
\usepackage{placeins}
\usepackage{amsmath}

\accepted{in ApJS}

\shorttitle{Survey overview}
\shortauthors{Dutta et al.}

\begin{document}

\title{ALMA Survey of Orion Planck Galactic Cold Clumps (ALMASOP)\\ 
 II. Survey overview: a first look at 1.3\,mm continuum maps and molecular outflows}

\correspondingauthor{Somnath Dutta, Tie Liu}
\email{sdutta@asiaa.sinica.edu.tw, liutie@shao.ac.cn}

\author[0000-0002-2338-4583]{Somnath Dutta}
\affiliation{Institute of Astronomy and Astrophysics, Academia Sinica, Roosevelt Rd, Taipei 10617, Taiwan, R.O.C.}

\author[0000-0002-3024-5864]{Chin-Fei Lee}
\affiliation{Institute of Astronomy and Astrophysics, Academia Sinica, Roosevelt Rd, Taipei 10617, Taiwan, R.O.C.}

\author[0000-0002-5286-2564]{Tie Liu}
\affiliation{Shanghai Astronomical Observatory, Chinese Academy of Sciences, 80 Nandan Road, Shanghai 200030, China}
\affiliation{Key Laboratory for Research in Galaxies and Cosmology, Chinese Academy of Sciences, 80 Nandan Road, Shanghai 200030 People’s Republic of China}

\author[0000-0001-9304-7884]{Naomi Hirano}
\affiliation{Institute of Astronomy and Astrophysics, Academia Sinica, Roosevelt Rd, Taipei 10617, Taiwan, R.O.C.}

\author[0000-0003-4603-7119]{Sheng-Yuan Liu}
\affiliation{Institute of Astronomy and Astrophysics, Academia Sinica, Roosevelt Rd, Taipei 10617, Taiwan, R.O.C.}

\author[0000-0002-8149-8546]{Ken'ichi Tatematsu}
\affil{Nobeyama Radio Observatory, National Astronomical Observatory of Japan, 
National Institutes of Natural Sciences, 
462-2 Nobeyama, Minamimaki, Minamisaku, Nagano 384-1305, Japan}
\affiliation{Department of Astronomical Science,
SOKENDAI (The Graduate University for Advanced Studies),
2-21-1 Osawa, Mitaka, Tokyo 181-8588, Japan}

\author[0000-0003-2412-7092]{Kee-Tae Kim}
\affil{Korea Astronomy and Space Science Institute (KASI), 776 Daedeokdae-ro, Yuseong-gu, Daejeon 34055, Republic of Korea}
\affil{University of Science and Technology, Korea (UST), 217 Gajeong-ro, Yuseong-gu, Daejeon 34113, Republic of Korea}

\author[0000-0001-8385-9838]{Hsien Shang}
\affiliation{Institute of Astronomy and Astrophysics, Academia Sinica, Roosevelt Rd, Taipei 10617, Taiwan, R.O.C.}

\author[0000-0002-4393-3463]{Dipen Sahu}
\affiliation{Institute of Astronomy and Astrophysics, Academia Sinica, Roosevelt Rd, Taipei 10617, Taiwan, R.O.C.}

\author[0000-0003-2011-8172]{Gwanjeong Kim}
\affil{Nobeyama Radio Observatory, National Astronomical Observatory of Japan, 
National Institutes of Natural Sciences, 
462-2 Nobeyama, Minamimaki, Minamisaku, Nagano 384-1305, Japan}

\author{Anthony Moraghan}
\affiliation{Institute of Astronomy and Astrophysics, Academia Sinica, Roosevelt Rd, Taipei 10617, Taiwan, R.O.C.}

\author{Kai-Syun Jhan}
\affiliation{Institute of Astronomy and Astrophysics, Academia Sinica, Roosevelt Rd, Taipei 10617, Taiwan, R.O.C.}

\author{Shih-Ying Hsu}
\affiliation{Institute of Astronomy and Astrophysics, Academia Sinica, Roosevelt Rd, Taipei 10617, Taiwan, R.O.C.}

\author{Neal J. Evans}
\affiliation{Department of Astronomy The University of Texas at Austin 2515 Speedway, Stop C1400 Austin, TX 78712-1205, USA}

\author[0000-0002-6773-459X]{Doug Johnstone}
\affiliation{NRC Herzberg Astronomy and Astrophysics, 5071 West Saanich Rd., Victoria, BC V9E 2E7, Canada}
\affiliation{Department of Physics and Astronomy, University of Victoria, Victoria, BC V8P 1A1, Canada}

\author[0000-0003-1140-2761]{Derek Ward-Thompson}
\affiliation{Jeremiah Horrocks Institute, University of Central Lancashire, Preston PR1  2HE, UK}

\author[0000-0002-4336-0730]{Yi-Jehng Kuan}
\affiliation{Department of Earth Sciences, National Taiwan Normal University, Taipei, Taiwan, R.O.C. \& Institute of Astronomy and Astrophysics, Academia Sinica, Roosevelt Rd, Taipei 10617, Taiwan, R.O.C.}

\author{Chang Won Lee}
\affil{Korea Astronomy and Space Science Institute (KASI), 776 Daedeokdae-ro, Yuseong-gu, Daejeon 34055, Republic of Korea}
\affil{University of Science and Technology, Korea (UST), 217 Gajeong-ro, Yuseong-gu, Daejeon 34113, Republic of Korea}

\author{Jeong-Eun Lee} 
\affil{School of Space Research, Kyung Hee University, Yongin-Si, Gyeonggi-Do 17104, Republic of Korea}

\author{Alessio Traficante}
\affil{IAPS-INAF, via Fosso del Cavaliere 100, I-00133, Rome, Italy}

\author[0000-0002-5809-4834]{Mika Juvela}
\affiliation{Department of Physics, P.O.Box 64, FI-00014, University of Helsinki, Finland}

\author{Charlotte Vastel}
\affiliation{IRAP, Université de Toulouse, CNRS, UPS, CNES, 31400, Toulouse, France}

\author{Zhang, Qizhou}
\affiliation{Center for Astrophysics | Harvard \&amp; Smithsonian, 60 Garden Street, Cambridge, MA 02138, USA}

\author[0000-0002-7125-7685]{Patricio Sanhueza} 
\affiliation{National Astronomical Observatory of Japan, National Institutes of Natural Sciences, 2-21-1 Osawa, Mitaka, Tokyo 181-8588, Japan}
\affiliation{Department of Astronomical Science,
SOKENDAI (The Graduate University for Advanced Studies),
2-21-1 Osawa, Mitaka, Tokyo 181-8588, Japan}

\author[0000-0002-6386-2906]{Archana Soam}
\affil{SOFIA Science Center, Universities Space Research Association, NASA Ames Research Center, Moffett Field, California 94035, USA}

\author{Woojin Kwon}
\affil{Department of Earth Science Education, Seoul National University, 1 Gwanak-ro, Gwanak-gu, Seoul 08826, Republic of Korea}
\affil{Korea Astronomy and Space Science Institute, 776 Daedeokdae-ro, Yuseong-gu, Daejeon 34055, Republic of Korea}

\author[0000-0002-9574-8454]{Leonardo Bronfman}
\affil{Departamento de Astronomía, Universidad de Chile, Casilla 36-D, Santiago, Chile}

\author{David Eden}
\affiliation{Astrophysics Research Institute, Liverpool John Moores University, IC2, Liverpool Science Park, 146 Brownlow Hill, Liverpool, L3 5RF, UK}

\author{Paul F. Goldsmith}
\affiliation{Jet Propulsion Laboratory, California Institute of Technology, 4800 Oak Grove Drive, Pasadena, CA 91109, USA}

\author[0000-0002-3938-4393]{Jinhua He}
\affil{Yunnan Observatories, Chinese Academy of Sciences, 396 Yangfangwang, Guandu District, Kunming, 650216, P. R. China}
\affiliation{Chinese Academy of Sciences South America Center for Astronomy, National Astronomical Observatories, CAS, Beijing 100101, China}
\affiliation{Departamento de Astronom{\'i}a, Universidad de Chile, Casilla 36-D, Santiago, Chile}

\author{Yuefang Wu}
\affiliation{Department of Astronomy, Peking University, 100871 Beijing, People's Republic of China}

\author{Veli-Matti Pelkonen}
\affiliation{Institut de Ci\`{e}ncies del Cosmos, Universitat de Barcelona, IEEC-UB, Mart\'{i} i Franqu\`{e}s 1, E08028 Barcelona, Spain}

\author{Sheng-Li Qin}
\affiliation{Department of Astronomy, Yunnan University, and Key Laboratory of Particle Astrophysics of Yunnan Province, Kunming, 650091, People's Republic of China}

\author[0000-0003-1275-5251]{Shanghuo Li}
\affiliation{Korea Astronomy and Space Science Institute (KASI), 776 Daedeokdae-ro, Yuseong-gu, Daejeon 34055, Republic of Korea}

\author[0000-0003-3010-7661]{Di Li}
\affiliation{National Astronomical Observatories, Chinese Academy of Sciences, Beijing 100101, People’s Republic of China}
\affiliation{NAOC-UKZN Computational Astrophysics Centre, University of KwaZulu-Natal, Durban 4000, South Africa}





\begin{abstract} 
Planck Galactic Cold Clumps (PGCCs) are contemplated to be the ideal targets to probe the early phases of star formation. We have conducted a survey of 72 young dense cores inside PGCCs in the Orion complex with the Atacama Large Millimeter/submillimeter Array (ALMA) at 1.3\,mm (band 6) using three different configurations (resolutions $\sim$ 0$\farcs$35, 1$\farcs$0, and 7$\farcs$0) to statistically investigate their evolutionary stages and sub-structures. We have obtained images of the 1.3\,mm continuum and molecular line emission ($^{12}$CO, and SiO) at an angular resolution of $\sim$ 0$\farcs$35 ($\sim$ 140\,au) with the combined arrays. We find 70 substructures within 48 detected dense cores with median dust-mass $\sim$ 0.093\,M$_{\sun}$ and deconvolved size $\sim$ 0$\farcs$27. Dense substructures are clearly detected within the central 1000\,au of four candidate prestellar cores. The sizes and masses of the substructures in continuum emission are found to be significantly reduced with protostellar evolution from Class\,0 to Class\,I.  We also study the evolutionary change in the outflow characteristics through the course of protostellar mass accretion. A total of 37 sources exhibit CO outflows, and 20 ($>$50\%) show high-velocity jets in SiO. The CO velocity-extents ($\Delta$Vs) span from 4 to 110 km/s with outflow cavity opening angle width at  400\,au ranging from $[\Theta_{obs}]_{400}$ $\sim$ 0$\farcs$6 to 3$\farcs$9, which corresponds to 33$\fdg$4$-$125$\fdg$7. For the majority of the outflow sources, the $\Delta$Vs show a positive correlation with $[\Theta_{obs}]_{400}$, suggesting that as protostars undergo gravitational collapse, the cavity opening of a protostellar outflow widens and the protostars possibly generate more energetic outflows.

\end{abstract}


\section{Introduction} \label{sec:intro}
Stars form within dense cores (typical size $\sim$ 0.1 pc, density $\sim$ $10^4$ cm$^{-3}$, and temperature $\sim$ 10 K) in the clumpy and filamentary environment of molecular clouds \citep{Myers_1983,2000prpl.conf...97W}. In past decades, observations revealed the presence of embedded protostars within dense cores, which has also led to the classification of ``prestellar" and ``protostellar" phases of dense cores \citep{1986ApJ...307..337B,2007ARA&A..45..339B}. The puzzle begins with the understanding of how a prestellar core condenses to form a star or multiple system and how a protostar accumulates its central mass from the surrounding medium during its evolution. Studies of extremely young dense cores at different evolutionary phases offer the best opportunity to probe the core formation under diverse environmental conditions, as well as determine the transition phase from prestellar to protostellar cores, protostellar evolution and, investigate the outflow/jet launching scenario and physical changes with the protostellar evolution.

 In addition, a significant fraction of stars are found in multiple systems. Thus, our understanding of star formation must account for the formation of multiple systems.
In one popular star formation theory, the ``turbulent fragmentation" theory, turbulent fluctuations
in a dense core become Jeans unstable and collapse faster than the background core \citep[e.g.,][]{Padoan2002,Fisher2004,Goodwin2004}, forming multiple systems. Turbulent fragmentation is likely the dominant mechanism for wide-binary systems \citep{Chen2013,Tobin2016,Lee2017NatAs}. Observations indicate that the multiplicity fraction and the companion star
fraction are highest in Class 0 protostars and decrease in more evolved protostars \citep{Chen2013,Tobin2016}, confirming that multiple systems form in the very early phase.

The ``turbulent fragmentation" theory predicts that the fragmentation begins in the starless core stage \citep{Offner10}. Small scale fragmentation/coalescence processes have been detected within 0.1 pc scale regions of some starless cores in nearby molecular clouds \citep{Ohashi2018,Tatematsu2020,Tokuda2020}. To shed light on the formation of multiple stellar systems, however, we ultimately need to study the internal structure and gas motions within the central 1000 AU of starless cores. Over the past few years, several attempts have been made to detect the very central regions and possible substructures of starless cores \citep[e.g.,][]{Schnee2010,Schnee2012,Dunham2016,Kirk2017,Caselli2019}. However, no positive results on the fragmentation within the central 1000 AU of starless cores have been collected so far. Probing substructures of a statistically significant sample of starless cores at the same distance will put this theoretical paradigm (``turbulent fragmentation") to a stringent observational test. If no substructure is detected, this will raise serious questions to our current understanding of this framework. Irrespective of the theoretical framework, these observations will empirically constrain, at high resolution, the starless core structure at or near collapse.

After the onset of star formation, a (Keplerian) rotating disk is formed, feeding a central protostar. However, the detailed process of the disk formation and evolution (growth) is unclear. In theory, material in a collapsing core will be guided by magnetic field lines towards the midplane, forming an infalling-rotating flattened envelope called a ``pseudodisk" \citep{Allen03}. A rotating disk is then formed in the innermost ($<$100 au) part of the pseudodisk. In the pseudodisk, magnetic braking may be efficient, affecting the formation and growth of the disk \citep{Galli06}. Therefore, high-resolution ($\times$10 au) dust polarization and molecular line observations of Class 0 protostars (the youngest known accreting protostars) and their natal cores are key to constrain theoretical models for the formation of protostellar disks by unveiling their magnetic fields and gas kinematics.

However, disks in young Class 0 protostars have largely remained elusive to date. We have lacked the observational facilities capable of probing this regime in these extremely young objects.  As a consequence, we do not know when disks form or what they look like at formation. Recently, large high-resolution continuum surveys have revealed several tens of Class 0 disk candidates \citep{Tobin2020}. So far, however, only several Class 0 protostars (e.g., VLA\,1623, HH\,212, L\,1527, and L\,1448-NB) have been suggested
to harbor Keplerian-like kinematics at scales $40<r<100$ AU \citep{Murillo13,Codella14,Ohashi14,Tobin16}.
The most convincing case for a resolved Class 0 protostellar disk was found in the HH\,212 Class 0 protostar, evidenced by  an equatorial dark dust lane with a radius of $\sim$60 AU at submillimeter wavelengths  \citep{Lee17}.
A systematic high-resolution continuum (polarization) and molecular line survey of Class 0 protostars is urgently needed to search for more Class 0 disk candidates and study disk formation. Collimated bipolar outflows together with fattened continuum emission (pseudodisk) can help identify Class 0 disk candidates.

Low-velocity bipolar outflows are nearly ubiquitous in accreting, rotating, and magnetized protostellar systems \citep{1980ApJ...239L..17S,1992A&A...261..274C,1996A&A...311..858B,2014ApJ...783...29D,2015A&A...576A.109Y,2019ApJS..240...18K}. The lower transitions of CO are the most useful tracers of molecular outflows since their low energy levels are easily populated by collisions with H$_2$ and He molecules at the typical densities and temperatures of molecular clouds \citep{2016ARA&A..54..491B,2020A&ARv..28....1L}. The outflows appear as bipolar from the polar regions along the axis of rotation at the early collapsing phase, driven by the first core \citep{1969MNRAS.145..271L}, and remain active throughout the journey of protostellar accretion from the outer pseudodisk region \citep{1998ApJ...508L..95B,2000ApJ...531..350M,2002ApJ...575..306T,2014MNRAS.438.2278M,2020A&ARv..28....1L}.
As protostars evolve, the physical properties of outflow components diversify significantly based on the natal environment. Both numerical simulations and observations have revealed that the opening angle  of the outflow cavity widens with time as more material is evacuated from the polar region and the equatorial pseudodisk grows \citep{2007prpl.conf..245A,2008ApJ...675..427S,2014prpl.conf..451F,2016ApJ...832...40K}. Typically, sources in the Class 0 phase exhibit CO outflow opening angles of 20$\degr$ $-$ 50$\degr$, which increase for Class I (80$\degr$ $-$ 120$\degr$) and Class II (100$\degr$ $-$ 160$\degr$). The outflow velocity is also expected to increase with time as the mass loss increases with accretion rate \citep{2009ApJ...705.1388H,2016ARA&A..54..491B}.

A significant number of Class 0, I, and early II protostars are observed to exhibit extremely high velocity (EHV) collimated molecular jets (or typically high-density knots) within the wide-angle low-velocity outflow cavities. These high-velocity jets mainly originate from the inner edges of the disk and jet velocities increase with the evolutionary stage of the protostars in the range of $\sim$ 100 km s$^{-1}$ to a few $\sim$ 100 km s$^{-1}$ in the later phases \citep{2007AJ....133.2799A,2011ApJ...736...29H,2019ApJ...876..149M}. The gas content of the jets also transitions from molecular predominant to mostly atomic \citep{2016ARA&A..54..491B,2020A&ARv..28....1L}. The jets in the younger sources, like Class 0, are mainly detectable in molecular gas, e.g., CO, SiO, and SO at (sub)millimeter and H$_2$ in the infrared wavelength. Conversely, in the older population like evolved Class I and Class II sources, the jets are mainly traceable in atomic and ionized gas e.g., O, H$\alpha$, and S II \citep{2001ARA&A..39..403R,2016ARA&A..54..491B,2020A&ARv..28....1L}.

To summarize, more high resolution observations are needed to study the fragmentation and structures (e.g., disks, outflows) of dense cores in the earliest phases of star formation, i.e., from prestellar cores to the youngest protostellar (Class 0) cores.

\subsection{Observations of Planck Galactic Cold Clumps in the Orion complex}

The low dust temperatures ($\sim$ 14 K) of the Planck Galactic Cold Clumps (PGCCs)  make them ideal targets for investigating the initial conditions of star formation \citep{2016A&A...594A..28P}. Through observations of $\sim$1000 Planck Galactic Cold Clumps in the JCMT large survey program ``SCOPE: SCUBA-2 Continuum Observations of Pre-protostellar Evolution" (PI: Tie Liu), we have cataloged nearly 3500 cold (T$_d\sim$6-20 K) dense cores, most of which are either starless or in the earliest phase of star formation \citep{2018ApJS..234...28L,Eden2019}. This sample of ``SCOPE" dense cores represents a real goldmine for investigations of the very early phases of star formation.

The Orion complex contains the nearest giant molecular clouds (GMCs) that harbor high-mass star formation sites. As a part of the SCOPE survey, all the dense PGCCs (average column density $>$ 5 $\times$ 10$^{20}$ cm$^{-2}$) of the Orion complex (Orion A, B, and $\lambda$ Orionis GMCs) were observed at 850 $\mu$m using the SCUBA-2 instrument at the JCMT 15 m telescope \citep{2018ApJS..234...28L,2018ApJS..236...51Y}. A total of 119 dense cores were revealed inside these PGCCs, which includes protostars and gravitationally unstable starless cores \citep[][]{2018ApJS..236...51Y}. This sample represents the dense cores of mass spectrum in the range 0.2 - 14 M$_{\sun}$ with a median mass of $\sim$ 1.4 M$_{\sun}$ and mean radius $\sim$ 0.05 pc as estimated from SCUBA-2 850 $\mu$m continuum observations \citep{2018ApJS..236...51Y}. Their centrally peaked emission features in the SCUBA-2 850 $\mu$m continuum attribute them to likely be gravitationally unstable and possible for imminent collapse \citep{2016MNRAS.463.1008W}.

These Orion dense cores were further investigated in multiple molecular lines (e.g., N$_2$D$^+$, DCO$^+$, DNC in J=1-0 transitions) with the NRO 45-m telescope \citep[][]{2020ApJS..249...33K,Tatematsu2020}. This follow-up molecular line survey toward 113 of these 119 SCUBA-2 objects with the Nobeyama Radio Observatory (NRO) 45m telescope revealed nearly half of these SCUBA-2 objects showing strong emission from young, cold, and dense gas tracers, such as N$_2$D$^+$, DCO$^+$, DNC \citep[][]{2020ApJS..249...33K,Tatematsu2020}.

In particular, high spatial resolution observations with interferometers have already reported very young stellar objects inside some of these SCUBA-2 dense cores. With the Submillimeter Array (SMA), \cite{Liu2016} reported the detection of an extremely young Class 0 protostellar object and a proto-brown dwarf candidate in the bright-rimmed clump PGCC G192.32-11.88 located in the $\lambda$ Orionis cloud. Very recently, \citet{Tatematsu2020} observed a star-forming core (PGCC G210.82-19.47 North1; hereafter, G210) and a starless core (PGCC G211.16-19.33 North3; hereafter, G211) in the Orion A cloud with the 7m Array of the Atacama Compact Array (ACA) of the Atacama Large Millimeter/submillimeter Array (ALMA). The two cores show a relatively high deuterium fraction in single-pointing observations with the Nobeyama 45 m radio telescope. In ACA observations, the starless core G211 shows a clumpy structure with several sub-cores, which in turn show chemical differences. In contrast, the star-forming core G210 shows an interesting spatial feature of two N$_2$D$^+$ peaks of similar intensity and radial velocity located symmetrically with respect to the single dust continuum peak, suggesting the existence of an edge-on pseudo-disk.

All of the previous observations indicate that those Orion SCUBA-2 cores inside PGCCs are ideal for investigating the initial conditions of star formation in a GMC environment. 

\subsection{ALMASOP: ALMA Survey of Orion PGCCs} \label{sec:sampleSelection}

\begin{figure}
\fig{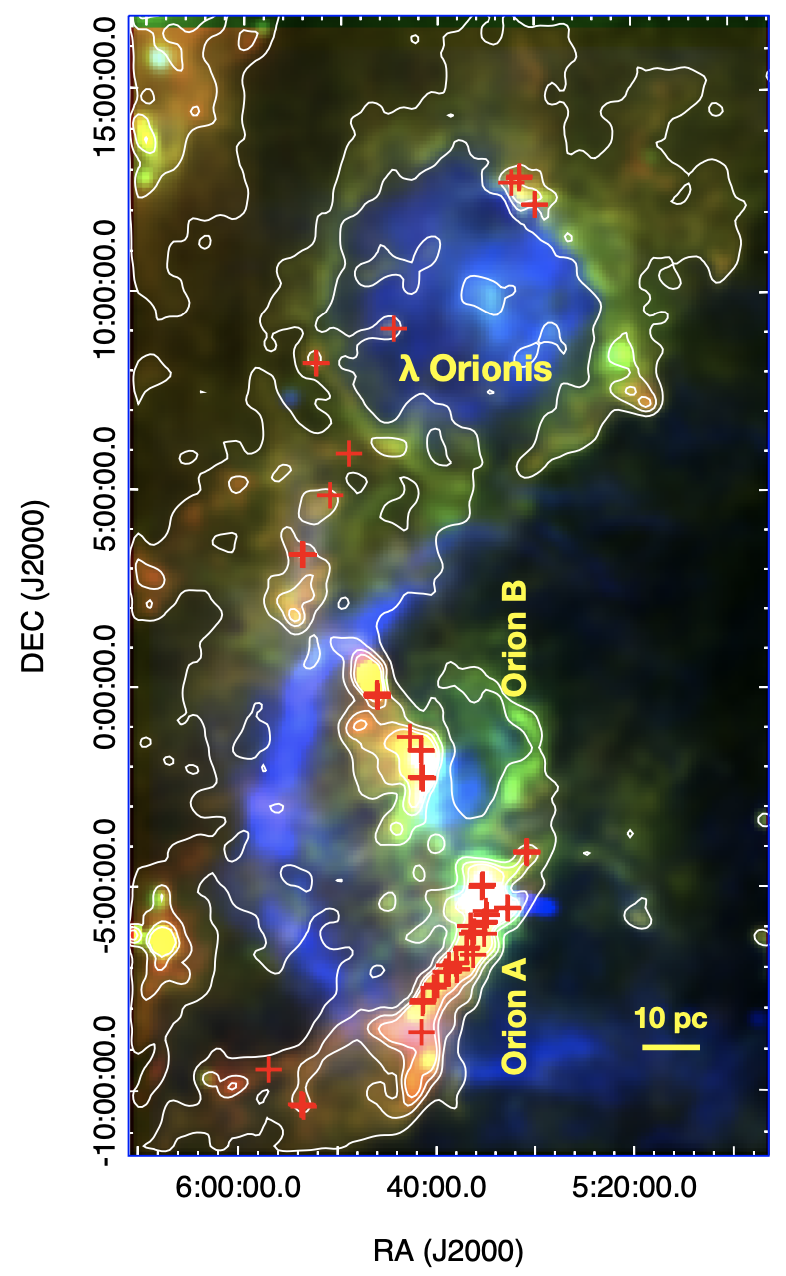}{0.5\textwidth}{}
\caption{Spatial distribution of the observed cores (red ``+") on the three-color composite image (red: Planck 857 GHz; green: IRAS 100 $\micron$; blue: H$_{\alpha}$) of the Orion complex. The images are smoothed with a gaussian kernel. The white contours represent the flux density of Planck 857 GHz continuum emission. The contour levels are 14.8, 29.7, 44.5 and 59.4 MJy/sr.}
\label{fig:spatial_core}
\end{figure}

In ALMA cycle 6, we initiated a survey-type project (ALMASOP: ALMA Survey of Orion PGCCs) to systematically investigate the fragmentation of starless cores and young protostellar cores in Orion PGCCs with ALMA. We selected 72 extremely cold young dense cores from \citet{2018ApJS..236...51Y}, including 23 starless core candidates and 49 protostellar core candidates. We call them candidates because they were classified mainly based on the four WISE bands (3.4-22 $\micron$) in \citet{2018ApJS..236...51Y}. In this work, we will further classify them with all available infrared data (e.g. Spitzer, Herschel) as well as our new ALMA data. All 23 starless core candidates of this sample show high-intensity N$_2$D$^+$(1-0) emission with peak brightness temperature higher than 0.2 K in 45m NRO observations \citep[][]{2020ApJS..249...33K,Tatematsu2020}, a signpost for the presence of a dense core on the verge of star formation. Intense N$_2$D$^+$ emission was also observed in 21 protostellar core candidates \citep[][]{2020ApJS..249...33K,Tatematsu2020}. The remaining 28 protostellar core candidates were not detected in N$_2$D$^+$ \citep[][]{2020ApJS..249...33K,Tatematsu2020}, suggesting they are more evolved than those detected in N$_2$D$^+$. These dense cores, therefore, design a unique sample to probe the onset of star formation and the early evolution of dense cores. The observed target names and coordinates are listed in columns 1, 2, and 3, respectively, in Table \ref{tab:targetobserved} and their spatial distribution is shown in Figure \ref{fig:spatial_core}.

In this paper, we present an overview of the ALMASOP survey including the observations and data products, and mostly qualitative previews of the results from forthcoming papers. We have incorporated some perspectives of  detection of multiplicity in protostellar systems and the physical characteristics of their outflow lobes. More detailed quantitative results of multiplicity formation in prestellar to protostellar phase, outflow and jet characteristics, disk formation, astrochemical changes from the prestellar to protostellar phases will be presented in the forthcoming papers. Section \ref{sec:dataObservations} discusses the details of the observations in the survey and data analyses. In section \ref{sec:resultsContOutflow}, the science goals of this survey and early results are described.  Section \ref{sec:discussionEvolutionCont_EvolutionOuflow} delineates the discussion on the evolution of dense cores and protostellar outflows. Section \ref {sec:summary_conclusion} deals with the summary and conclusions of this study.

\section{Observations} \label{sec:dataObservations}

The ALMA observations of ALMASOP (Project ID:2018.1.00302.S.; PI: Tie Liu) were carried out with ALMA band 6 in Cycle 6 toward the 72 extremely young dense cores, during 2018 October to 2019 January. The observations were executed in four blocks in three different array configurations: 12m C43-5 (TM1), 12m C43-2 (TM2) \& 7m ACA. The execution blocks, date of observations, array configurations, number of antennas, exposure times on the targets, and unprojected baselines are listed in Table~\ref{tab:obslog}. For observations in the C43-5, C43-2, and compact 7m ACA, the unprojected baseline lengths range from 15 to 1398, 15 to 500, and 9 to 49 m, respectively. The resulting maximum recoverable scale was 25$\arcsec$. 

The ALMA band 6 receivers were utilized to simultaneously capture four spectral windows (SPWs), as summarized by the correlator setup in Table \ref{tab:spectralsetup}. The ALMA correlator was configured to cover several main targeted molecular line transitions (e.g., J=2-1 of CO and C$^{18}$O; J=3-2 of N$_2$D$^+$, DCO$^+$ and DCN; and SiO J=5-4) simultaneously. A total bandwidth of 1.875 GHz was set up for all SPWs. The velocity resolution is about 1.5 km~s$^{-1}$. Different quasars were observed to calibrate the bandpass, flux, and phase, as tabulated in Table \ref{tab:calibrator} with their flux densities. 

In this paper, we present the results of the cold dusty envelope+disk emission tracer 1.3\,mm continuum, low-velocity outflow tracer CO J=2-1 (230.462 GHz) and high-velocity jet tracer SiO J=5-4 (217.033 GHz) line emission. The acquired visibility data were calibrated using the standard pipeline in CASA 5.4 \citep{2007ASPC..376..127M} for different scheduling blocks (SB) separately. We then separated visibilities for all 72 sources, each with their three different observed configurations. For each source, we generated both 1.3\,mm continuum and spectral visibilities by selecting all line-free channels, fitting, and subtracting continuum emission in the visibility domain. Imaging of the visibility data was performed with the TCLEAN task in CASA 5.4, using a threshold of 3$\sigma$ theoretical sensitivity, and  ``hogbom" deconvolver. We applied Briggs weighting with robust $+$2.0 (natural weighting) to obtain a high sensitivity map to best suit the weak emission at the outer envelope, and it does not degrade the resolution much in comparison with robust $+$0.5. We generated two sets of continuum images. One set includes all configurations TM1+TM2+ACA to obtain continuum maps with a synthesized beam of $\sim$ 0$\farcs$38 $\times$ 0$\farcs$33 and typical sensitivity ranging from 0.01 to 0.2 mJy beam$^{-1}$; where TM1, TM2 configurations contributed to improve resolution, and the compact ACA configuration improves the missing flux problem. 
For the large scale structures, we also obtained a second set of continuum images from only the 7-m ACA configuration visibilities with a synthesized beam of 7$\farcs$6 $\times$ 4$\farcs$1 and typical sensitivity of 0.6 to 2.0 mJy beam$^{-1}$. The detections of dense cores are listed in combined configurations (column 5), with rms (column 6), plus in ACA only (column 7) with rms (column 8) in Table \ref{tab:targetobserved}.
 
On the other hand, since CO J=2-1 and SiO J=5-4 emission are strong, a robust weighting factor of +0.5 was used to generate CO and SiO channel maps using a combination of three visibilities (i.e., TM1+TM2+ACA) with typical synthesized beam sizes of $\sim$ 0$\farcs$41 $\times$ 0$\farcs$35 and  $\sim$ 0$\farcs$44 $\times$ 0$\farcs$37, respectively. We binned the channels with a velocity resolution of 2 km s$^{-1}$ to improve the signal-to-noise ratio and thus we obtained typical sensitivity ranging 0.02 to 0.2 mJy beam$^{-1}$.

\section{Science goals and Early results} \label{sec:resultsContOutflow}
\subsection{Continuum emission at 1.3\,mm}\label{sec:resultsContOutflowCONT}

\begin{figure*}
\fig{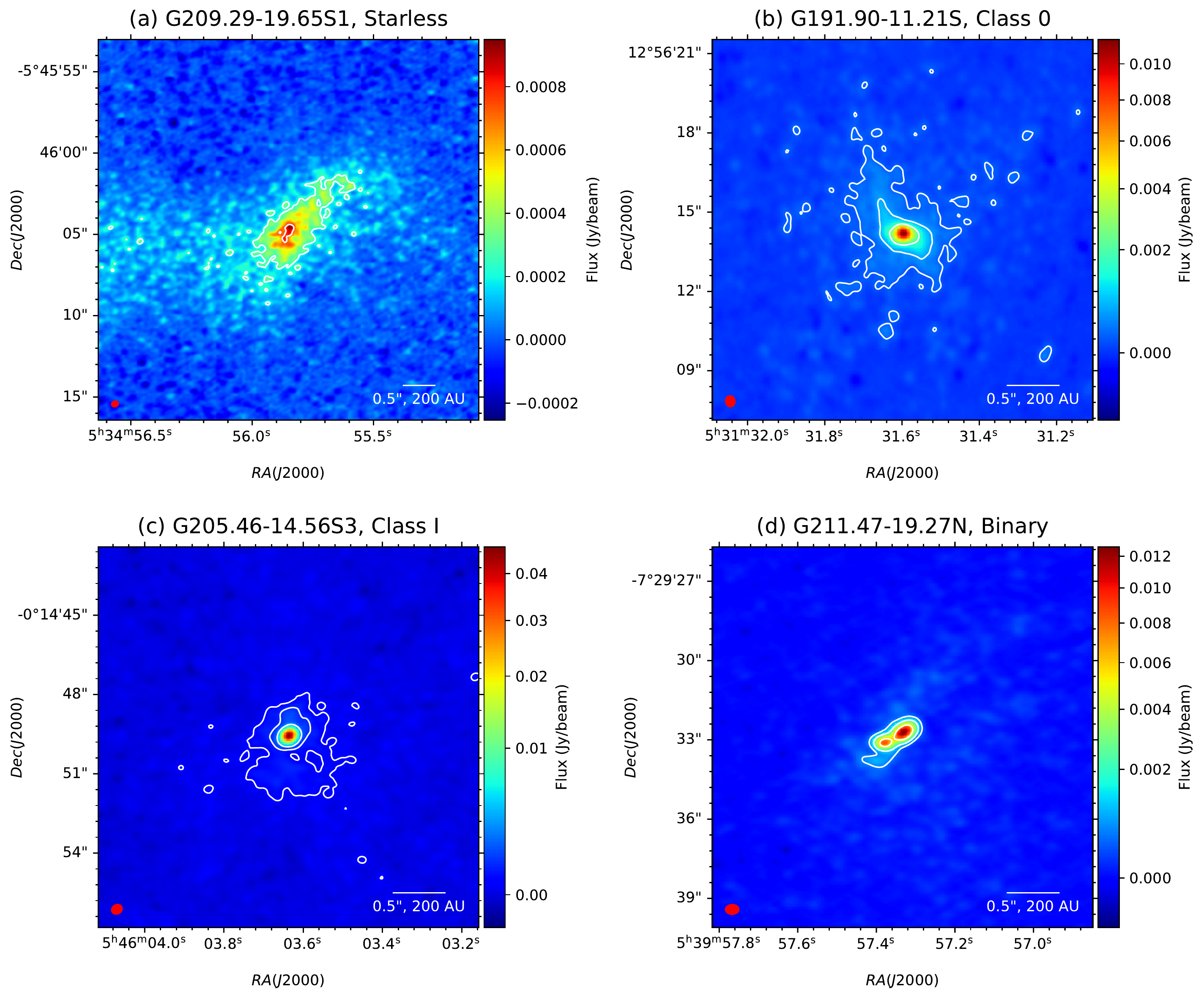}{0.9\textwidth}{}
\caption{Example images of ALMA 1.3 mm continuum toward selected dense cores. The typical beam sizes $\sim$ 0$\farcs$35 are drawn in the lower left of each panel in red ellipse. The contour levels are at 5$\times$(1, 2, 10)$\sigma$. The source sequences are (a) starless core G209.29-19.65S1, where $\sigma$ = 5 $\times$ 10$^{-5}$ Jy beam$^{-1}$, (b) Class 0 system G191.90-11.21S, where $\sigma$ = 4 $\times$ 10$^{-5}$ Jy beam$^{-1}$, (c) Class I system G205.46-14.53S3, where $\sigma$ = 6 $\times$ 10$^{-5}$ Jy beam$^{-1}$ (d) binary system G211.47-19.27N, where $\sigma$ = 12 $\times$ 10$^{-5}$ Jy beam$^{-1}$. Notice that the extended emission turns more compact as we evolve from starless, Class 0 to Class I, interestingly the peak emission is also increasing on the same sequence (see text for more details). 
 }
\label{fig:example_continuum}
\end{figure*}

\begin{figure*}
\fig{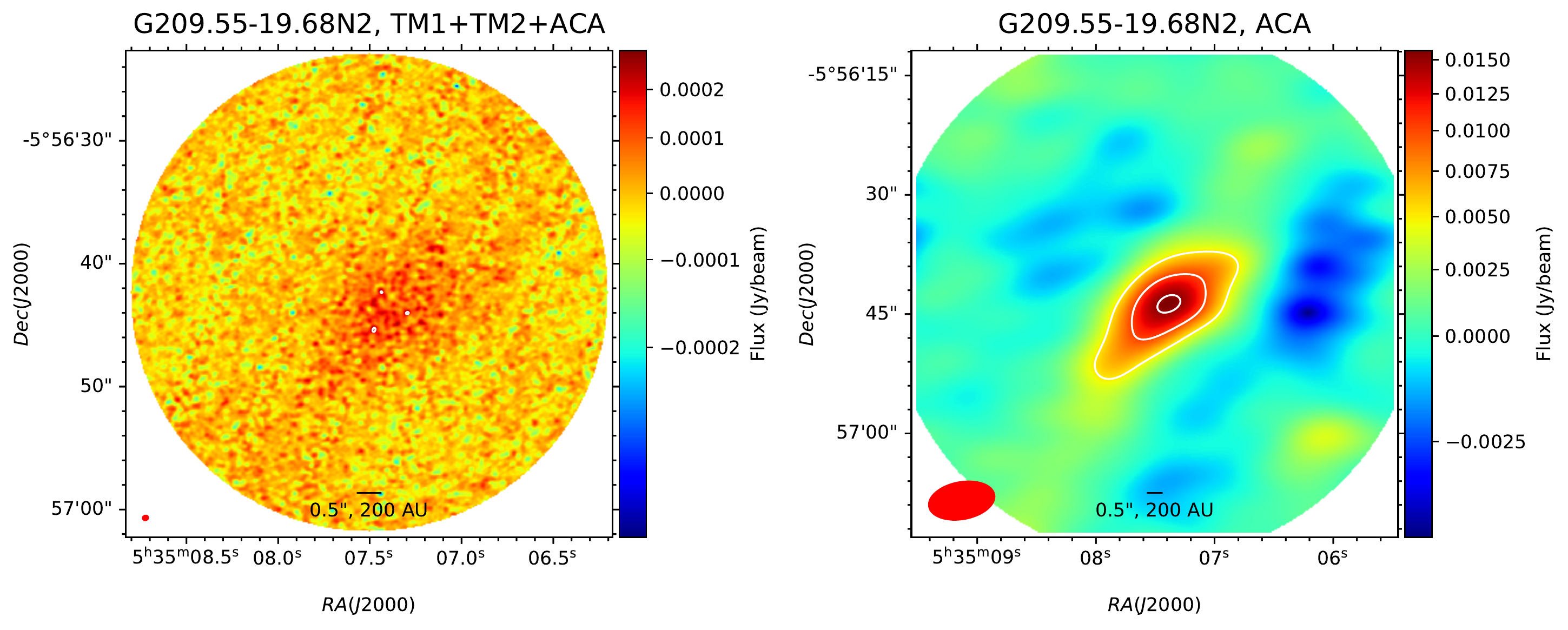}{0.9\textwidth}{}
\caption{Example images of 1.3 mm continuum observations for combined TM1+TM2+ACA in the left panel, and ACA only in the right panel are shown. Typical beam sizes are shown at the lower left in each panel with the red ellipses. The combined resolution resolved out the emission, whether a compact structure is clearly seen in ACA only with contour levels 5$\times$(1, 2, 10)$\sigma$, where $\sigma$ = 0.001 Jy beam$^{-1}$. 
}
\label{fig:example_ContComp_COMB_ACA}
\end{figure*}

\begin{figure}
\fig{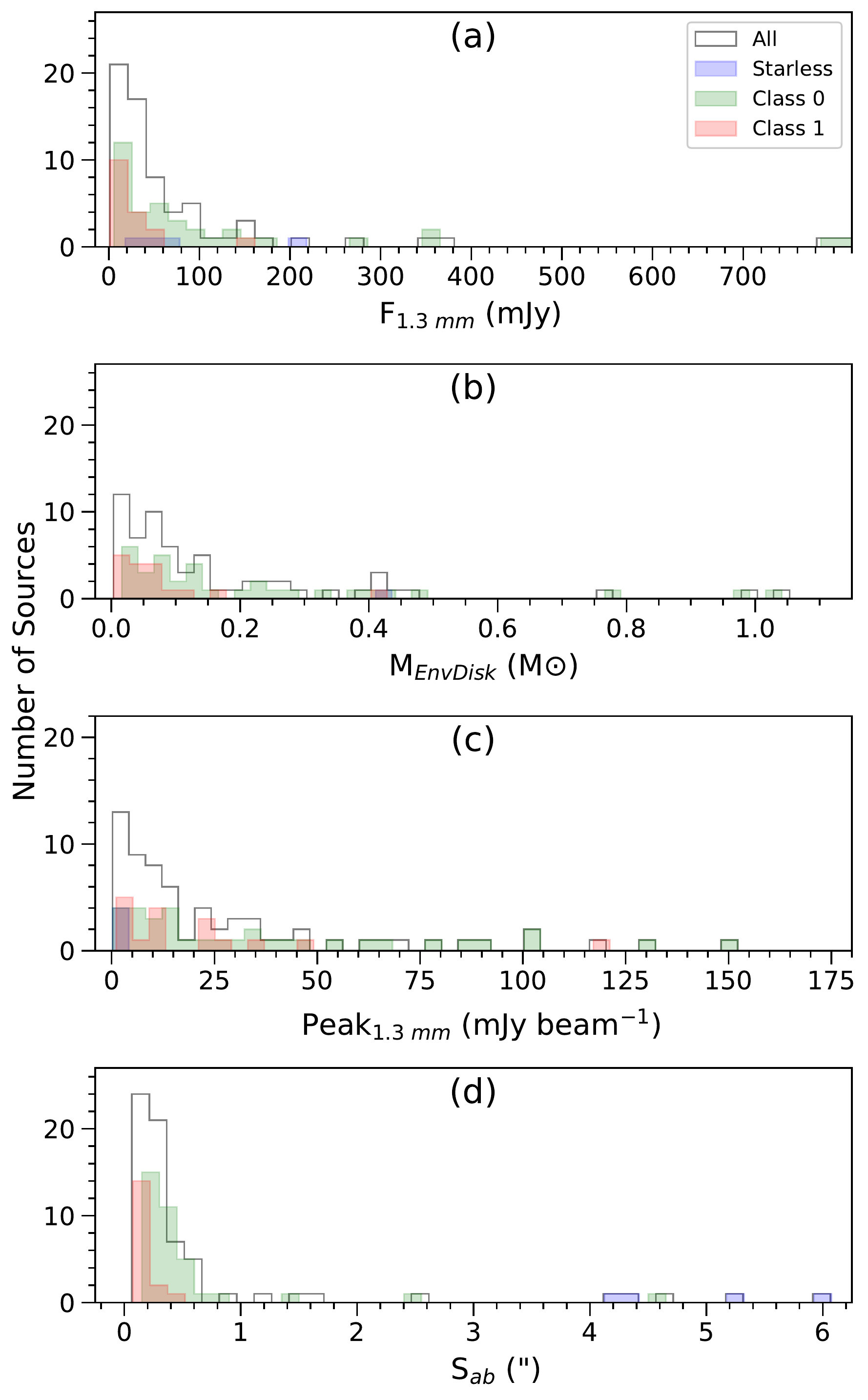}{0.45\textwidth}{}
\caption{Histograms of (a) integrated flux densities (b) envelope+disk mass (c) peak emission and (d) geometrical sizes derived with 2-D Gaussian fitting of 1.3 mm continuum emission for all the sources (black steps), which includes starless (blue), Class 0 (green), Class I (red) and unclassified sources. 
 }
\label{fig:continuum_statistics}
\end{figure}

The main science goal of the ALMASOP project is to study the fragmentation of these extremely young dense cores with high resolution 1.3\,mm continuum data from ALMA. We will investigate the substructures of starless cores and the multiplicities of protostellar cores. In this work, we only present the 1.3\,mm continuum images and briefly discuss the properties of the detected cores. We leave the detailed discussions of the substructures of starless cores and the multiplicities of protostellar cores to forthcoming papers.

Figure \ref{fig:example_continuum}a-d shows some selected examples of the 1.3\,mm continuum maps toward the dense cores with a typical resolution of $\sim$ 0$\farcs$35 ($\sim$ 140 au). The continuum maps reveal diverse morphologies of the dense cores. For example, Figure \ref{fig:example_continuum}(a) displays 1.3 mm continuum emission of G209.29-19.65S1, which is a candidate prestellar core. It shows an extended envelope that contains a dense blob-like structure. In Figure \ref{fig:example_continuum}(b), the compact core of G191.90-11.21S is likely protostar with a much brighter peak than the candidate prestellar core G209.29-19.65S1 (Figure \ref{fig:example_continuum}a) as it is surrounded by extended emission, and this source was later classified as Class 0 (section \ref{sec:sedProtostars}). Whereas Figure \ref{fig:example_continuum}(c) contains the compact emission of G205.46-14.56S3 with a relatively fainter surrounding envelope than typical Class 0, and this source was later found to be a Class I source (section \ref{sec:sedProtostars}). Some protostellar continuum structures exhibit close multiplicity on the present observed scale, as shown in Figure \ref{fig:example_continuum}(d). 

The full 1.3\,mm continuum images for targets in $\lambda$-Orionis, Orion A and Orion B GMCs are presented in Figures \ref{cont1}, \ref{cont2} and \ref{cont3}, respectively.

Out of 72 targets, 48 have been detected in the combined 3-configurations ($\sim$ 66 \%), where a total of 70 compact cores have been revealed including the multiple systems. In the other 24 targets, there is either no emission or only 3$\sigma$ level emission in the combined TM1+TM2+ACA continuum maps, where the dense cores may have larger sizes than the maximum recoverable size (MRS $\sim$ 14$\arcsec$) of the combined data, although they could be detected in ACA maps (MRS  $\sim$ 25$\arcsec$).
As an example, Figure \ref{fig:example_ContComp_COMB_ACA} (left panel) does not display significant emission in its combined map, although we can see significant emission in ACA only (right panel of Figure \ref{fig:example_ContComp_COMB_ACA}). We, therefore, checked those targeted positions in ACA only (see Figure \ref{ACAcont}), which reveals an additional 10 detections ($>$ 5$\sigma$). Thus from the present survey, we are able to detect the emission of 80\% of the targeted sources (58, out of 72).  

We performed one component two-dimensional Gaussian fitting in TM1+TM2+ACA maps within the 5-sigma contour level to those 70 core structures detected in the combined configurations. Here we do not compare the measurement from ACA-only detections due to different resolutions and these results of ACA configurations will be presented in a separate paper. The fitting parameters are listed in the Table \ref{tab:ContinuumProtoStellarProperties}, 
 which includes deconvolved major axis, minor axis, position angle, integrated flux density (F$_{1.3~mm}$), and peak flux (Peak$_{1.3 ~ mm}$). The source sizes\footnote{Here, these sizes are analogous to diameters of the sources.} (S$_{ab}$) were obtained from the geometrical mean of major and minor axes (i.e., S$_{ab}$ = $\sqrt{major \times minor}$).

Assuming optically thin emission, the (gas and dust) mass of the envelope+disk can be roughly estimated using the formula
\begin{equation}
M_{EnvDisk} \sim \frac{D^2 F_\nu}{B_\nu (T_{dust}) \kappa_\nu}
\end{equation}
where D is the distance to the sources, which is $\sim$ 389 $\pm$ 3, 404 $\pm$ 5, and 404 $\pm$ 4 pc for Orion A, Orion B and $\lambda$-Ori sources, respectively \citep{2018AJ....156...84K}. B$_\nu$ is the Planck blackbody function at the dust temperature T$_{dust}$, F$_\nu$ observed flux density, and $\kappa_\nu$ is the mass opacity per gram of the dust mass. We assume the dust temperature as 25 K for candidate protostellar disk-envelopes\footnote{The protostellar systems may show different dust temperatures of the envelope+disk system based on the stellar luminosity. If these sources also have an extended but colder envelope, the mass of the cold envelope will be underestimated by this assumption of warm temperature. For instance, if we vary the temperature from 15  to 100 K of the protostars, the masses will change by a factor of 1.7 to 0.25 times of the present estimated masses at 25 K.} \citep{Tobin2020} and 6.5 K\footnote{Due to the heating effect from the environment, the temperature of the starless core is relatively higher ($\sim$ 10 K) than the denser inner part \citep[e.g.,][]{2007ARA&A..45..339B,2019MNRAS.487.1269S}. When the starless cloud collapses and density increases at the central region (as in the prestellar core) then the temperature can reach as low as $\sim$ 6.5 K at the central dense portion.} for candidate starless cores \citep{2007A&A...470..221C,2019ApJ...874...89C}.
 Taking a gas-to-dust mass ratio of 100, the theoretical dust mass opacity at 1.3 mm is considered as  $\kappa_\nu$ = 0.00899($\nu$/231 GHz)$^\beta$  cm$^2$ g$^{-1}$ \citep{2018ApJ...863...94L} in the early phase for coagulated dust particles with no ice mantles  \citep[see also, OH5: column 5 of][]{1994A&A...291..943O}, where  we assume the dust opacity spectral index, $\beta$ = 1.5 for this size scale.
Table \ref{tab:ContinuumProtoStellarProperties} lists the estimated masses from these analyses.

Figure \ref{fig:continuum_statistics}a-d (black steps) shows the distribution of all the measured $F_{1.3 \rm mm}$, M$_{EnvDisk}$, Peak$_{1.3 ~ mm}$ and S$_{ab}$ with median values of 32.10 mJy, 0.093 M$_{\sun}$, 14.33 mJy beam$^{-1}$, 0$\farcs$27, respectively. More than 80\% of this sample have 1.3 mm flux densities $<$ 100 mJy, peak fluxes $<$ 50 mJy beam$^{-1}$ and average sizes $<$ 0$\farcs$6. Note that the ALMA emission peaks (Table \ref{tab:ContinuumProtoStellarProperties}) 
 are shifted from JCMT peaks (Table \ref{tab:targetobserved}), which is mainly due to resolution difference of the two telescopes.

\subsection{Outflow and Jet profiles}

\begin{figure}
\fig{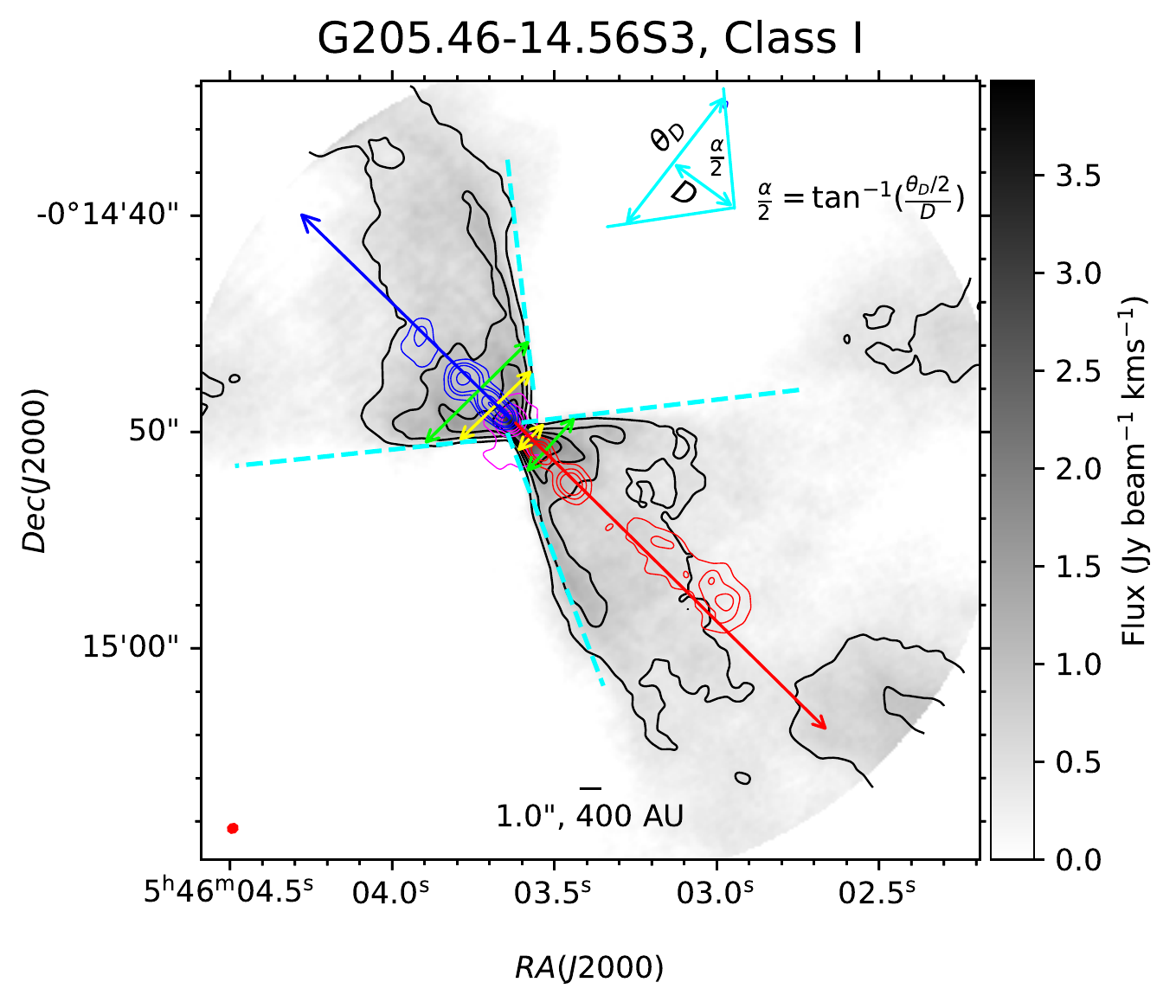}{0.5\textwidth}{}
\caption{Example of molecular outflow detected at ALMA $^{12}$CO(2-1) (grey) is shown for a Class I source G205.46-14.56S3. The black contours are at 3n$\sigma$, where n = 1, 2, ..... and $\sigma$ = 0.14 Jy/beam km/s. The blue and red arrows indicate the blueshifted and redshifted emissions, respectively. The magenta contours are 1.3 mm continuum emission at levels 6$\times$(1, 3, 8, 16)$\sigma$, where $\sigma$ = 6 $\times$ 10$^{-5}$ Jy beam$^{-1}$. The blue and red contours are blue- and redshifted integrated SiO(5-4) emission at 3$\times$(1, 2, 3, 6, 9)$\sigma$, where $\sigma$ = 0.03 Jy beam$^{-1}$.  The average tangents through the 3$\sigma$ outermost contours at $\sim$ 1$\arcsec$ and $\sim$ 2$\arcsec$ from the continuum peak  are drawn in cyan dashed lines. The yellow and green double headed arrows indicate the opening angle width $[\Theta_{obs}]_{400}$ and $[\Theta_{obs}]_{800}$, respectively, which are at different distance of $\sim$ 1$\arcsec$ and $\sim$ 2$\arcsec$ from the continuum peak, respectively.  A schematic of opening angle ($\alpha$) measurement is also shown (see text for details). 
}
\label{fig:example_outflow}
\end{figure}

\begin{figure}
\fig{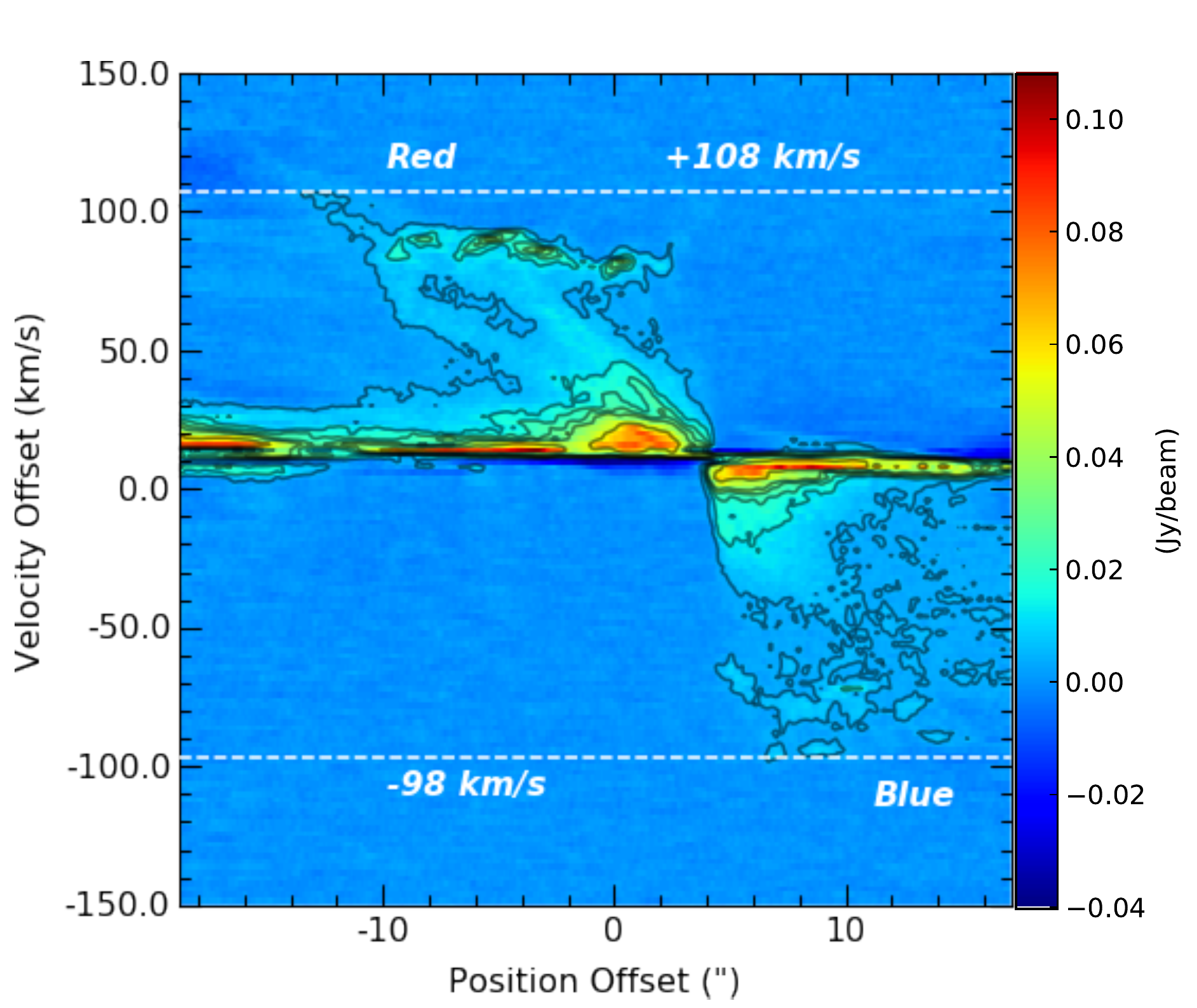}{0.5\textwidth}{}
\caption{Position-Velocity diagram of $^{12}$CO molecular outflow emission along jet-axis for G205.46-14.56S3. The black contour levels are at 3$\times$(1, 2, 3, 4, 6, 10, 15)$\sigma$, where $\sigma$ = 0.001 Jy beam$^{-1}$. The systemic velocity of the source is $\sim$ +12$\pm$4 km s$^{-1}$. Prominent nearly-continuous emission can be seen up to $-$98 and $+$108 km s$^{-1}$ in the blue- and redshifted lobe, respectively. Including the near-source overlapping blue- and resdhifted emission, the velocity extents are obtained as $\Delta$V$_B$ $\sim$ 114 km s$^{-1}$ and $\Delta$V$_R$ $\sim$ 106 km s$^{-1}$ for blue- and redshifted lobes, respectively. 
}
\label{fig:example_PositionVelocity}
\end{figure}

\begin{figure*}
\fig{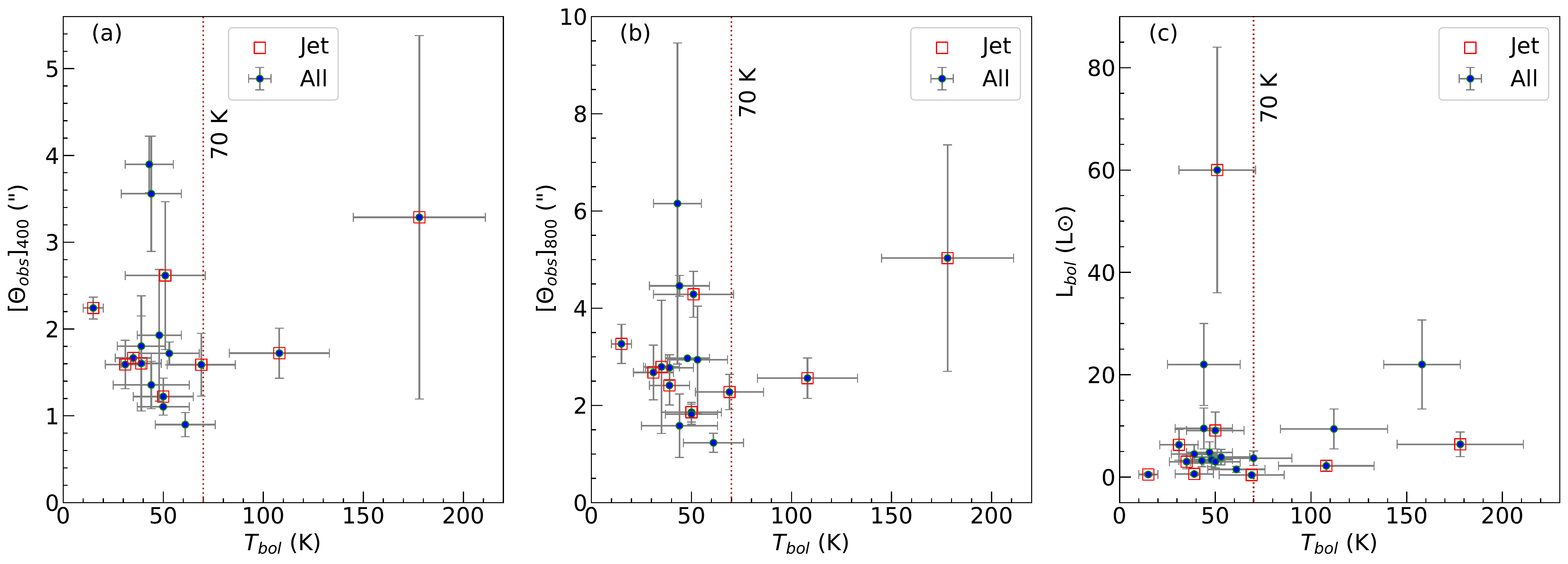}{0.99\textwidth}{}
\caption{ Opening angle $\Theta_{op}$ ($\arcsec$), that is the average width of blue- and redshifted outflow cavity (a) at $\sim$ 400 AU and (b) at $\sim$ 800 AU from continuum peak, as a function of T$_{bol}$ (K) for the protostars of the survey sample. (c) L$_{bol}$(L$_{\sun}$) as function of T$_{bol}$ (K). The blue data points with grey error bars represent all the outflow sources having a good detection in both, blue and redshifted, outflow lobes. The red squares indicate the sources with SiO knot detection (i.e., jet emission). The dotted vertical lines in all three panels are indicating T$_{bol}$ = 70 K, a boundary between Class 0 to Class I sources (see text for details).}
\label{fig:openingAngle_temperature}
\end{figure*}

\begin{figure}
\fig{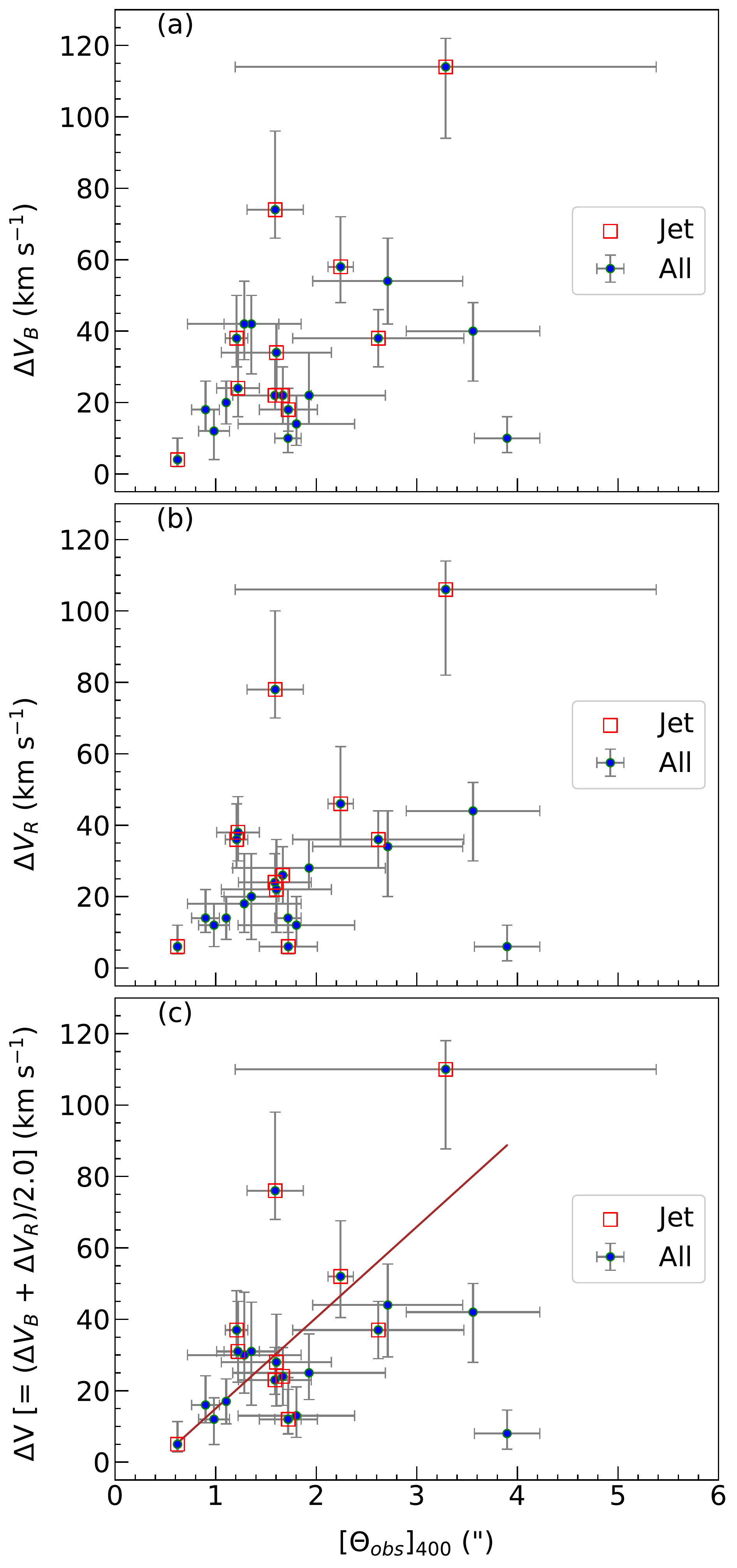}{0.4\textwidth}{}
\caption{Maximum outflow velocity ($\Delta$V) for (a) blueshifted, (b) redshifted and (c) average of both velocity components as a function of $[\Theta_{obs}]_{400}$ ($\arcsec$). The symbols are same as Figure \ref{fig:openingAngle_temperature}. The linear regression is shown with brown line in panel (c).
}
\label{fig:VelocityDispersion_OpAngle}
\end{figure}

\begin{figure}
\fig{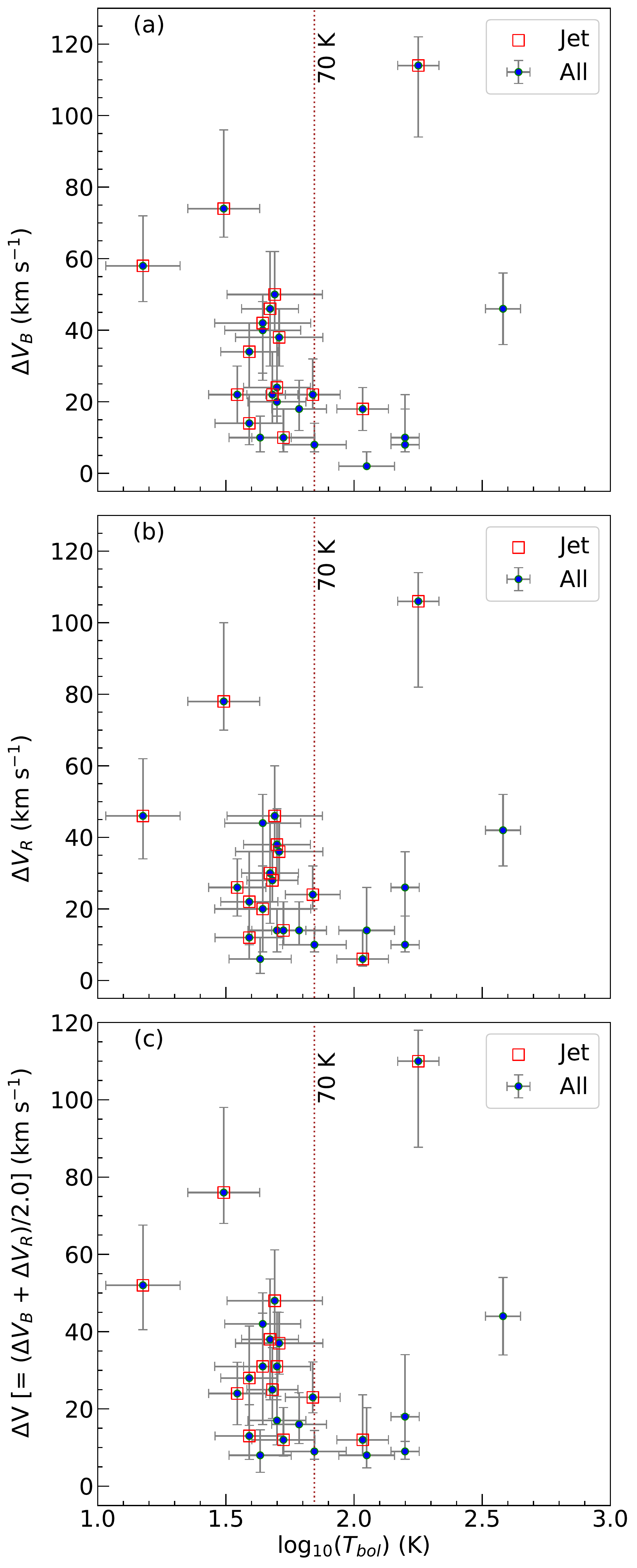}{0.38\textwidth}{}
\caption{Maximum outflow velocity  ($\Delta$V) for (a) blueshifted, (b) redshifted and (c) average of both velocity components as a function of T$_{bol}$. The symbols are same as Figure \ref{fig:openingAngle_temperature}. The majority of the Class 0 sources (i.e., T$_{bol}$ $<$ 70 K) follow an increasing trend in all three panels.}
\label{fig:VelocityDispersion_BolT}
\end{figure}

\begin{figure*}
\fig{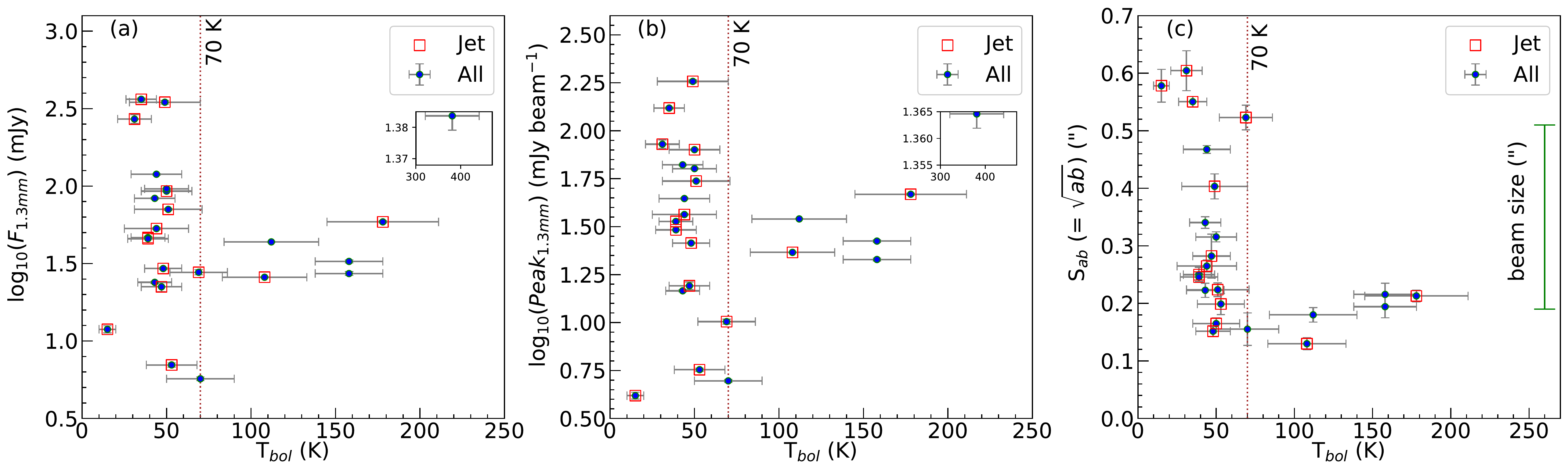}{0.98\textwidth}{}
\caption{(a) Flux densities, (b) peak, (c) deconvolved size at 1.3 mm from 2-D gaussian fitting as a function of T$_{bol}$. The symbols are same as Figure \ref{fig:openingAngle_temperature}. The y-axis error bars are shown at inset-figures in the panels (a) and (b).  A typical beam size is shown in panel (c).
}
\label{fig:BolT_int_peak_size}
\end{figure*}

The ALMASOP project will investigate the jet launching mechanisms and the evolution of outflows in the earliest phases, i.e., Class 0 stage, of star formation. Using the $^{12}$CO(2-1) and SiO(5-4) transitions at $\sim$ 0$\farcs$35 ($\sim$ 140 AU) angular resolution, we have performed a systematic search for low-velocity outflow components and high-velocity collimated jet components driven by protostellar objects.

\subsubsection{Outflow components from CO emission}\label{sec:resultsOutflowComponentsCOemission}

One common way to distinguish young protostars from a sample of the dense cores embedded in the molecular cloud is to identify the molecular outflowing gas in the lower rotational transition $^{12}$CO (2-1). We have traced such blue- and redshifted outflow wings through visual inspection of velocity channel maps and their spectra. An example of a bipolar $^{12}$CO outflow total intensity map integrated over the full blueshifted and redshifted velocity range is shown in Figure \ref{fig:example_outflow} for the source G205.46-14.53S3. The blue- and redshifted components (grey color and black contours) shows V-shaped structures toward the NE and SW directions, respectively. The 1.3 mm continuum (magenta contours) exhibits a compact continuum with its continuum inner core ($>$ 20$\sigma$ in Figure \ref{fig:example_outflow}) nearly elongated in a direction nearly perpendicular to the outflow axis.

The velocity extents of the blue- and redshifted lobes are selected from the channel where it appears for the first time at 3$\sigma$ level, to the channel of disappearance at the same the 3$\sigma$ limit \citep[e.g.,][]{1992A&A...261..274C,2015A&A...576A.109Y}. As an example, Figure \ref{fig:example_PositionVelocity} shows the position-velocity (PV) diagram, derived along the outflow axis. The object systemic velocity is likely 12 $\pm$ 4 km s$^{-1}$. The maximum outflow velocity or extent of the blue component is estimated as $\Delta$V$_B$ = 114$^{+8}_{-24}$ km s$^{-1}$, where the redshifted components have a velocity extent of $\Delta$V$_R$ = 106$^{+8}_{-24}$ km s$^{-1}$, without any inclination correction. The average velocity extent ($\Delta$V) is estimated from both components. We have identified 37 outflow sources with CO emission wings. The extents of both blue- and redshifted lobes observed in CO are tabulated in Table \ref{tab:ContinuumProtoStellarProperties}. 
However, these $\Delta$Vs are the lower limits in small field-of-view (FOV) of our combined configuration maps, and we do not know the actual spatial extension  of the outflow wings. The CO outflow images for all the protostellar samples are shown in Figure \ref{alloutflows}.

These velocity extents are different for blue- and redshifted lobes with high uncertainties, which could be due to the missing short velocity-spacing on both ends of the lobes in the present poor velocity-resolution observations, unknown inclination angle, complex gas dynamics of ambient clouds, or global infall in the protostar bearing filaments. In some cases, such as G211.01-19.45S, the outflow is identified as monopolar where the other part could be disregarded due to low velocities, or confused with emission from other sources. Estimated $\Delta$Vs range from 4 to 110 km s$^{-1}$, with a median value 26.5 km s$^{-1}$. In some cases, complex structures are observed, where it is difficult to distinguish the outflow wings from the complex cloud environment (marked ``cx" in Table \ref{tab:ContinuumProtoStellarProperties}).
These sources can not be ruled out from the outflow candidates, and further investigations are needed at high velocity and spatial resolution with numerical analysis to extract their features from the cloud dynamics.

\subsubsection{Identification of high velocity knots}\label{sec:resultsjetComponentsSiOemission}
The large impact of the Orion cloud kinematics on the outflows makes it difficult to elucidate the original outflow morphology in CO(2-1) tracer. SiO(5-4) has been found to provide more insights into the outflow chemistry \citep{2016A&A...595A.122L}. The excitation conditions of the SiO(5-4) emission line have a high critical density of (5-10) $\times$ 10$^6$ cm$^{-3}$ \citep{2020arXiv200205720N}, which could be reached in high-density knot components. The collimated jets frequently appear as a series of knots, which are interpreted as made by the internal shocks originated by episodic accretion/ejection at the protostellar mass-loss rate \citep[][]{1991A&A...251..639B}. 
 An example of blue- and redshifted SiO emission is shown in Figure \ref{fig:example_outflow}. The identification of the jet-components is marked in Table \ref{tab:ContinuumProtoStellarProperties} (column 14), and these sources are considered as jet sources throughout the paper.

 Out of 37 outflow sources, 18 ($\sim$ 50\%) are detected having knots in the SiO line emission within the CO outflow cavities. Additionally, two non CO emitting sources are also identified with SiO emission, where CO emission is possibly non-detectable due to complex cloud environment, as discussed above. High-mass molecular clumps are reported to have $\sim$ 50-90\% jet detection in low-angular resolution surveys in SiO(2-1), (3-2), (5-4) emission lines \citep[e.g.,][]{2016A&A...586A.149C,2019ApJ...878...29L,2020arXiv200205720N}. It is to be noted that the high-density shock components could also be detected in more high-density tracers e.g., SiO (8-7). So the higher transitions of SiO could reveal more knot ejecting sources. Additionally, the knot tracers may vary with the evolution of the protostars \citep{2020A&ARv..28....1L}.

\subsubsection{Outflow Opening Angle}
Among the main characteristics of outflows, opening angle ($\alpha$) is one of the less-explored observational parameters to date. In the low-velocity regime, the CO delineates two-cavity walls open in the blue and redshifted directions.  Measuring the $\alpha$ is quite complicated for the sources with no well-defined cavity walls throughout the full observed extent due to the presence of a complex cloud environment (e.g., G200.34-10.97N, G205.46-14.56S1, G209.55-19.68S1), or secondary outflows (e.g., G209.55-19.68N1) (see Appendix, Figure \ref{alloutflows}).
For both, blue- and redshifted directions, if the conical structures appear to be symmetrical, then one can find the apex by extrapolating the cavity boundaries \citep[e.g.,][]{2014ApJ...780...49W}. However, the real complexity of finding the apex position appears for asymmetrical outflow lobes, even if we assume the continuum peak to be the apex position, the tangent will be needed to  allow us to trace back to that apex location. Hence we may miss a significant fraction of the cavity-width near the source. In that case, we also do not know the outflow-launching radius for the source, which essentially varies from source to source. Thus, we adopt a consistent approach for all the sources, where the outflow cavity width ($\Theta_{obs}$) is measured perpendicular to the outflow axis.

Firstly, the outflow axis of each lobe is derived from their knot structures in SiO emission (Figure \ref{fig:example_outflow}).
For the sources having no SiO emission, CO-jets are utilized to find the jet-axis from the dense CO-emission near the middle of the outflow cavity walls. Some of the sources show neither SiO knots nor CO jets; in those cases, their outflow axis was assumed to be in the middle of the outflow cavity. Secondly, we draw an average tangent at the outermost 3$\sigma$ contours at the local point of consideration (cyan dashed lines in Figure \ref{fig:example_outflow}). Now, the width perpendicular to jet-axis of the 3$\sigma$ cavity wall at 1$\arcsec$ (i.e., $[\Theta_{obs}]_{400}$ at $\sim$ 400 au; yellow double headed arrow) and 2$\arcsec$ (i.e., $[\Theta_{obs}]_{800}$  at  $\sim$ 800 au; green double headed arrow) distance from continuum peak represents the opening angle at the corresponding distance from the stellar core.
 As shown in the schematic diagram on top of Figure \ref{fig:example_outflow}, if the opening angle width is measured as $[\Theta_{obs}]_{D}$ at a distance D from the continuum peak, from right angle trigonometry the half of opening angle is, $\frac{\alpha}{2}=\tan^{-1} (\frac{[\theta_{obs}]_{D}/2}{D})$. We also measured $[\theta_{obs}]_{D}$ at distances $>$ 2$\arcsec$, and found that $\alpha$ measurements are quite consistent for the outflows with well-defined cavity walls. However, we prefer to present $[\theta_{obs}]_{D}$ close to the source, i.e. at 1$\arcsec$ and 2$\arcsec$, for all the sources to minimize the environmental effects on the measurements, and as shown in Figure \ref{fig:openingAngle_temperature}a-b, the overall trends of $[\theta_{obs}]_{D}$ with T$_{bol}$ remain the same for both the distances. The exact envelope boundaries and other environment effects towards each of the outflow lobes are also unknown, which could lead to unequal deformation on both the outflow lobes. Thus, we have taken an average of blue- and redshifted opening angles to measure the final $\Theta_{obs}$ to reduce the unknown contamination. 
From the present analyses, we are able to estimate $[\Theta_{obs}]_{D}$ of 22 outflow sources, and the values of the final $[\Theta_{obs}]_{D}$ are listed in Table \ref{tab:ContinuumProtoStellarProperties}.

The CO outflow cavities have an opening angle width at 1$\arcsec$ ($\sim$ 400 AU) ranging from 0$\farcs$6$-$3$\farcs$9 (i.e., typically $\alpha$ = 33$\fdg$4 $-$ 125$\fdg$7 near the source) with a median value 1$\farcs$64. The median value for 19 Class 0 sources is 1$\farcs$60 and 3 Class I sources is 2$\farcs$70 (see section \ref{sec:sedProtostars} for objects classification). 

These measured quantities of opening angles are not corrected for inclination angle, $i$.
 As in Figure \ref{fig:example_outflow}, the continuum emission is apparently shifted towards the the blueshifted lobes, which is most probably an inclination effect, and at the same distance from the continuum peaks, the blue lobes appear wider than the red lobes.
Measuring inclination angle needs well-defined outflow cavity walls with their full spatial extent. We, therefore, need high-velocity resolution and wide field-of-view for the outflows, which we lack in the present datasets. Note that, we need to define the exact shell structure to estimate the real-age opening angle, for a rotating outflow it is complex to search the corresponding shell cavity in low-velocity resolution observations. In such cases, we assume the outer boundary as the outflow shell, which introduces error in the $\Theta_{obs}$. 
Thus, theoretical models are necessary to reduce the environmental effects of complex cloud dynamics, envelope emission, and interacting outflows.  Further high velocity resolution and single dish observations are also very important to determine the envelope boundary and inclination angle.

\subsection{Protostellar Signatures}\label{sec:sedProtostars}

\subsubsection{Multiwavelength catalog}
 The surrounding envelopes are dissipated during protostellar  evolution. They gradually appear from sub-mm, mid-infrared (MIR) to near-infrared (NIR) wavelengths hence they become less sensitive to 1.3 mm emission. Thus, we searched for the sub-mm, MIR and NIR  counterpart of each dense core in the archived Two-Micron All-Sky Survey \citep[2MASS;][]{2003tmc..book.....C}, UKIRT Infrared Deep Sky Survey \citep[UKIDSS;][]{2007MNRAS.379.1599L}, {\it Spitzer Space Telescope} survey of Orion A-B \citep{2012AJ....144..192M}, Wide-field Infrared Survey Explorer \citep[WISE;][]{2010AJ....140.1868W}, AKARI \citep[][]{2015PASJ...67...50D}, Herschel Orion Protostellar survey \citep[HOPS; ][]{2013ApJ...767...36S,2015ApJ...798..128T}, Atacama Pathfinder Experiment \cite[APEX;][]{2013ApJ...767...36S}, the 850 $\mu$m JCMT  \citep[][]{2018ApJS..236...51Y}. In addition to these catalogs, we include our present ALMA 1.3 mm emission to estimate a more accurate bolometric temperature (T$_{bol}$)  and luminosity (L$_{bol}$) than that of \cite{2018ApJS..236...51Y}.

 The final multiwavelength catalogue was obtained by cross-matching all the catalogues described above. Initially, we adopted a matching radius of r$_{m}$ $\sim$ 3$\arcsec$ for all the catalogues \citep[see also][for details]{2015MNRAS.454.3597D}, which best suits the relatively high resolution catalogues, 2MASS, UKIDSS, $Spitzer$ and ALMA. For the relatively poor resolution catalogues, WISE, AKARI, Herschel, APEX, JCMT, we further  checked the images within their corresponding resolution limits to consider the counterpart of an object.  For the possible close binary in the present analysis,  with the available observations, it is difficult to determine the exact source of infrared emission since the binary system is embedded in a common envelope. We therefore assigned the same  measurements to both protostars. The final cross-matched catalogue is presented in Table \ref{tab:sedcrossmatch}. Finally, the objects with good photometric accuracy  (signal-to-noise ratio: SNR  $>$ 10 for 2MASS, UKIDSS, $Spitzer$-IRAC and $Spitzer$-MIPS; SNR $>$ 20 for WISE and ALMA; SNR $>$ 50 for AKARI, JCMT, Herschel, APEX) were utilized for the further analyses \citep[e.g.,][]{2018ApJ...864..154D}. For the HOPS fluxes, we adopted the uncertainty flags as provided in \citet[][]{2016ApJS..224....5F}.

The T$_{bol}$ and L$_{bol}$ were estimated with trapezoid-rule integration over the available fluxes, assuming the distance as $\sim$ 389 $\pm$ 3, 404 $\pm$ 5, and 404 $\pm$ 4 pc for Orion A, Orion B and $\lambda$-Ori sources, respectively \citep{2018AJ....156...84K}, and the measured values are listed in Table \ref{tab:ContinuumProtoStellarProperties}. Following \cite{1993ApJ...413L..47M}, the flux weighted mean frequencies in the observed spectral energy distributions (SEDs) were utilized to obtain T$_{bol}$.  We assume T$_{bol}$ = 70 K as a quantitative transition temperature from Class 0 to Class I \citep[e.g.,][]{1995ApJ...445..377C}.  Our distribution of  T$_{bol}$ and L$_{bol}$ are close to the measured values of the HOPS catalog \citep[][]{2016ApJS..224....5F}, the HOPS IDs are marked in column 18 of Table \ref{tab:ContinuumProtoStellarProperties}. Some differences are expected since we are using [additional] mid infrared data not included in the HOPS catalog. For some sources, the mid-infrared observations (e.g., AKARI and Herschel) are not available, therefore our measurements should give the lower limit for those sources \citep[][]{2012AJ....144...31K}.

The distribution of T$_{bol}$ can be seen in Figure \ref{fig:openingAngle_temperature}(a), (b) (see also Figure \ref{fig:VelocityDispersion_BolT} and Figure \ref{fig:BolT_int_peak_size}). 
Figure \ref{fig:openingAngle_temperature}(c) shows the distribution of L$_{bol}$ with the T$_{bol}$ of our protostellar sample. Two separate wings are prominent in Figure \ref{fig:openingAngle_temperature}(c), where the nearly horizontal wing represents the increment from Class 0 to Class I sources. The nearly vertical wing possibly originates from the combined luminosity of multiple stellar components, since they possess a common envelope and the present available infrared resolution is not enough to distinguish their emission components. We estimated the bolometric temperature of 53 sources, those having 5 or more wavelength detections, which also includes all sources in multiple systems.

\subsubsection{Outflows in protostellar candidates}
 The detection of infrared emission could be biased by the high-background emission from the ambient cloud. In addition, Herschel does not have coverage of all the Orion dense cores. Hence, some of the protostars in this ALMASOP sample could not be detected from the infrared only catalog. Outflows are another potential tool to identify protostars. As such, eight sources (G192.32-11.88N, G205.46-14.56M1$\_$B, G205.46-14.56S1$\_$B, G208.68-19.20N3$\_$A, G208.89-20.04W, G209.55-19.68N1$\_$A, G211.47-19.27N$\_$B, G215.87-17.62M$\_$A) are not listed in the infrared catalog, however they have bipolar CO outflows. We consider these sources are likely young Class 0 sources. However, the complex cloud dynamics prevent the detection of less extended and evolved outflows in CO (2-1), which are marked as ``cx" in Table \ref{tab:ContinuumProtoStellarProperties}.

Finally, we classify 56 sources based on T$_{bol}$ estimation and outflow detection. Out of them, 19 are candidate Class I sources, the other 37 sources are candidate Class 0 sources. However, higher resolution multi-band infrared observations would more effectively refine the classification. For some sources in multiple systems (e.g., G196.92-10.37$\_$C, G205.46-14.56M2$\_$A and G206.93-16.61E2$\_$A - D), we obtain T$_{bol}$, but there are no clear signatures of outflows. The infrared emission for those sources are also confusing with others. These sources are not classified in this paper.

\subsection{Candidates for Class 0 Keplerian-like disks}
 
\begin{figure*}
\fig{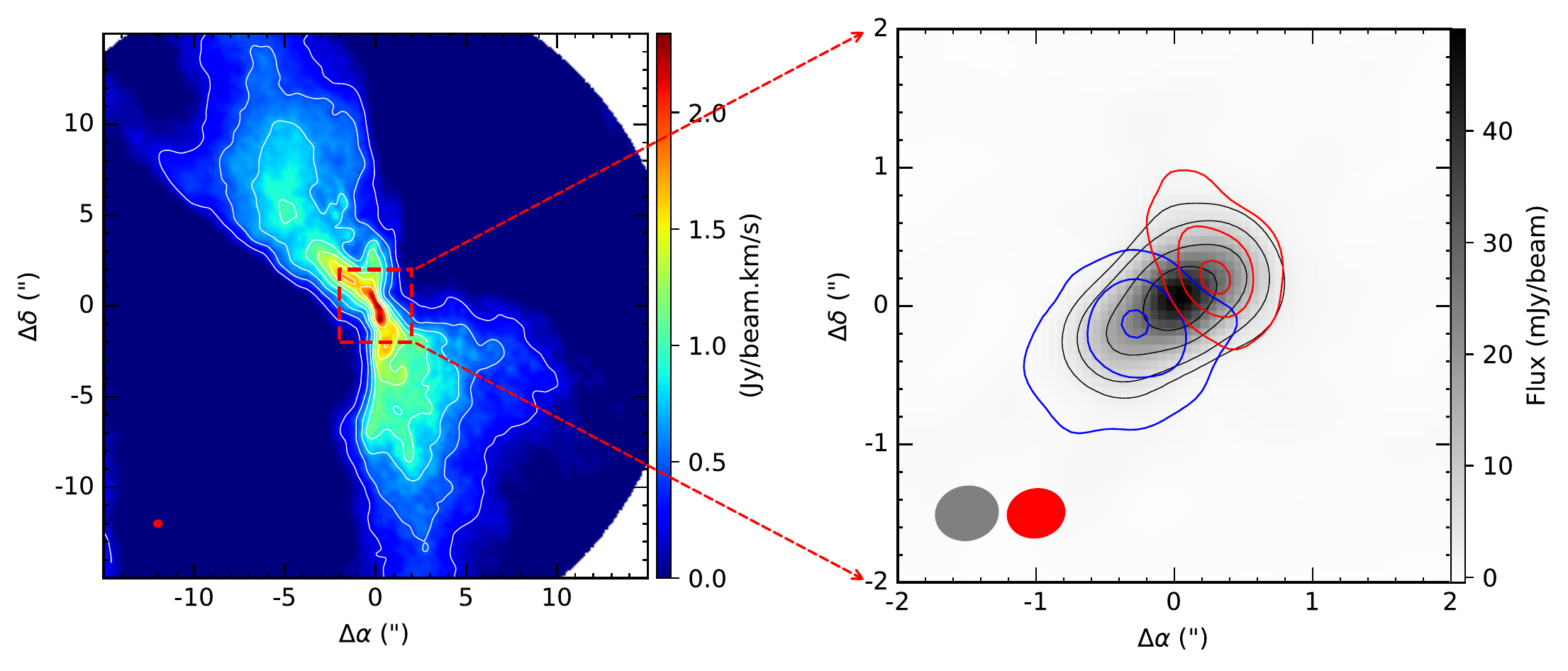}{0.98\textwidth}{}
\caption{(left panel) ALMA $^{12}$CO(2$–$1) integrated intensity (moment zero) color-scale map of the source G192.12-11.10.  The white contours start from 10\% to 70\% in steps of 10\% of the intensity peak. The CO intensity peak
is 2.3 Jy beam$^{-1}$ km s$^{-1}$. The synthesized beam size is shown in the bottom left corner in red.   (right panel) A zoomed view of the central part. The blueshifted component (blue contours), and the redshifted component (red contours) of C$^{18}$O(2$–$1) emission are overplotted  on top of 1.3\,mm continuum images. The blue- and red-contours are at 10, 20, 30$\sigma$, where the noise level is $\sigma$ $\sim$  0.017 Jy beam$^{-1}$ km s$^{-1}$. The gray scale is the 1.3\,mm continuum emission with contour levels at (n$^2$+1)$\times$50 $\sigma$, with $\sigma$ =  0.06  mJy beam$^{-1}$. The synthesized beam sizes are shown in the bottom left corner in gray (continuum) and red (C$^{18}$O). 
}
\label{fig:G19212_1110c18Odisk}
\end{figure*}

The ALMASOP project also aims to search for Keplerian-like disks surrounding Class 0 protostars. Figure \ref{fig:G19212_1110c18Odisk} presents a candidate for Keplerian-like disk surrounding a Class 0 protostar, G192.12-11.10. Its $^{12}$CO J=2-1 emission reveals a collimated bipolar outflow (see left panel of Figure \ref{fig:G19212_1110c18Odisk}). As shown in the right panel of Figure \ref{fig:G19212_1110c18Odisk}, the 1.3 mm continuum emission of G192.12-11.10 shows a flattened structure, that may be a candidate disk. The redshifted and blueshifted C$^{18}$O J=2-1 emission clearly shows a rotation pattern of the disk-like structure. We have identified a handful of disk candidates surrounding Class 0 protostars such as G192.12-11.10. The properties of these disk candidates will be discussed in a forthcoming paper (Dutta et al., in preparation).

\subsection{Chemical Signatures}

As illustrated in Table \ref{tab:spectralsetup}, the four spectral windows cover a suite of molecular species and transitions, most of which are of importance for the chemical diagnostics of young star-forming regions. The successful detection and imaging of these tracers enables the
analysis of chemical compositions of our diverse sample of objects from starless to young Class 0 and Class I protostellar cores.

\begin{figure*}
\fig{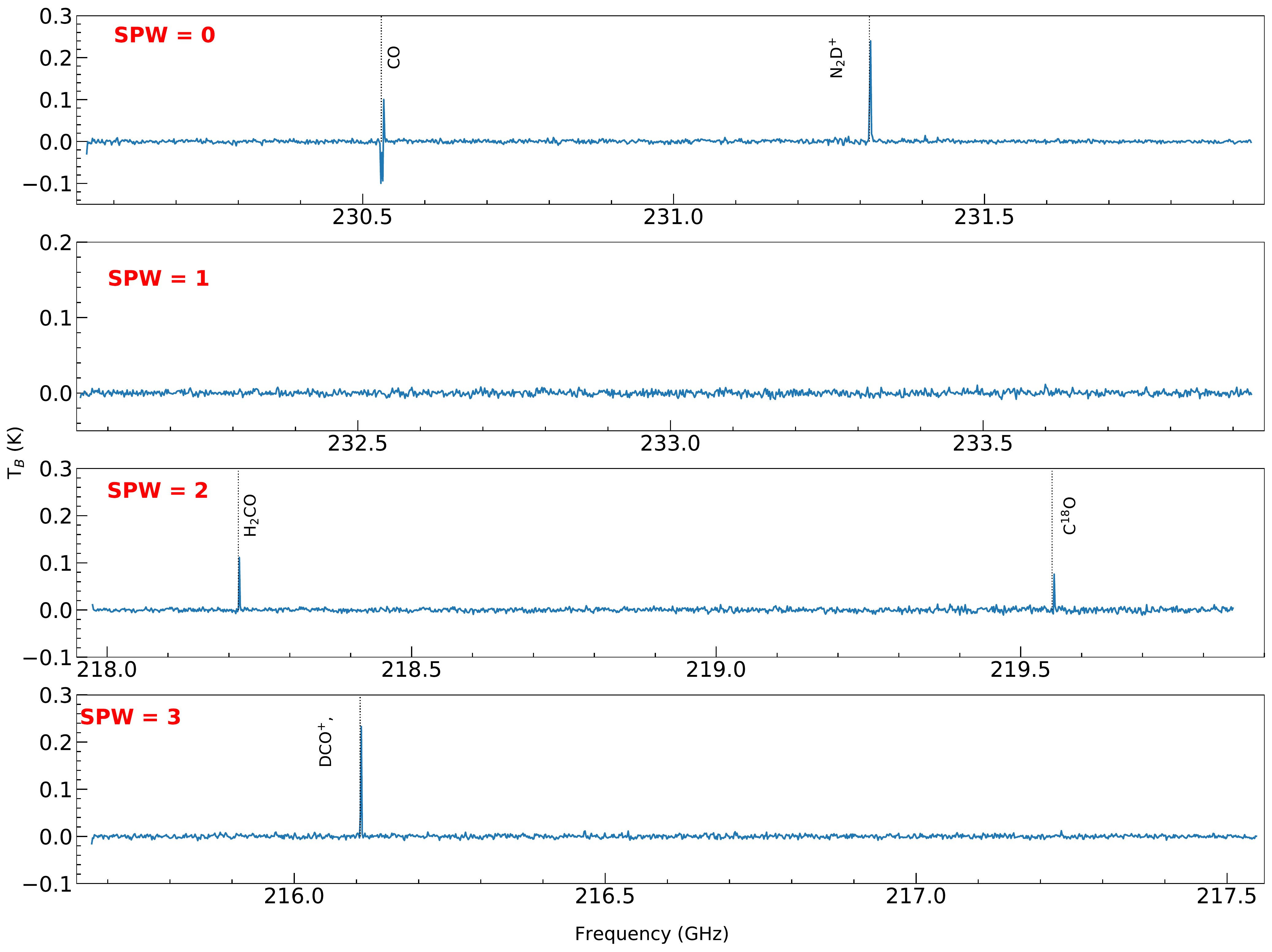}{0.7\textwidth}{}
\caption{Spectra of the starless dense core G209.29-19.65S1 in 4 spectral windows (SPWS; see Table \ref{tab:spectralsetup}) observed in combined TM1+TM2+7m-ACA configurations. Various identified lines are marked.
}
\label{fig:G20929_1965S1spectra_starless}
\end{figure*}

\begin{figure*}
\fig{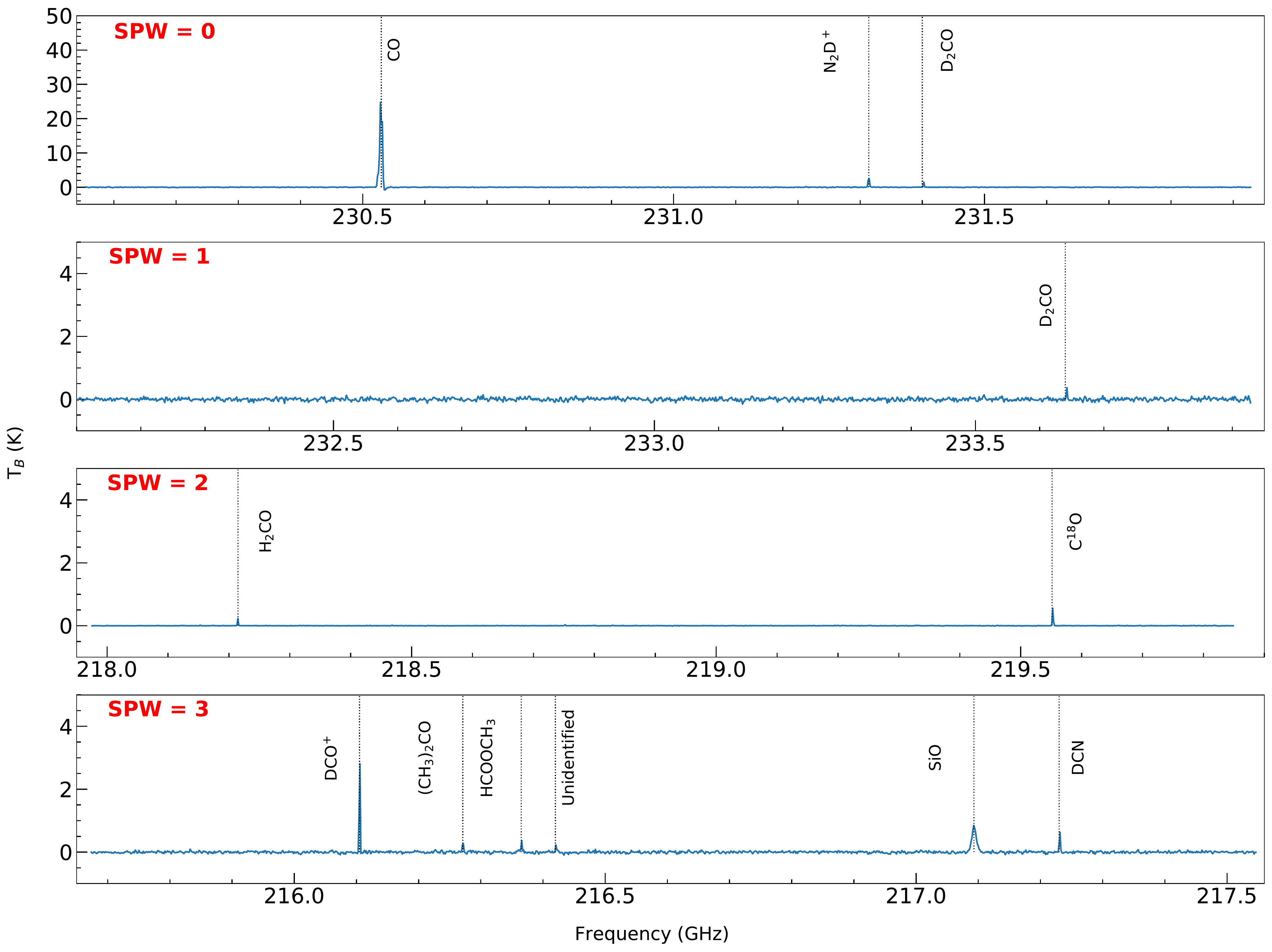}{0.7\textwidth}{}
\caption{Spectra of the protostellar object G191.90-11.21S. All SPWs, array configuration, line-identification are same as Figure \ref{fig:G20929_1965S1spectra_starless}
}
\label{fig:G19190_1121SspectraProtostars}
\end{figure*}

It has been suggested that the deuterium fraction increases at the cold starless core phase and then decreases as the protostar warms up the surrounding material in the protostellar phase \citep[e.g.,][]{2019ApJ...870...81T,Tatematsu2020}. As shown in Figure \ref{fig:G20929_1965S1spectra_starless} and \ref{fig:G19190_1121SspectraProtostars}, N$_{2}$D$^{+}$ and DCO$^{+}$ are detected toward both starless and protostellar cores. The emission morphology will aid in diagnosing their thermal structure and history, which will be discussed in forthcoming papers (Sahu Dipen et al. in preparation; Liu Sheng-Yuan et al. in preparation).

\begin{figure*}
\fig{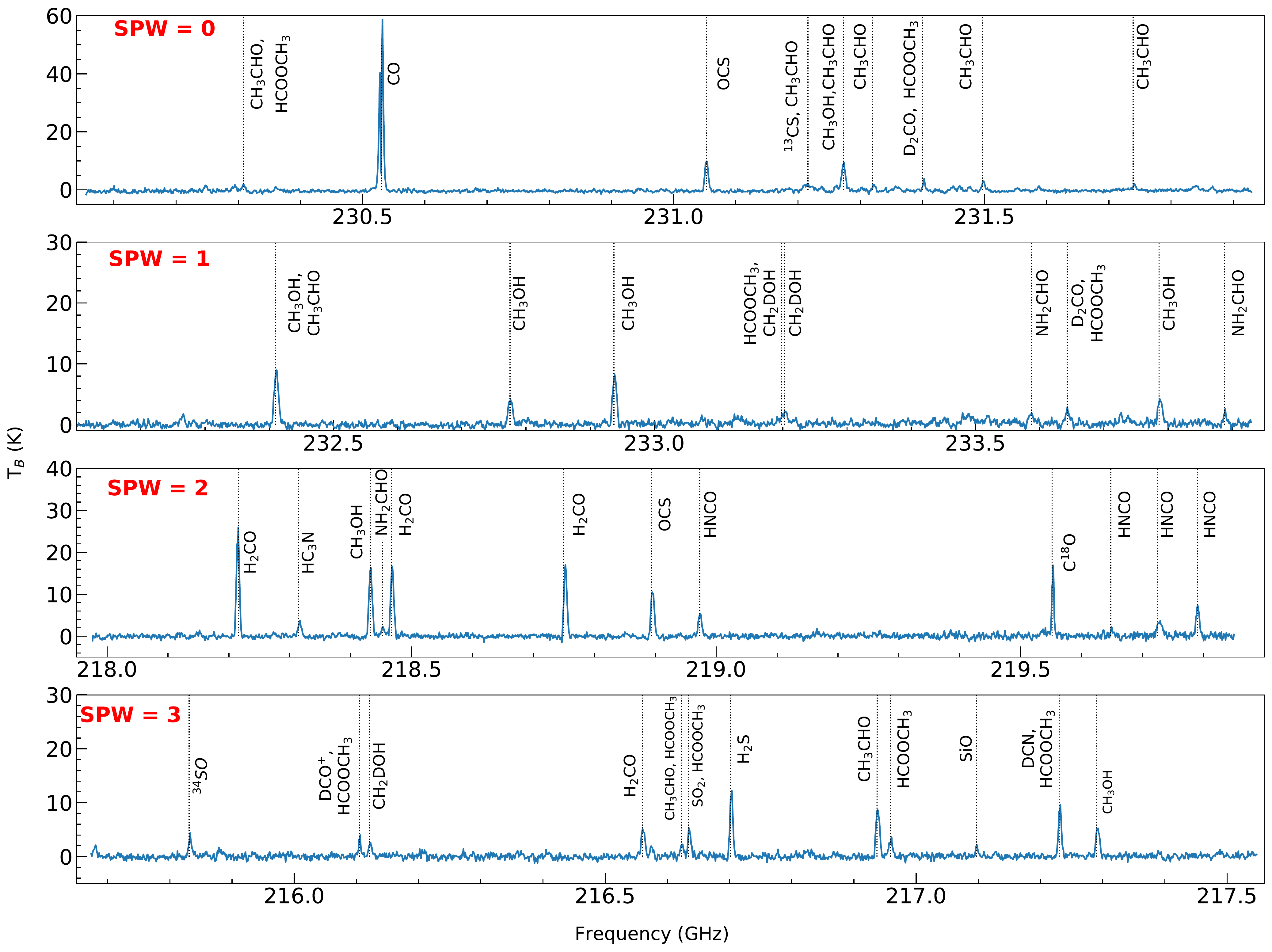}{0.7\textwidth}{}
\caption{Spectra of a line rich protostellar object G192.12-11.10 (``hot corino") as identified with 7m-ACA configuration. All SPWs, line-identification are same as Figure \ref{fig:G20929_1965S1spectra_starless}
}
\label{fig:G19212_1110spectra_hotcorino}
\end{figure*}

Some low- to intermediate mass Class 0/I protostars, dubbed ``hot corinos", exhibit considerably abundant saturated complex organic molecules (COMs: CH$_3$OH, H$_2$CO, HCOOCH$_3$, HCOOH) in the compact ($<$ 100 au) and warm ($\sim$ 100 K) regions immediately surrounding the YSO \citep[e.g.,][]{2004ApJ...616L..27K,2004ASPC..323..195C}, as shown in Figure \ref{fig:G19212_1110spectra_hotcorino}.  By utilizing our ACA 7m data, \cite{Hsu2020} has readily identified four new hot corino candidates (G192.12-11.10, G211.47-19.27S, G208.68-19.20N1, and G210.49-19.79W) in the sample. A more detailed study of hot corinos with high resolution 12-m array data will be presented in a forthcoming paper (Hsu Shih-Ying et al. in preparation)

As discussed in section \ref{sec:resultsOutflowComponentsCOemission} and \ref{sec:resultsjetComponentsSiOemission}, the outflow and jet components and their interaction with the core can be traced both in position and velocity by $^{12}$CO and SiO line emission. The other molecular species such as CS, C$^{18}$O, CH$_3$OH, C$_3$H$_2$, OCS, HCO$^{+}$ could be utilized to trace the dense structures underlying protostellar winds \citep[e.g.,][]{2005A&A...442..527M,2004A&A...415.1021J,2005MNRAS.361..244C,2007prpl.conf..245A,2020A&ARv..28....1L}. The molecular species available in observed spectra are displayed in Figures \ref{fig:G19190_1121SspectraProtostars} and \ref{fig:G19212_1110spectra_hotcorino}. The shock chemistry with ALMASOP data will be presented in a forthcoming paper (Liu Sheng-Yuan et al. in preparation).

\section{Discussion} \label{sec:discussionEvolutionCont_EvolutionOuflow}
\subsection{Evolution of the dense cores} \label{sec:discussionEvolutionDenseCores}

From the 1.3 mm continuum morphology of the 70 dense cores and their infrared counterpart, we perceived three categories; one, 48 dense cores are relatively compact in 1.3\,mm continuum with protostellar signatures, either low-velocity outflow, high-velocity jet, or infrared detections. In the second category, 4 dense starless cores exhibit extended emission and compact blobs (see Table \ref{tab:ContinuumProtoStellarProperties}). They are likely prestellar cores with substructures, and deserve detailed investigation. The physical and chemical properties of these 4 cores will be further discussed in forthcoming papers (Sahu et al., in preparation; Hirano et al., in preparation). In the third category, another 16 dense cores are not classified due to their complex cloud dynamics and confusing infrared detection. Moreover, out of 72 targeted JCMT positions, 24 show no emission in the combined TM1+TM2+ACA continuum maps. They are likely the starless cores with low density and with sizes larger than the maximum recoverable size, 
as discussed above (see section \ref{sec:resultsContOutflowCONT} and Figure \ref{fig:example_ContComp_COMB_ACA}). However, 10 out of the 24 starless cores are detected with ACA alone. The detailed properties of all the starless cores will be presented in a forthcoming paper (Sahu et al., in preparation).

Figure \ref{fig:continuum_statistics}a-d shows the histogram distribution of all types of sources, which includes starless, Class 0, Class I, and unclassified sources. The starless, Class 0, and Class I have median values of F$_{1.3 ~ mm}$ $\sim$ 59.65, 46.42, and 14.96 mJy, respectively, whereas the median values of M$_{EnvDisk}$ are 1.34, 0.13, 0.04 M$_{\sun}$, respectively. 
 A similar sequence was observed at 4.1 cm and 6.1 cm fluxes in \cite{2018ApJS..238...19T}, where Class 0 sources exhibit larger flux than Class I in both wavelengths. The geometrical sizes, S$_{ab}$ of the starless cores (deconvolved median size $\sim$ 4$\farcs$77) are found to be larger than Class 0 (median deconvolved size $\sim$ 0$\farcs$32) and Class I (median deconvolved size $\sim$ 0$\farcs$18). The Gaussian 2-D integrated flux and sizes of the dense cores basically depend on the power-law indexes, which vary from starless, Class 0 to Class I \citep[e.g.,][]{2019ApJ...879..101L}. So, the above outcomes could be interpreted as varying density profiles \citep[e.g.,][]{2019ApJ...887..209A}. The starless cores have a flat density distribution in the inner regions, so we get larger sizes and hence larger masses. On the other hand, the small sizes from Class 0 to Class I sources suggest that pseudodisk/disks are dominating the 1.3\,mm fluxes and the apparent mass-supplying radius of the continuum reduces with the evolution from Class 0 to Class I (see also Figure \ref{fig:BolT_int_peak_size}c, section 4.2.3). 
These decreasing sizes and masses findings from Class 0 to Class I could also indicate the dissipation of the envelope due to accretion and ejection activity of the protostars from Class 0 to Class I evolution.  
 Although, our present analyses of 1-component 2D-Gaussian fitting could not infer to the presence of secondary sources within the common envelope. Therefore, the actual envelope size of the  individual sources could not be specified, in those cases two or more component 2D-Gaussian fittings are required. 
It is also not clear only from our present sample consisting of a small fraction of Class I sources whether these are the intrinsic correlations of dense core evolution or biased by the sample selection, more statistical studies may explain this more comprehensively.

Likewise, if we compare the Peak$_{1.3 ~ mm}$, the Class 0 sources have larger values of peak emission (median $\sim$ 28.20 mJy beam$^{-1}$) than Class I (median $\sim$ 10.41 mJy beam$^{-1}$) and starless cores (median $\sim$ 0.52 mJy beam$^{-1}$). This result suggests a possible evolutionary trend of the dense cores, where the starless cores exhibit a lower peak and as they form a Class 0 system, their emission heats up the surrounding disk-envelope material and making them brightest in this wavelength. On the other hand, as they evolve to the Class I system, their surrounding material may also dissipate and the stellar core becomes more luminous towards the shorter wavelength regime, hence they tend to show a fainter peak in the 1.3\,mm wavelength. However, it could be also an interferometric effect. As starless cores are more diffuse, so the emission is resolved out. Protostellar cores are denser with a different density profile, that can be recovered by the interferometer because they are compact.

Figure \ref{fig:BolT_int_peak_size}a-b display the distribution of 1.3\,mm flux densities and peak flux, respectively as a function of T$_{bol}$ . The Class I  (i.e., T$_{bol}$ $> $ 70 K) sources are mostly concentrated at $\log(F_{1.3~mm})$  $\sim$ 1.3 to 1.8 mJy and $\log(Peak_{1.3 mm})$ $\sim$ 1.25 to 1.70 mJy beam$^{-1}$, whereas the Class 0 flux densities and peaks are wide spread.  
Figure \ref{fig:BolT_int_peak_size}c shows the decreasing size distribution of 2D Gaussian fitting with T$_{bol}$. Despite of fewer Class I sources and unresolved disk-scale geometry, one can see it is significantly smaller sizes than Class 0.  
Figure \ref{fig:BolT_int_peak_size}c points towards a transition from Class 0 to Class I  at T$_{bol}$ = 60$-$70 K for envelope+disk size $<$ 0$\farcs$2 (i.e. 80 au) in this sample, which is also an empirical boundary temperature between Class 0 to Class I sources. These findings also support either the possible density variation according to power-law index or envelope dissipation with protostellar evolution could contribute towards such flux, peak, and size variation from Class 0 to Class I.

\subsection{Evolution of protostellar outflows}
The bolometric temperature and luminosity derived from SED analyses can be somewhat questionable due to inconsistent multiwavelength data catalogs and misidentification due to multiplicity. Rather than exclusively depending on the SED results,  we also searched for possible evolutionary trends from the physical appearance of the outflows from their opening angle, and maximum outflow velocity in the ISM.

\subsubsection{Time Sequence Outflow Opening Angle} \label{sec:Time_sequence_Outflow_opening_angle}

Protostellar jets and winds propagate into the envelope as its immediate environment. As the protostars evolve, the collapsing material settles into the equatorial pseudodisk along with the magnetic field lines. With the growing size of pseudodisk, the matter is evacuated by the magnetic field from the polar region. It is to be noted that the envelope mass declines typically a few orders of magnitudes during the evolution from Class 0 to Class I \citep{1996A&A...311..858B,2006ApJ...646.1070A}. The excavated surroundings set off the widening opening of wind-blown outflow lobe with time \citep[e.g., ][]{1999ASIC..540..227B,2006ApJ...646.1070A,2006ApJ...649..845S}.

The outflow opening angle remains narrower than 20$\degr$ independent of the launching protostar's properties (e.g., mass of the protostars, ejection to accretion mass ratio) during the early stages \citep[][]{2016ApJ...832...40K}, and the low-velocity outflow appears from the first core \citep{1969MNRAS.145..271L}, without any high-velocity component. The high-velocity jet catches up to the outflow after a few hundred years, and the jet speed increases with time \citep[e.g.,][]{2019ApJ...876..149M}. With the emergence of the jet, a strong radiation pressure pushes the outflow material outward \citep[][]{2016ApJ...832...40K,2019ApJ...876..149M}. The observed opening angles are observed to span over 20$\degr$ in  early accretion phases and up to 160$\degr$ at later phases \citep[][]{2005ASSL..324..105B,2014prpl.conf..451F}. For example, HH\,211 is among the youngest known Class 0 protostars with narrow opening angle \citep[][]{1999ASIC..540..227B}, while the evolved Class 0 or embedded Class I systems \citep[e.g., HH\,46/47;][]{2009A&A...501..633V} have relatively  wider opening angles of their outflow cavity \citep[][]{2009A&A...501..633V}.  
The older outflow cavities driven by Class I sources, such as L\,43, L\,1551, and B5 \citep[][]{2000prpl.conf..867R}, appear characteristically with low-velocity CO outflows from wider opening cavities up to 90$\degr$ \citep[][]{2002ApJ...576..294L,2006ApJ...646.1070A}. 
 Observations of a large number of outflows at different evolutionary stages from Class 0 and Class I to Class II, revealed a systematic widening of opening angle with the stellar evolution \citep[][]{2006ApJ...646.1070A,2014ApJ...783....6V,2017AJ....153..173H}.

In Figure \ref{fig:openingAngle_temperature}a-b, the opening angles are plotted as a function of the T$_{bol}$. 
The Class I sources exhibit a higher opening angle range (median $[\Theta_{obs}]_{400}$ $\sim$ 2$\farcs$7) than Class 0 ($[\Theta_{obs}]_{400}$ $\sim$ 1$\farcs$6).   However, from the present scattered distribution, a linear regression suggests a minor correlation only, which may be due to a limited number of opening angle measurements at $>$ 70 K (i.e., only three in Class I and none in Class II), high uncertainty in T$_{bol}$ estimation, and/or unknown inclination of the outflow axis. 
Additional observations of more Class I and early-Class II are required to obtain the evolutionary changes of opening angle accurately, as observed in \citet[][]{2006ApJ...646.1070A,2014ApJ...783....6V,2017AJ....153..173H}.

\subsubsection{Age Dispersal velocity Distribution}

Several outflow models have been proposed to demonstrate the formation of molecular outflow driven by protostars and how they propagate in the ambient cloud environment (See review by Arce et al. 2007; Frank et al 2014). The two more broadly accepted are (a) disk-wind model \citep[e.g.,][]{2000prpl.conf..759K}, where a wind-driven outflow launched from the entire protostellar disk surface and, (b) a two-component protostellar wind model or X-wind model \citep[e.g.,][]{2000prpl.conf..789S}, initiated from the innermost region of the disk. 
In the X-wind model, the disk wind could drive a slow wide-angle outflow along with a collimated central fast-moving jet-component. This model also predicts that the wide-opening angle near outflow launching protostar could escalate a large radial velocity extent \citep[][]{2006ApJ...649..836P,2009ApJ...705.1388H}. One potential interesting constraint from Figure \ref{fig:example_outflow} is that a fraction of blueshifted emission occurs on the redshifted side, and similarly, a fraction of redshifted emission occurs on the blueshifted side. This could be explained either by the wider line width produced by disk wind \citep[][]{2004A&A...416L...9P} or inclination angle of the outflow axis. 

We can infer something about the flow plateau with the velocity extent, assuming that all the outflow wings provide consistent measurements for equal FOV (see also section \ref{sec:resultsOutflowComponentsCOemission}). The outflow velocity V$_{real}$  = V$_{obs}$/cos(i), where V$_{obs}$ is the observed radial velocity. The velocity extent of the outflow caused by the observed opening angle ($\Theta_{obs}$), $\Delta$V = $V_{real} \sin(i)\Theta_{obs}$; implying $\Delta$V = $V_{obs} \tan(i) \Theta_{obs}$, where $\Theta_{obs} = \Theta_{real}\sin(i)$. Thus, to establish a correlation between $\Delta$V and $\Theta_{obs}$, we need a reliable estimation of inclination angle, which we are lacking. Moreover, if we assume a random distribution of inclination angles, the mean value is given by, $\bar{i} = \int_0^{\pi/2} \;i \; sin(i) \mathrm{d}i = 1 \;rad = 57.3^\circ$, it will lead to a homogeneous projection effects. So, we adhere to the observed value of velocity extent and $\Theta_{obs}$ to search for a correlation.

Figure \ref{fig:VelocityDispersion_OpAngle}  displays that the $\Delta V_{obs}$ increases with $\Theta_{obs}$. A linear regression provides: 
\[ \Delta V_{obs} = 25.45(\pm 6.55)\; [\Theta_{obs}]_{400} -10.45 (\pm 11.63), \]
It can be explained by considering the opening angle as an age indicator (see also section \ref{sec:Time_sequence_Outflow_opening_angle}). In the early stages of the protostars, the outflow is detected in small velocity ranges around the systemic velocity. With protostellar evolution, the central mass of the protostars keeps growing, and then higher energetic outflows/jets are likely to originate from a deeper gravitational potential well, thus one can expect a higher $\Delta V_{obs}$. In Figure \ref{fig:VelocityDispersion_OpAngle}, two non-jet sources, G192.12-11.10 and G212.10-19.15S, exhibit smaller $\Delta V_{obs}$ with higher $\Theta_{obs}$, which are possibly evolved Class 0 sources ejecting weak disk winds. However, they deserve to be probed at evolved outflow tracers and more high-density jet tracers like higher transitions of SiO.

Such a correlation could be largely  contributed from the unknown inclination angle of the observable parameters. In absence of proper inclination measurements, we have applied the major-to-minor axis aspect ratio of the 1.3\,mm continuum emission as a proxy to the inclination correction, and the above correlation is found to be more scattered although the overall increasing trend remains the same. However, this aspect ratio could also show larger value for geometrically thick disk-envelope systems \citep[e.g.,][]{2018ApJ...863...94L}.

In Figure \ref{fig:VelocityDispersion_BolT}, the $\Delta V_{obs}$s for Class 0 sources are found to be distributed from 4 - 110 km s$^{-1}$, whereas evolved Class I sources show mostly toward smaller CO $\Delta V_{obs}$. Additionally, all jet sources have higher values of $\Delta V_{obs}$ (median $\sim$ 24 km s$^{-1}$) than the non-jet sources (median $\sim$ 16 km s$^{-1}$), suggesting more active accretion and mass-loss rate of jet sources in comparison to non-jet sources. One exception occurs for the source G208.89-20.04E, which is located in a complex cloud environment and it also has overlapping blue- and redshifted velocity channels, possibly indicating a high inclination angle to the line-of-sight.

In summary, as the protostar evolves, the outflow cavity opening widens and the protostar ejects more energetic outflowing material, as expected if outflow originates from a deeper gravitational potential well of an evolved protostellar.

\section{Summary and Conclusion}\label{sec:summary_conclusion}
We have conducted a survey toward 72 dense cores in the Orion A, B, and $\lambda$ Orionis molecular clouds with ALMA 1.3\,mm continuum in  three different resolutions (TM1 $\sim$ 0$\farcs$35, TM2 $\sim$ 1$\farcs$0 and ACA $\sim$ 7$\farcs$0). This unique combined configuration survey enables us to characterize the dense cores at unprecedented high sensitivity at this high resolution. The main outcomes are as follows:

\begin{itemize}
\item We are able to detect emission in 44 protostellar cores and 4 candidate prestellar cores in the combined three configurations, where another 10 starless cores have detection in the individual ACA array configurations.  The starless, Class 0 and Class I sources have continuum median deconvolved size of $\sim$ 4$\farcs$77, 0$\farcs$32, and 0$\farcs$18, respectively decreasing with dense core evolution.  The peak emission of Class 0, Class I, and starless cores are 28.20, 10.41, 0.52 mJy beam$^{-1}$, respectively, suggesting that with protostellar formation, the envelope is heated up in Class 0 and the envelope loses material while transitioning from Class 0 to Class I.

\item A total of 37 sources show CO outflow emission and 18 ($\sim$ 50\%) of them also show high velocity jets in SiO. The CO velocity extends from 4 to 110 km s$^{-1}$, with a median velocity of 26.5 km s$^{-1}$. The CO outflow cavities have opening angle widths at 1$\arcsec$ ($\sim$ 400 au) ranging from $[\Theta_{obs}]_{400}$ $\sim$ 0$\farcs$6 - 3$\farcs$9 (i.e., 33$\fdg$4 $-$ 125$\fdg$7 near the source) with a median value 1$\farcs$64. The median value of $[\Theta_{obs}]_{400}$ for 19 Class 0 sources is 1$\farcs$60 and 3 Class I sources 2$\farcs$70.

\item  From the present analysis, the outflow opening angle shows a weak correlation with bolometric temperature in our limited sample observations. 

\item The $\Delta$Vs exhibit a correlation with $[\Theta_{obs}]_{400}$.  As the protostar evolves, the envelope depletes from the polar region and  the cavity opening widens, the outflow material possibly becomes more energetic.

\item The 2D Gaussian fitted 1.3 mm continuum size is found to be reduced in Class I (i.e., beyond the Class 0 to Class I transition region, T$_{bol}$ = 60-70 K), which could be due to either varying density profiles depending on power-law indexes or envelope dissipation with protostellar evolution.  The overall mass distribution of Class 0 (median $\sim$ 0.13 M$_{\sun}$) and Class I (median $\sim$ 0.04 M$_{\sun}$) also supports the same conclusion.

\item Potential pseudo-disks are revealed in 1.3\,mm continuum, and C$^{18}$O line emission in some Class 0 sources (e.g., G192.12-11.10). Further investigation in higher spatial and higher velocity resolutions are required to probe the Keplerian rotation.

\item The spectral coverage of this survey incorporates a suit of important diagnostic molecular transitions from the astrochemical perspective. Emission from deuterated species such as N$_{2}$D$^{+}$ and DCO$^{+}$ are detected and serves, for example, as a particularly useful tracer for highlighting the transition from starless to protostellar phases. A subset of protostellar objects with rich features of CH$_3$OH, H$_2$CO, and other COMs like HCOOCH$_3$ and CH$_3$CHO signifies the presence of hot corinos. Broad CO and SiO spectral lines seen towards protostellar sources further delineate active outflows and shocked gas.

\end{itemize}

This survey provides statistical studies performed to explore the correlation between envelope material, outflow opening angle, and outflow velocity extent with the evolution of protostars. The spectral coverage comprise  the importance of astrochemical diagnosis molecular species for tracing the transition from starless to protostellar phases.  Further high-angular and high-velocity resolutions observations covering different evolutionary stages can apprise  these observational findings. In addition, numerical simulations of protostellar outflows launching from variable envelope sizes are definitely required to proceed beyond the qualitative hints given by this analysis. \\

\acknowledgments
We thank the anonymous referee for the constructive comments on our paper. This paper makes use of the following ALMA data:  ADS/JAO.ALMA$\#$2018.1.00302.S. ALMA is a partnership of ESO (representing its member states), NSF (USA) and NINS (Japan), together with NRC (Canada), NSC and ASIAA (Taiwan), and KASI (Republic of Korea), in cooperation with the Republic of Chile. The Joint ALMA Observatory is operated by ESO, AUI/NRAO and NAOJ. 
S.D. and C.-F.L. acknowledge grants from the Ministry of Science and Technology of Taiwan (MoST 107-2119-M- 001- 040-MY3) and the Academia Sinica (Investigator Award AS-IA-108-M01).
Tie Liu is supported by international partnership program of Chinese academy of sciences grant No.114231KYSB20200009 and the initial fund of scientific research for high-level talents at Shanghai Astronomical Observatory. 
DJ is supported by the National Research Council of Canada and by a Natural Sciences and Engineering Research Council of Canada (NSERC) Discovery Grant. 
P.S. was partially supported by a Grant-in-Aid for Scientific Research (KAKENHI Number 18H01259) of Japan Society for the Promotion of Science (JSPS).
L.B. acknowledges support from CONICYT project Basal AFB-170002.
J.He thanks the National Natural Science Foundation of China under grant Nos. 11873086 and U1631237 and support by the Yunnan Province of China (No.2017HC018).
This work is sponsored (in part) by the Chinese Academy of Sciences (CAS), through a grant to the CAS South America Center for Astronomy (CASSACA) in Santiago, Chile.
CWL is supported by the Basic Science Research Program  through the National Research Foundation of Korea (NRF) funded by the Ministry of Education, Science and Technology (NRF-2019R1A2C1010851). 
VMP acknowledges support by the Spanish MINECO under project AYA2017-88754-P. 
S.L. Qin is supported by the Joint Research Fund in Astronomy (U1631237) under cooperative agreement between the National Natural Science Foundation of China (NSFC) and Chinese Academy of Sciences (CAS)

\FloatBarrier
\startlongtable
\begin{deluxetable*}{llllcccc}
\tablenum{1}
\tablecaption{Details of targeted dense cores in the Orion Complex\label{tab:targetobserved}}
\tablewidth{6pt}
\addtolength{\tabcolsep}{-3pt}
\tablehead{
\colhead{ALMA} & \colhead{RA (J2000)}  & \colhead{Dec (J2000)}& \colhead{JCMT} & \colhead{Detection} & \colhead{rms} &  \colhead{Detection} & \colhead{rms}\\
\colhead{Targets} & \colhead{(h:m:s)}  & \colhead{(d:m:s)}& \colhead{name} & \colhead{(TM1+TM2+ACA)} & \colhead{(mJy beam$^{-1}$)} &  \colhead{(ACA only)} & \colhead{(mJy beam$^{-1}$)} \\ 
}
\decimalcolnumbers
\startdata
\multicolumn{7}{c}{$\lambda$-Orionis}\\
\cline{1-8} \\
  G191.90-11.21N & 05:31:28.99 & +12:58:47.16 & G191.90-11.21N & NO & 0.03 & NO (weak?) & 0.24\\
  G191.90-11.21S & 05:31:31.73 & +12:56:14.99 & G191.90-11.21S & YES & 0.04 & YES & 3.3\\
  G192.12-11.10 & 05:32:19.54 & +12:49:40.19 & G192.12-11.10 & YES & 0.06 & YES & 2.1\\
  G192.32-11.88N & 05:29:54.47 & +12:16:56 & G192.32-11.88N & YES & 0.08 & YES & 1.0\\
  G192.32-11.88S & 05:29:54.74 & +12:16:32 & G192.32-11.88S & YES & 0.03 & YES & 1.0\\
  G196.92-10.37 & 05:44:29.6 & +09:08:54 & G196.92-10.37 & YES & 0.04 & YES & 1.8\\
  G198.69-09.12N1 & 05:52:29.61 & +08:15:37 & G198.69-09.12N1 & NO & 0.06 & NO & 0.3\\
  G198.69-09.12N2 & 05:52:25.3 & +08:15:09 & G198.69-09.12N2 & NO & 0.06 & NO (weak?) & 0.4\\
  G200.34-10.97N & 05:49:03.71 & +05:57:56 & G200.34-10.97N & YES & 0.04 & YES & 1.0\\
\cline{1-8} \\
\multicolumn{6}{c}{Orion A}\\
\cline{1-8} \\
  G207.36-19.82N1 & 05:30:50.94 & -04:10:35.6 & G207.36-19.82N1 & YES & 0.06 & YES & 1.2\\
  G207.36-19.82N2 & 05:30:50.853 & -04:10:13.641 & G207.36-19.82N2 & NO & 0.04 & YES & 1.2\\
  G207.36-19.82N4 & 05:30:44.546 & -04:10:27.384 & G207.36-19.82N4 & NO (weak?) & 0.035 & YES & 0.5\\
  G207.36-19.82S & 05:30:47.199 & -04:12:29.734 & G207.36-19.82S & NO & 0.04 & NO & 0.4\\
  G208.68-19.20N1 & 05:35:23.486 & -05:01:31.583 & G208.68-19.20N1 & YES & 0.45 & YES & 4.0\\
  G208.68-19.20N2 & 05:35:20.469 & -05:00:50.394 & G208.68-19.20N2 & YES & 0.14 & YES & 6.0\\
  G208.68-19.20N3 & 05:35:18.02 & -05:00:20.7 & G208.68-19.20N3 & YES & 0.2 & YES & 6.0\\
  G208.68-19.20S & 05:35:26.32 & -05:03:54.393 & G208.68-19.20S & YES & 0.1 & YES & 7.0\\
  G208.89-20.04E & 05:32:48.262 & -05:34:44.335 & G208.89-20.04E & YES & 0.1 & YES & 2.5\\
  G208.89-20.04Walma\tablenotemark{$\dagger$} & 05:32:28.03 & -05:34:26.69 & $---$ & YES & 0.04 & YES & 1.8\\
  G209.29-19.65N1 & 05:35:00.379 & -05:39:59.741 & G209.29-19.65N1 & NO (weak?) & 0.04 & YES (weak?) & 2.2\\
  G209.29-19.65S1 & 05:34:55.991 & -05:46:04 & G209.29-19.65S1 & YES & 0.05 & YES & 3.3\\
  G209.29-19.65S2 & 05:34:53.809 & -05:46:17.627 & G209.29-19.65S2 & NO (weak?) & 0.04 & NO (weak?) & 1.5\\
  G209.55-19.68N1 & 05:35:08.9 & -05:55:54.4 & G209.55-19.68N1 & YES & 0.09 & YES & 4.0\\
  G209.55-19.68N2 & 05:35:07.5 & -05:56:42.4 & G209.55-19.68N2 & NO (weak?) & 0.04 & YES & 0.9\\
  G209.55-19.68S1 & 05:35:13.476 & -05:57:58.646 & G209.55-19.68S1 & YES & 0.2 & YES & 4.2\\
  G209.55-19.68S2 & 05:35:09.076 & -05:58:27.378 & G209.55-19.68S3\tablenotemark{*} & YES & 0.08 & YES & 1.9\\
  G209.77-19.40E2 & 05:36:31.977 & -06:02:03.765 & G209.77-19.40E2 & NO & 0.05 & NO & 0.5\\
  G209.77-19.40E3 & 05:36:35.9 & -06:02:42.165 & G209.77-19.40E3 & YES & 0.04 & YES & 0.7\\
  G209.79-19.80W & 05:35:10.696 & -06:13:59.318 & G209.79-19.80W & NO & 0.04 & NO (weak?) & 0.7\\
  G209.94-19.52N & 05:36:11.55 & -06:10:44.76 & G209.94-19.52N & YES & 0.09 & YES & 2.0\\
  G209.94-19.52S1 & 05:36:24.96 & -06:14:04.71 & G209.94-19.52S1 & NO & 0.05 & YES (weak?) & 1.0\\
  G210.37-19.53N & 05:36:55.03 & -06:34:33.19 & G210.37-19.53N & NO & 0.04 & YES & 1.0\\
  G210.37-19.53S & 05:37:00.55 & -06:37:10.16 & G210.37-19.53S & YES & 0.05 & YES & 2.3\\
  G210.49-19.79W & 05:36:18.86 & -06:45:28.035 & G210.49-19.79W & YES & 0.7 & YES & 4.0\\
  G210.82-19.47N2 & 05:37:59.989 & -06:57:15.462 & G210.82-19.47N2 & NO (weak?) & 0.05 & YES & 1.0\\
  G210.82-19.47S & 05:38:03.677 & -06:58:24.141 & G210.82-19.47S & YES & 0.07 & YES & 0.5\\
  G210.97-19.33S2 & 05:38:45.3 & -07:01:04.41 & G210.97-19.33S2 & YES & 0.05 & YES & 1.0\\
  G211.01-19.54N & 05:37:57.469 & -07:06:59.068 & G211.01-19.54N & YES & 0.07 & YES & 2.3\\
  G211.01-19.54S & 05:37:59.007 & -07:07:28.772 & G211.01-19.54S & YES & 0.05 & YES & 0.8\\
  G211.16-19.33N2 & 05:39:05.831 & -07:10:41.515 & G211.16-19.33N2 & YES & 0.04 & YES & 0.5\\
  G211.16-19.33N4 & 05:38:55.68 & -07:11:25.9 & G211.16-19.33N4 & NO & 0.05 & YES (weak) & 0.7\\
  G211.16-19.33N5 & 05:38:46 & -07:10:41.9 & G211.16-19.33N5 & NO (other?) & 0.07 & YES & 0.7\\
  G211.47-19.27N & 05:39:57.18 & -07:29:36.082 & G211.47-19.27N & YES (Close Binary?) & 0.12 & YES & 2.0\\
  G211.47-19.27S & 05:39:56.097 & -07:30:28.403 & G211.47-19.27S & YES & 0.25 & YES & 11.0\\
  G211.72-19.25S1alma\tablenotemark{$\dagger$} & 05:40:21.21 & -07:36:08.79 & $---$ & NO & 0.05 & NO & 1.0\\
  G212.10-19.15N1 & 05:41:21.34 & -07:52:26.92 & G212.10-19.15N1 & YES & 0.04 & YES & 1.0\\
  G212.10-19.15N2 & 05:41:24.03 & -07:53:47.51 & G212.10-19.15N2 & YES & 0.04 & YES & 1.0\\
  G212.10-19.15S & 05:41:26.446 & -07:56:52.547 & G212.10-19.15S & YES & 0.25 & YES & 3.0\\
  G212.84-19.45N & 05:41:32.146 & -08:40:10.45 & G212.84-19.45N & YES & 0.12 & YES (weak?) & 4.5\\
  G215.44-16.38 & 05:56:58.45 & -09:32:42.3 & G215.44-16.38 & NO & 0.04 & YES (weak?) & 0.7\\
  G215.87-17.62M & 05:53:32.4 & -10:25:05.99 & G215.87-17.62M & YES & 0.04 & YES & 2.0\\
  G215.87-17.62N & 05:53:41.89 & -10:24:02 & G215.87-17.62N & YES & 0.04 & YES & 0.8\\
  G215.87-17.62S & 05:53:26.249 & -10:27:29.473 & G215.87-17.62S & NO (other?) & 0.04 & YES (weak?) & 0.8\\
 \cline{1-8} \\
\multicolumn{6}{c}{Orion B}\\
\cline{1-8} \\
  G201.52-11.08 & 05:50:59.01 & +04:53:53.1 & G201.52-11.08 & YES & 0.03 & YES & 0.5\\
  G203.21-11.20E1 & 05:53:51.004 & +03:23:07.3 & G203.21-11.20E1 & NO (weak?) & 0.03 & YES & 1.0\\
  G203.21-11.20E2 & 05:53:47.483 & +03:23:11.3 & G203.21-11.20E2 & NO & 0.04 & NO (weak?) & 0.4\\
  G203.21-11.20W1 & 05:53:42.702 & +03:22:35.3 & G203.21-11.20W1 & YES & 0.04 & YES & 3.0\\
  G203.21-11.20W2 & 05:53:39.492 & +03:22:24.9 & G203.21-11.20W2 & YES & 0.04 & YES & 0.3\\
  G205.46-14.56M1 & 05:46:08.053 & -00:10:43.712 & G205.46-14.56N3\tablenotemark{*} & YES & 0.5 & YES & 2.0\\
  G205.46-14.56M2 & 05:46:07.9 & -00:10:01.82 & G205.46-14.56N2\tablenotemark{*} & YES & 0.08 & YES & 2.0\\
  G205.46-14.56M3 & 05:46:05.66 & -00:09:33.64 & G205.46-14.56N1\tablenotemark{*} & YES & 0.05 & YES & 1.0\\
  G205.46-14.56N1 & 05:46:09.75 & -00:12:16.45 & G205.46-14.56M1\tablenotemark{*} & YES & 0.15 & YES & 1.0\\
  G205.46-14.56N2 & 05:46:07.4 & -00:12:21.84 & G205.46-14.56M2\tablenotemark{*} & YES & 0.15 & YES & 2.5\\
  G205.46-14.56S1 & 05:46:07.048 & -00:13:37.777 & G205.46-14.56S1 & YES & 0.15 & YES & 4.0\\
  G205.46-14.56S2 & 05:46:04.49 & -00:14:18.81 & G205.46-14.56S2 & YES & 0.08 & YES & 1.5\\
  G205.46-14.56S3 & 05:46:03.385 & -00:14:51.715 & G205.46-14.56S3 & YES & 0.06 & YES & 2.0\\
  G206.12-15.76 & 05:42:45.358 & -01:16:13.262 & G206.12-15.76 & YES & 0.3 & YES & 12.0\\
  G206.21-16.17N & 05:41:39.544 & -01:35:52.212 & G206.21-16.17N & NO (weak?) & 0.04 & YES & 1.0\\
  G206.21-16.17S & 05:41:36.373 & -01:37:43.61 & G206.21-16.17S & NO (weak?) & 0.03 & YES & 0.4\\
  G206.93-16.61E2 & 05:41:37.31 & -02:17:18.135 & G206.93-16.61E2 & YES & 0.15 & YES & 4.0\\
  G206.93-16.61W2 & 05:41:25.132 & -02:18:06.455 & G206.93-16.61W3\tablenotemark{*} & YES & 0.15 & YES & 10.0\\
  G206.93-16.61W4 & 05:41:28.77 & -02:20:04.3 & G206.93-16.61W5\tablenotemark{*} & NO & 0.04 & NO & 3.0\\
 \enddata
\tablecomments{In column 5 \& 7, {\it weak} emission detections are marked, whereas the $\sim$ 3$\sigma$ level emissions or questionable detections are marked with {\it weak?}. These are not included in the final detection count. In few targeted positions, no emission detected around the dense core coordinates but some other compact emission detected. They are marked with {\it other?}.}
\tablenotetext{\dagger}{In ALMA archive, they are listed as G208.89-20.04W and G211.72-19.25S1, respectively.  These objects are different than JCMT dense cores catalog in \citet{2018ApJS..236...51Y}, with the same names. These objects are selected directly from JCMT images for ALMA observations.}
\tablenotetext{*}{Note that, the ALMA archive names are different than the JCMT dense core names in \citet{2018ApJS..236...51Y}.}

\end{deluxetable*}
\FloatBarrier

\begin{deluxetable*}{cccccDlc}[h]
\tablenum{2}
\tablecaption{Log of Observations\label{tab:obslog}}
\tablewidth{0pt}
\tablehead{
\colhead{Scheduling} &\colhead{Number of} & \colhead{Date} & \nocolhead{Common} & \colhead{Array} &\multicolumn2c{Number of} & \colhead{Time on} & \colhead{Unprojected} \\
\colhead{Block} & \colhead{Execution} &\colhead{} & \nocolhead{Name} & \colhead{Configuration} &\multicolumn2c{Antennas} & \colhead{Target (S) } & \colhead{Baselines (m)}
}
\decimalcolnumbers
\startdata
 1              & 1   & 2018 Oct 24 &  & C43-5 & 48 & 3430 &  15-1398 \\
                 & 2   & 2018 Dec 21 & & C43-2 & 46 & 1394 & 15-500\\
                 & 3   & 2018 Nov 19 & & ACA.  & 12 &4590 & 9-49\\
\hline
 2              & 1   & 2018 Oct 29  &  &C43-5 & 47 & 4569 & 15-1398\\
                 & 2   & 2018 Nov 01 & &  C43-5 & 44 & 4654 & 15-1358\\
                 & 3   & 2018 Nov 01 & &  C43-5 &  44 & 4655 & 15-1358\\
                 & 4   & 2019 Jan 16 & &   C43-2&46  &3542 & 15-313\\
                 & 5   & 2018 Nov 21 & &  ACA &12 &5324 & 9-49\\
                 & 6   & 2018 Nov 27 & &  ACA &12 &5201 & 9-49\\    
                 & 7   & 2018 Nov 27 & &  ACA &12 &5185 & 9-49\\  
                 & 8   & 2018 Nov 27 & &  ACA &12 &5320 & 9-49\\ 
                 & 9   & 2018 Nov 28 & &  ACA &11 &5200 & 9-49\\
\hline                                               
3               & 1   & 2018 Oct 29  &  &C43-5 & 47 & 1918 & 15-1398\\
                 & 2   & 2019 Mar 05  &  &C43-2& 48 & 1086 & 15-360\\
                 & 3   & 2018 Nov 21  &  &ACA & 12 & 2634 & 9-49\\
                 & 4   & 2018 Nov 26  &  &ACA & 12 & 2635 & 9-49\\
\hline                  
4               & 1   & 2018 Oct 25  &  & C43-5 & 47 & 3134 & 15-1398\\ 
                 & 2   & 2019 Jan 24  &  & C43-2& 51 & 1252 & 15-360\\                 
                 & 3   & 2018 Nov 21  &  &ACA & 12 & 4330 & 9-49\\  
                 & 4   & 2018 Nov 26  &  &ACA & 12 & 4048 & 9-49\\ 
\enddata
\tablecomments{This table is organised according to execution block and Array configuration, not with date of observations.}
\end{deluxetable*}

\begin{deluxetable*}{ccccc}[h]
\tablenum{3}
\tablecaption{Correlator Setup\label{tab:spectralsetup}}
\tablewidth{0pt}
\tablehead{
\colhead{Spectral} & \colhead{Central} & \colhead{Main molecular Lines}  & \colhead{Bandwidth} & \colhead{Velocity} \\
\colhead{Window} & \colhead{Frequency} & \colhead{} & \colhead{} & \colhead{Resolution}\\
\colhead{} & \colhead{(GHz)} & \colhead{} & \colhead{(GHz)} & \colhead{(km s$^{-1}$)}\
}
\decimalcolnumbers
\startdata
0 & 231.000000 & $^{12}$CO J=2-1; N$_2$D$^+$ J=3-2 &1.875 & 1.465 \\
1 & 233.000000 &  CH$_3$OH transitions  & 1.875 &1.453 \\
2 & 218.917871 & C$^{18}$O J=2-1; H$_2$CO transitions  & 1.875 & 1.546 \\
3 & 216.617675 & SiO J=5-4; DCN J=3-2; DCO$^{+}$ J=3-2  & 1.875 & 1.563 \\
\enddata
\end{deluxetable*}

\begin{deluxetable*}{ccccc}[h]
\tablenum{4}
\tablecaption{Calibrators and Their Flux Densities\label{tab:calibrator}}
\tablewidth{0pt}
\tablehead{
\colhead{Scheduling} & \colhead{Date} & \colhead{Bandpass Calibrator} & \colhead{Flux Calibrator}  & \colhead{Phase Calibrator} \\
\colhead{Block} & \colhead{} & \colhead{(Quasar, Flux Density)} & \colhead{(Quasar, Flux Density)} & \colhead{(Quasar, Flux Density)} 
}

\decimalcolnumbers
\startdata
 1 &  2018 Oct 24 & J0423$-$0120, 2.68 Jy  & J0423$-$0120, 2.68 Jy & J0607$-$0834, 0.78 Jy\\
     & 2018 Dec 21 & J0522$-$3627, 3.65 Jy &  J0522$-$3627, 3.65 Jy & J0542$-$0913, 0.47 Jy \\
     & 2018 Nov 19 & J0522$-$3627, 4.91 Jy &  J0522$-$3627, 4.91 Jy & J0607$-$0834, 0.78 Jy\\
\hline
  2 &   2018 Oct 29  & J0423$-$0120, 2.53 Jy & J0423$-$0120, 2.53 Jy & J0541$-$0211, 0.095 Jy \\
     &   2018 Nov 01 & J0423$-$0120, 2.53 Jy & J0423$-$0120, 2.53 Jy & J0541$-$0211, 0.095 Jy \\
     &   2018 Nov 01 & J0423$-$0120, 2.53 Jy & J0423$-$0120, 2.53 Jy & J0541$-$0211, 0.095 Jy \\
     &   2019 Jan 16 & J0522$-$3627, 3.14 Jy &  J0522$-$3627, 3.14 Jy & J0542$-$0913, 0.47 Jy  \\
     &   2018 Nov 21 & J0854$+$2006, 2.77 Jy &  J0854$+$2006, 2.77 Jy & J0607$-$0834, 0.78 Jy \\
     &   2018 Nov 27 & J0423$-$0120, 2.30 Jy &  J0423$-$0120, 2.30 Jy & J0542$-$0913, 0.47 Jy \\    
     &   2018 Nov 27 & J0522$-$3627, 4.39 Jy&  J0522$-$3627, 4.39 Jy & J0542$-$0913, 0.47 Jy \\  
     &   2018 Nov 27 & J0854$+$2006, 3.06 Jy &  J0854$+$2006, 3.06 Jy & J0607$-$0834, 0.78 Jy\\ 
     &   2018 Nov 28 & J0423$-$0120, 2.29 Jy &  J0423$-$0120, 2.29 Jy & J0542$-$0913, 0.47 Jy  \\
\hline                                               
  3 &  2018 Oct 29  & J0510$+$1800, 1.40 Jy & J0510$+$1800, 1.40 Jy & J0530$+$1331, 0.31 Jy  \\
     &  2019 Mar 05  & J0750$+$1231, 0.65 Jy & J0750$+$1231, 0.65 Jy & J0530$+$1331, 0.30 Jy  \\
     &  2018 Nov 21  & J0423$-$0120, 2.40 Jy & J0423$-$0120, 2.29 Jy & J0530$+$1331, 0.30 Jy  \\
     &  2018 Nov 26  & J0423$-$0120, 2.40 Jy & J0423$-$0120, 2.29 Jy & J0530$+$1331, 0.30 Jy  \\
\hline                  
4   &  2018 Oct 25  & J0510$+$1800, 1.54 Jy & J0510$+$1800, 1.54 Jy & J0552$+$0313, 0.35 Jy  \\ 
     &  2019 Jan 24  & J0423$-$0120, 2.68 Jy & J0423$-$0120, 2.68 Jy &  J0552$+$0313, 0.35 Jy \\                 
     &  2018 Nov 21  & J0522$-$3627, 5.07 Jy & J0522$-$3627, 5.07 Jy & J0532$+$0732, 1.13 Jy \\  
     &  2018 Nov 26  & J0423$-$0120, 2.40 Jy & J0423$-$0120, 2.40 Jy & J0532$+$0732, 1.13 Jy \\ 
 \enddata
\end{deluxetable*}

\movetabledown=15mm
\FloatBarrier
\begin{longrotatetable}
\begin{deluxetable*}{l@{\extracolsep{3pt}}c@{\extracolsep{3pt}}ccccccc@{\extracolsep{3pt}}c@{\extracolsep{3pt}}c@{\extracolsep{3pt}}c@{\extracolsep{6pt}}cc@{\extracolsep{3pt}}c@{\extracolsep{2pt}}c@{\extracolsep{1pt}}cc}
\tablenum{5}
\tablecaption{Continuum and and line-emission properties of all objects\label{tab:ContinuumProtoStellarProperties}}
\tablewidth{0pt}
\addtolength{\tabcolsep}{-2pt}
\tabletypesize{\scriptsize}
\tablewidth{6pt}
\addtolength{\tabcolsep}{-4pt}
\tablehead
{
 \multicolumn{9}{c}{1.3\,mm Continuum (TM1+TM2+ACA)}&  \multicolumn{5}{c}{CO \& SiO}&   \multicolumn{3}{c}{Infrared}\\
\cline{2-9}  \cline{10-14} \cline{15-16}\\ 
\colhead{Source}&
  \multicolumn{1}{c}{RA}&  \multicolumn{1}{c}{Dec}&   \multicolumn{1}{c}{Maj}&  \multicolumn{1}{c}{Min}&   \multicolumn{1}{c}{PA} &  \multicolumn{1}{c}{F$_{1.3 mm}$} &   \multicolumn{1}{c}{Peak$_{1.3 mm}$} &  \multicolumn{1}{c}{Mass} &  \multicolumn{1}{c}{$\Delta$V$_B$\tablenotemark{}} &  \multicolumn{1}{c}{$\Delta$V$_R$\tablenotemark{}} & \multicolumn{1}{c}{$\Theta_{400}$}&  \multicolumn{1}{c}{$\Theta_{800}$} &   \multicolumn{1}{c}{SiO\tablenotemark{*}} & \multicolumn{1}{c}{T$_{bol}$}&  \multicolumn{1}{c}{L$_{bol}$}&  \multicolumn{1}{c}{Class\tablenotemark{\dag}} &  \multicolumn{1}{c}{HOPS}\\
 \multicolumn{1}{c}{} & \multicolumn{1}{c}{(h:m:s)}&  \multicolumn{1}{c}{(d:m:s)}&   \multicolumn{1}{c}{$\arcsec$}&  \multicolumn{1}{c}{$\arcsec$}&   \multicolumn{1}{c}{($\degr$)} &  \multicolumn{1}{c}{(mJy)} &   \multicolumn{1}{c}{(mJy/beam)} &  \multicolumn{1}{c}{(M$_{\sun}$)} &  \multicolumn{1}{c}{(km/s)} &  \multicolumn{1}{c}{(km/s)} & \multicolumn{1}{c}{($\arcsec$)}&  \multicolumn{1}{c}{($\arcsec$)} &  \multicolumn{1}{c}{(Y/N)} &  \multicolumn{1}{c}{(K)}&  \multicolumn{1}{c}{(L$_{\sun}$)} &  \multicolumn{1}{c}{} &  \multicolumn{1}{c}{}\\
}
\decimalcolnumbers
\startdata
G191.90-11.21S&05:31:31.60&+12:56:14.15&0.693$\pm0.031$&0.395$\pm0.027$&79.93$\pm3.76$&27.77$\pm1.08$&10.11$\pm0.30$&0.079$\pm0.034$&22$^{+10}_{-4}$&24$^{+8}_{-4}$&1.59$\pm0.36$&2.28$\pm0.36$&    Y    &69$\pm17$&0.4$\pm0.2$&0&$--$\\
G192.12-11.10&05:32:19.37&+12:49:40.92&0.792$\pm0.009$&0.276$\pm0.007$&121.96$\pm0.52$&119.24$\pm1.25$&44.32$\pm0.35$&0.340$\pm0.145$&40$^{+8}_{-14}$&44$^{+8}_{-14}$&3.56$\pm0.66$&4.46$\pm0.22$&    N    &44$\pm15$&9.5$\pm4.0$&0&$--$\\
G192.32-11.88N&05:29:54.15&+12:16:52.99&0.276$\pm0.005$&0.237$\pm0.006$&65.27$\pm5.77$&143.22$\pm0.82$&102.72$\pm0.38$&0.408$\pm0.174$&20$^{+8}_{-4}$&4$^{+10}_{-2}$&    na    &    na    &    N    &    na    &    na    &0&$--$\\
G192.32-11.88S&05:29:54.41&+12:16:29.68&5.374$\pm0.361$&4.001$\pm0.269$&23.85$\pm9.12$&34.99$\pm2.35$&0.27$\pm0.02$&0.100$\pm0.043$&    cx    &    cx    &    na    &    na    &    N    &60$\pm13$&0.1$\pm0.1$&0&$--$\\
G196.92-10.37$\_$A&05:44:29.26&+09:08:52.18&0.459$\pm0.023$&0.375$\pm0.022$&17.68$\pm13.47$&24.06$\pm0.62$&12.72$\pm0.23$&0.069$\pm0.029$&54$^{+12}_{-12}$&34$^{+10}_{-14}$&2.71$\pm0.75$&4.75$\pm1.30$&    N    &    na    &    na    &0&$--$\\
G196.92-10.37$\_$B&05:44:30.02&+09:08:57.30&0.234$\pm0.011$&0.067$\pm0.042$&86.46$\pm3.96$&14.82$\pm0.16$&12.96$\pm0.08$&0.042$\pm0.018$&    na    &    na    &    na    &    na    &    N    &143$\pm28$&3.5$\pm2.0$&1&$--$\\
G196.92-10.37$\_$C\tablenotemark{$\xi$}&05:44:29.98&+09:08:56.25&0.000$\pm0.000$&0.000$\pm0.000$&0.00$\pm0.00$&1.62$\pm0.10$&1.84$\pm0.06$&0.005$\pm0.002$&    na    &    na    &    na    &    na    &    N    &143$\pm28$&3.5$\pm2.0$&1&$--$\\
G200.34-10.97N&05:49:03.35&+05:57:58.11&0.361$\pm0.011$&0.321$\pm0.016$&142.61$\pm20.97$&23.92$\pm0.42$&14.64$\pm0.17$&0.068$\pm0.029$&18$^{+8}_{-6}$&14$^{+8}_{-4}$&0.90$\pm0.14$&1.23$\pm0.20$&    N    &43$\pm10$&1.5$\pm0.6$&0&$--$\\
G201.52-11.08&05:50:59.15&+04:53:49.65&0.673$\pm0.010$&0.182$\pm0.014$&124.81$\pm0.80$&21.17$\pm0.28$&10.02$\pm0.09$&0.060$\pm0.026$&    cx    &    cx    &    na    &    na    &    N    &263$\pm55$&0.3$\pm0.2$&1&$--$\\
G203.21-11.20W1&05:53:42.59&+03:22:34.97&0.395$\pm0.010$&0.176$\pm0.010$&73.37$\pm1.84$&32.09$\pm0.40$&22.33$\pm0.18$&0.091$\pm0.039$&12$^{+6}_{-8}$&12$^{+6}_{-6}$&0.98$\pm0.15$&1.33$\pm0.10$&    N    &    na    &    na    &0&$--$\\
G203.21-11.20W2&05:53:39.51&+03:22:23.85&0.777$\pm0.051$&0.430$\pm0.032$&64.50$\pm4.50$&11.89$\pm0.64$&4.16$\pm0.17$&0.034$\pm0.015$&58$^{+14}_{-10}$&46$^{+16}_{-12}$&2.24$\pm0.13$&3.27$\pm0.40$&    Y    &15$\pm5$&0.5$\pm0.3$&0&$--$\\
G205.46-14.56M1$\_$A&05:46:08.60&-00:10:38.49&0.314$\pm0.052$&0.254$\pm0.054$&123.35$\pm77.51$&22.38$\pm1.58$&15.54$\pm0.71$&0.064$\pm0.028$&46$^{+16}_{-16}$&30$^{+14}_{-14}$&    na    &    na    &    Y    &47$\pm12$&4.8$\pm2.1$&0&317\\
G205.46-14.56M1$\_$B&05:46:08.38&-00:10:43.54&1.268$\pm0.033$&0.582$\pm0.018$&84.49$\pm1.25$&788.03$\pm19.41$&148.79$\pm3.11$&2.245$\pm0.960$&36$^{+14}_{-14}$&8$^{+10}_{-4}$&    na    &    na    &    N    &    na    &    na    &0&$--$\\
G205.46-14.56M2$\_$A&05:46:07.85&-00:10:01.30&0.147$\pm0.027$&0.098$\pm0.052$&68.01$\pm30.45$&12.93$\pm0.26$&11.84$\pm0.14$&0.037$\pm0.016$&    na    &    na    &    na    &    na    &    N    &112$\pm27$&9.4$\pm3.9$&111&387\\
G205.46-14.56M2$\_$B&05:46:07.84&-00:09:59.60&0.269$\pm0.008$&0.121$\pm0.016$&40.52$\pm4.64$&43.62$\pm0.41$&34.70$\pm0.20$&0.124$\pm0.053$&2$^{+4}_{-0}$&14$^{+12}_{-8}$&    na    &    na    &    N    &112$\pm28$&9.4$\pm3.9$&1&387\\
G205.46-14.56M2$\_$C&05:46:08.48&-00:10:03.04&0.135$\pm0.009$&0.071$\pm0.024$&49.23$\pm9.08$&31.87$\pm0.25$&29.79$\pm0.14$&0.091$\pm0.039$&    cx    &    cx    &    na    &    na    &    N    &163$\pm34$&21.0$\pm8.0$&111&386\\
G205.46-14.56M2$\_$D&05:46:08.43&-00:10:00.50&0.569$\pm0.052$&0.400$\pm0.051$&53.48$\pm13.21$&10.00$\pm0.67$&4.22$\pm0.21$&0.028$\pm0.012$&    cx    &    cx    &    na    &    na    &    Y    &163$\pm34$&21.0$\pm8.0$&1&386\\
G205.46-14.56M2$\_$E&05:46:08.92&-00:09:56.12&0.079$\pm0.053$&0.054$\pm0.026$&125.45$\pm34.83$&3.75$\pm0.12$&3.66$\pm0.07$&0.011$\pm0.005$&    na    &    na    &    na    &    na    &    N    &    na    &    na    &111&$--$\\
G205.46-14.56M3&05:46:05.97&-00:09:32.69&5.652$\pm0.247$&4.751$\pm0.207$&108.72$\pm10.21$&55.16$\pm2.40$&0.37$\pm0.02$&1.240$\pm0.532$&    na    &    na    &    na    &    na    &    N    &    na    &    na    &-1&$--$\\
G205.46-14.56N1&05:46:10.03&-00:12:16.88&0.382$\pm0.005$&0.254$\pm0.006$&57.13$\pm1.88$&166.75$\pm1.05$&103.98$\pm0.44$&0.475$\pm0.203$&    cx    &    cx    &    na    &    na    &    N    &29$\pm8$&0.6$\pm0.3$&0&402\\
G205.46-14.56N2&05:46:07.72&-00:12:21.27&0.445$\pm0.009$&0.332$\pm0.008$&139.79$\pm3.45$&78.31$\pm1.01$&41.79$\pm0.37$&0.223$\pm0.095$&    cx    &    cx    &    na    &    na    &    N    &32$\pm8$&0.8$\pm0.3$&0&401\\
G205.46-14.56S1$\_$A&05:46:07.26&-00:13:30.23&0.374$\pm0.011$&0.188$\pm0.014$&77.78$\pm2.84$&53.17$\pm0.77$&36.57$\pm0.34$&0.151$\pm0.065$&42$^{+8}_{-14}$&20$^{+12}_{-12}$&1.36$\pm0.27$&1.58$\pm0.65$&    Y    &44$\pm19$&22.0$\pm8.0$&0&358\\
G205.46-14.56S1$\_$B&05:46:07.33&-00:13:43.49&0.320$\pm0.006$&0.300$\pm0.008$&35.77$\pm19.70$&137.32$\pm1.19$&89.35$\pm0.51$&0.391$\pm0.167$&16$^{+8}_{-8}$&8$^{+8}_{-2}$&    na    &    na    &    N    &    na    &    na    &0&$--$\\
G205.46-14.56S2&05:46:04.77&-00:14:16.67&0.101$\pm0.014$&0.073$\pm0.032$&16.83$\pm64.27$&24.19$\pm0.25$&23.15$\pm0.14$&0.069$\pm0.029$&46$^{+10}_{-10}$&42$^{+10}_{-10}$&    na    &    na    &    N    &381$\pm60$&12.5$\pm4.7$&1&385\\
G205.46-14.56S3&05:46:03.63&-00:14:49.57&0.233$\pm0.015$&0.194$\pm0.014$&130.30$\pm21.59$&58.72$\pm1.02$&46.67$\pm0.50$&0.167$\pm0.072$&114$^{+8}_{-20}$&106$^{+8}_{-24}$&3.29$\pm2.09$&5.03$\pm2.33$&    Y    &178$\pm33$&6.4$\pm2.4$&1&315\\
G206.12-15.76&05:42:45.26&-01:16:13.94&0.625$\pm0.013$&0.485$\pm0.012$&166.34$\pm4.18$&363.35$\pm5.66$&131.29$\pm1.56$&1.035$\pm0.442$&22$^{+8}_{-8}$&26$^{+8}_{-8}$&1.67$\pm0.06$&2.79$\pm1.37$&    Y    &35$\pm9$&3.0$\pm1.4$&0&400\\
G206.93-16.61E2$\_$A&05:41:37.19&-02:17:17.34&0.300$\pm0.038$&0.228$\pm0.045$&156.79$\pm26.01$&98.22$\pm4.40$&69.25$\pm1.99$&0.280$\pm0.120$&    na    &    na    &    na    &    na    &    N    &198$\pm60$&36.0$\pm15.0$&111&298\\
G206.93-16.61E2$\_$B&05:41:37.04&-02:17:17.99&0.206$\pm0.037$&0.189$\pm0.045$&137.36$\pm76.95$&39.17$\pm1.53$&31.81$\pm0.76$&0.112$\pm0.048$&    na    &    na    &    na    &    na    &    N    &198$\pm60$&36.0$\pm15.0$&111&298\\
G206.93-16.61E2$\_$C&05:41:37.20&-02:17:15.97&1.186$\pm0.148$&1.063$\pm0.135$&32.30$\pm49.20$&76.91$\pm8.68$&9.05$\pm0.92$&0.219$\pm0.097$&    na    &    na    &    na    &    na    &    N    &198$\pm60$&36.0$\pm15.0$&111&298\\
G206.93-16.61E2$\_$D&05:41:37.15&-02:17:16.52&3.668$\pm0.319$&0.720$\pm0.068$&76.84$\pm1.35$&88.62$\pm7.23$&4.90$\pm0.38$&0.252$\pm0.110$&    na    &    na    &    na    &    na    &    N    &198$\pm60$&36.0$\pm15.0$&111&298\\
G206.93-16.61W2&05:41:24.93&-02:18:06.75&0.719$\pm0.056$&0.508$\pm0.043$&99.72$\pm10.31$&270.81$\pm17.70$&85.08$\pm4.35$&0.771$\pm0.333$&74$^{+22}_{-8}$&78$^{+22}_{-8}$&1.59$\pm0.28$&2.68$\pm0.56$&    Y    &31$\pm10$&6.3$\pm3.0$&0&399\\
G207.36-19.82N1$\_$A&05:30:51.23&-04:10:35.34&1.011$\pm0.026$&0.217$\pm0.013$&101.61$\pm0.57$&39.69$\pm0.88$&14.33$\pm0.24$&0.113$\pm0.048$&    cx    &    cx    &    na    &    na    &    N    &    na    &    na    &111&$--$\\
G207.36-19.82N1$\_$B&05:30:51.30&-04:10:32.22&0.139$\pm0.035$&0.058$\pm0.033$&101.42$\pm29.50$&3.74$\pm0.10$&3.54$\pm0.06$&0.011$\pm0.005$&    na    &    na    &    na    &    na    &    N    &    na    &    na    &111&$--$\\
G208.68-19.20N1&05:35:23.42&-05:01:30.60&0.563$\pm0.015$&0.522$\pm0.016$&171.29$\pm17.49$&811.53$\pm11.28$&299.26$\pm3.15$&2.312$\pm0.988$&    na    &    na    &    na    &    na    &    Y    &38$\pm13$&36.7$\pm14.5$&0&87\\
G208.68-19.20N2$\_$A&05:35:20.78&-05:00:55.67&14.642$\pm0.458$&2.422$\pm0.076$&118.87$\pm0.42$&212.56$\pm6.58$&1.07$\pm0.03$&4.777$\pm2.046$&    na    &    na    &    na    &    na    &    N    &    na    &    na    &-1&$--$\\
G208.68-19.20N2$\_$B\tablenotemark{$\xi$}&05:35:19.98&-05:01:02.59&0.000$\pm0.000$&0.000$\pm0.000$&0.00$\pm0.00$&3.16$\pm0.23$&3.40$\pm0.14$&0.009$\pm0.004$&    na    &    na    &    na    &    na    &    N    &112$\pm10$&2.1$\pm1.3$&0&89\\
G208.68-19.20N3$\_$A&05:35:18.06&-05:00:18.19&3.094$\pm0.251$&2.073$\pm0.169$&149.36$\pm8.06$&152.90$\pm12.32$&3.95$\pm0.31$&0.436$\pm0.189$&48$^{+12}_{-16}$&38$^{+12}_{-16}$&    na    &    na    &    Y    &    na    &    na    &0&$--$\\
G208.68-19.20N3$\_$B&05:35:18.34&-05:00:32.95&0.224$\pm0.023$&0.208$\pm0.031$&24.47$\pm62.39$&27.24$\pm0.78$&21.31$\pm0.38$&0.078$\pm0.033$&10$^{+12}_{-4}$&26$^{+10}_{-8}$&    na    &    na    &    N    &158$\pm20$&22.0$\pm8.7$&1&92\\
G208.68-19.20N3$\_$C&05:35:18.27&-05:00:33.93&0.208$\pm0.023$&0.181$\pm0.030$&173.14$\pm58.56$&32.63$\pm0.76$&26.59$\pm0.38$&0.093$\pm0.040$&8$^{+10}_{-2}$&10$^{+8}_{-2}$&    na    &    na    &    N    &158$\pm20$&22.0$\pm8.7$&1&92\\
G208.68-19.20S$\_$A&05:35:26.56&-05:03:55.11&0.251$\pm0.021$&0.124$\pm0.043$&169.90$\pm9.26$&147.84$\pm3.53$&119.82$\pm1.76$&0.421$\pm0.180$&    cx    &    cx    &    na    &    na    &    N    &96$\pm25$&49.0$\pm18.0$&1&84\\
G208.68-19.20S$\_$B&05:35:26.54&-05:03:55.71&0.283$\pm0.602$&0.255$\pm0.531$&40.05$\pm499.81$&14.96$\pm20.53$&10.41$\pm9.24$&0.043$\pm0.061$&    na    &    na    &    na    &    na    &    N    &96$\pm25$&49.0$\pm18.0$&1&84\\
G208.89-20.04E&05:32:48.12&-05:34:41.45&0.183$\pm0.009$&0.092$\pm0.014$&139.42$\pm5.50$&25.80$\pm0.22$&23.22$\pm0.12$&0.073$\pm0.031$&18$^{+6}_{-6}$&6$^{+8}_{-2}$&1.72$\pm0.29$&2.56$\pm0.42$&    Y    &108$\pm25$&2.2$\pm1.0$&1&$--$\\
G208.89-20.04Walma&05:32:28.26&-05:34:19.79&0.340$\pm0.061$&0.294$\pm0.065$&102.82$\pm89.60$&9.77$\pm0.80$&6.17$\pm0.34$&0.028$\pm0.012$&4$^{+6}_{-2}$&6$^{+6}_{-2}$&0.62$\pm0.01$&0.97$\pm0.19$&    Y    &    na    &    na    &0&$--$\\
G209.29-19.65S1&05:34:55.83&-05:46:04.75&6.581$\pm0.245$&2.585$\pm0.096$&136.90$\pm1.39$&64.14$\pm2.37$&0.66$\pm0.02$&1.442$\pm0.618$&    na    &    na    &    na    &    na    &    N    &    na    &    na    &-1&$--$\\
G209.55-19.68N1$\_$A&05:35:08.95&-05:55:54.98&0.376$\pm0.020$&0.195$\pm0.029$&53.73$\pm5.70$&49.57$\pm1.37$&32.94$\pm0.59$&0.141$\pm0.060$&42$^{+12}_{-10}$&18$^{+14}_{-8}$&1.28$\pm0.57$&1.91$\pm0.86$&    N    &    na    &    na    &0&$--$\\
G209.55-19.68N1$\_$B&05:35:08.63&-05:55:54.65&0.489$\pm0.083$&0.352$\pm0.076$&120.42$\pm71.17$&21.31$\pm2.45$&10.59$\pm0.86$&0.061$\pm0.027$&    cx    &    cx    &    na    &    na    &    N    &47$\pm13$&9.0$\pm3.7$&0&12\\
G209.55-19.68N1$\_$C&05:35:08.57&-05:55:54.54&1.603$\pm0.085$&1.262$\pm0.068$&83.11$\pm9.46$&46.30$\pm2.36$&3.62$\pm0.17$&0.132$\pm0.057$&    na    &    na    &    na    &    na    &    N    &47$\pm13$&9.0$\pm3.7$&0&12\\
G209.55-19.68S1&05:35:13.43&-05:57:57.89&0.167$\pm0.013$&0.163$\pm0.015$&23.17$\pm72.00$&92.75$\pm1.13$&79.66$\pm0.59$&0.264$\pm0.113$&24$^{+14}_{-8}$&38$^{+10}_{-8}$&1.22$\pm0.21$&1.86$\pm0.20$&    Y    &50$\pm15$&9.1$\pm3.6$&0&11\\
G209.55-19.68S2&05:35:09.05&-05:58:26.87&0.190$\pm0.012$&0.121$\pm0.012$&113.66$\pm8.35$&29.36$\pm0.40$&25.96$\pm0.21$&0.084$\pm0.036$&22$^{+12}_{-8}$&28$^{+8}_{-6}$&1.93$\pm0.76$&2.97$\pm0.01$&    Y    &48$\pm11$&3.4$\pm1.4$&0&10\\
G210.37-19.53S&05:37:00.43&-06:37:10.90&0.289$\pm0.016$&0.216$\pm0.019$&152.86$\pm10.12$&46.53$\pm0.85$&33.66$\pm0.39$&0.133$\pm0.057$&34$^{+8}_{-10}$&22$^{+14}_{-12}$&1.60$\pm0.55$&2.41$\pm0.40$&    Y    &39$\pm10$&0.6$\pm0.3$&0&164\\
G210.49-19.79W$\_$A&05:36:18.94&-06:45:23.54&0.263$\pm0.015$&0.191$\pm0.017$&75.98$\pm10.05$&70.71$\pm1.16$&54.66$\pm0.56$&0.201$\pm0.086$&38$^{+8}_{-8}$&36$^{+8}_{-8}$&2.62$\pm0.85$&4.29$\pm0.47$&    Y    &51$\pm20$&60.0$\pm24.0$&0&168\\
G210.49-19.79W$\_$B&05:36:18.50&-06:45:23.97&0.435$\pm0.036$&0.097$\pm0.060$&161.03$\pm5.53$&2.77$\pm0.14$&1.85$\pm0.06$&0.008$\pm0.003$&    na    &    na    &    na    &    na    &    N    &    na    &    na    &111&$--$\\
G210.82-19.47S$\_$B\tablenotemark{$\xi$}&05:38:03.43&-06:58:15.89&0.000$\pm0.132$&0.000$\pm0.072$&0.00$\pm0.00$&3.65$\pm0.10$&3.53$\pm0.05$&0.010$\pm0.004$& 4$^{+4}_{-2}$ & 4$^{+4}_{-2}$    &    na    &    na    &    N    &74$\pm12$&0.4$\pm0.2$&1&156\\
G210.97-19.33S2$\_$A&05:38:45.54&-07:01:02.02&0.218$\pm0.025$&0.182$\pm0.026$&101.22$\pm64.52$&7.00$\pm0.21$&5.68$\pm0.11$&0.020$\pm0.009$&10$^{+8}_{-4}$&14$^{+8}_{-4}$&1.72$\pm0.13$&2.94$\pm1.10$&    Y    &53$\pm15$&3.9$\pm1.5$&0&377\\
G210.97-19.33S2$\_$B&05:38:45.02&-07:01:01.68&0.122$\pm0.022$&0.107$\pm0.037$&20.34$\pm64.74$&6.16$\pm0.12$&5.70$\pm0.07$&0.018$\pm0.008$&    cx    &    cx    &    na    &    na    &    N    &82$\pm24$&4.1$\pm1.6$&1&144\\
G211.01-19.54N&05:37:57.02&-07:06:56.23&0.390$\pm0.005$&0.155$\pm0.009$&32.90$\pm1.19$&45.76$\pm0.35$&30.44$\pm0.15$&0.130$\pm0.056$&14$^{+8}_{-6}$&12$^{+8}_{-6}$&1.80$\pm0.58$&2.78$\pm0.27$&    Y    &39$\pm12$&4.5$\pm1.8$&0&153\\
G211.01-19.54S&05:37:58.75&-07:07:25.72&0.195$\pm0.020$&0.153$\pm0.019$&102.95$\pm31.20$&7.27$\pm0.16$&6.19$\pm0.08$&0.021$\pm0.009$&    na    &    $<$20$^{+8}_{-8}$    &    na    &    na    &    N    &52$\pm8$&0.9$\pm0.4$&0&152\\
G211.16-19.33N2&05:39:05.83&-07:10:39.29&0.172$\pm0.027$&0.140$\pm0.046$&0.94$\pm89.83$&5.70$\pm0.16$&4.97$\pm0.08$&0.016$\pm0.007$&8$^{+6}_{-2}$&10$^{+4}_{-2}$&    na    &    na    &    N    &70$\pm20$&3.7$\pm1.4$&0&133\\
G211.16-19.33N5&05:38:45.33&-07:10:56.03&0.184$\pm0.019$&0.088$\pm0.081$&31.60$\pm12.06$&5.32$\pm0.11$&4.71$\pm0.06$&0.015$\pm0.006$&    cx    &    cx    &    na    &    na    &    N    &112$\pm16$&1.3$\pm0.5$&1&135\\
G211.47-19.27N$\_$A&05:39:57.33&-07:29:32.73&0.599$\pm0.123$&0.210$\pm0.107$&133.25$\pm15.03$&23.65$\pm3.18$&13.04$\pm1.19$&0.067$\pm0.030$&    $<$30$^{+8}_{-8}$    &    $<$30$^{+8}_{-8}$    &    na    &    na    &    N    &48$\pm10$&1.9$\pm0.8$&0&290\\
G211.47-19.27N$\_$B&05:39:57.37&-07:29:33.10&0.462$\pm0.227$&0.268$\pm0.137$&110.05$\pm72.89$&13.82$\pm3.15$&8.58$\pm1.29$&0.039$\pm0.019$&    $<$30$^{+8}_{-8}$    &    $<$30$^{+8}_{-8}$    &    na    &    na    &    N    &48$\pm10$&2.1$\pm0.9$&0&290\\
G211.47-19.27S&05:39:56.00&-07:30:27.61&0.616$\pm0.025$&0.264$\pm0.026$&133.52$\pm3.02$&347.68$\pm9.56$&180.85$\pm3.44$&0.990$\pm0.424$&50$^{+12}_{-10}$&46$^{+14}_{-10}$&    na    &    na    &    Y    &49$\pm21$&180.0$\pm70.0$&0&288\\
G212.10-19.15N1&05:41:21.29&-07:52:27.44&5.813$\pm0.371$&3.260$\pm0.210$&139.70$\pm4.25$&18.26$\pm1.16$&0.20$\pm0.01$&0.410$\pm0.177$&    na    &    na    &    na    &    na    &    N    &    na    &    na    &-1&$--$\\
G212.10-19.15N2$\_$A&05:41:23.69&-07:53:46.74&0.193$\pm0.012$&0.176$\pm0.010$&76.00$\pm74.40$&11.84$\pm0.11$&10.17$\pm0.06$&0.034$\pm0.014$&    cx    &    cx    &    na    &    na    &    N    &114$\pm10$&1.1$\pm0.5$&1&263\\
G212.10-19.15N2$\_$B&05:41:23.99&-07:53:42.22&0.119$\pm0.024$&0.044$\pm0.055$&19.53$\pm29.51$&4.75$\pm0.08$&4.54$\pm0.05$&0.014$\pm0.006$&    na    &    na    &    na    &    na    &    N    &160$\pm30$&1.1$\pm0.5$&1&262\\
G212.10-19.15S&05:41:26.19&-07:56:51.93&0.251$\pm0.013$&0.198$\pm0.020$&29.02$\pm11.74$&83.33$\pm1.06$&66.42$\pm0.52$&0.237$\pm0.101$&10$^{+6}_{-4}$&6$^{+6}_{-4}$&3.90$\pm0.33$&6.15$\pm3.30$&    N    &43$\pm12$&3.2$\pm1.2$&0&247\\
G212.84-19.45N&05:41:32.07&-08:40:09.77&0.358$\pm0.009$&0.278$\pm0.014$&171.35$\pm6.91$&95.99$\pm1.13$&63.47$\pm0.49$&0.273$\pm0.117$&20$^{+6}_{-6}$&14$^{+6}_{-6}$&1.10$\pm0.01$&1.82$\pm0.21$&    N    &50$\pm13$&3.0$\pm1.2$&0&224\\
G215.87-17.62M$\_$A&05:53:32.52&-10:25:08.18&0.352$\pm0.019$&0.222$\pm0.021$&55.93$\pm6.79$&22.97$\pm0.44$&16.35$\pm0.20$&0.065$\pm0.028$&38$^{+12}_{-8}$&36$^{+10}_{-8}$&1.21$\pm0.11$&1.46$\pm0.02$&    Y    &    na    &    na    &0&$--$\\
G215.87-17.62N&05:53:42.56&-10:24:00.69&0.164$\pm0.032$&0.124$\pm0.038$&116.86$\pm44.11$&3.30$\pm0.07$&2.99$\pm0.04$&0.009$\pm0.004$&    na    &    na    &    na    &    na    &    N    &750$\pm193$&82.0$\pm40.0$&1&$--$\\
G215.87-17.62S$\_$off&05:53:25.07&-10:27:30.17&0.147$\pm0.048$&0.099$\pm0.043$&71.87$\pm89.43$&1.21$\pm0.04$&1.12$\pm0.02$&0.003$\pm0.001$&    na    &    na    &    na    &    na    &    N    &493$\pm60$&0.9$\pm0.5$&1&$--$\\
\enddata
\tablenotetext{*}{Y= detection and N= for non-detection in SiO}
\tablenotetext{\dag}{Starless= $-$1; Class 0 = 0; Class 1 = 1; Unclassified=111}
\tablenotetext{}{The `cx'  represents the complex structure; and 'na' is not estimated/found.}
\tablenotetext{\xi}{The objects are likely point sources and they are not resolved in deconvolved 2-D Gaussian fitting in combined TM1+TM2+ACA beam. Further investigation are required to confirm their candidacy.}
\end{deluxetable*}
\end{longrotatetable}

\movetabledown=15mm
\FloatBarrier
\begin{longrotatetable}
\begin{deluxetable*}{l@{\extracolsep{5pt}}c@{\extracolsep{10pt}} c@{\extracolsep{5pt}} c@{\extracolsep{5pt}} c@{\extracolsep{5pt}} c@{\extracolsep{5pt}} c@{\extracolsep{5pt}} c@{\extracolsep{5pt}} c@{\extracolsep{5pt}} c@{\extracolsep{5pt}} c@{\extracolsep{5pt}} c@{\extracolsep{5pt}} c@{\extracolsep{5pt}} c@{\extracolsep{5pt}} c@{\extracolsep{5pt}} c@{\extracolsep{5pt}} c@{\extracolsep{5pt}} c@{\extracolsep{5pt}} c@{\extracolsep{5pt}} c@{\extracolsep{5pt}}
}
\tablenum{6}
\tablecaption{SED Data for the continuum peak\label{tab:sedcrossmatch}}
\tablewidth{0pt}
\addtolength{\tabcolsep}{-2pt}
\tabletypesize{\scriptsize}
\tablewidth{6pt}
\addtolength{\tabcolsep}{-4pt}
\tablehead
{
\colhead{Source}&
\multicolumn{1}{c}{$...$} & \multicolumn{1}{c}{K$_{2mass}$} & \multicolumn{1}{c}{eK$_{2mass}$} &  \multicolumn{1}{c}{$...$} & \multicolumn{1}{c}{Wise1} & \multicolumn{1}{c}{eWise1}&  \multicolumn{1}{c}{$...$} &  \multicolumn{1}{c}{pacs1} &   \multicolumn{1}{c}{epacs1} & \multicolumn{1}{c}{$...$} & \multicolumn{1}{c}{akari09} & \multicolumn{1}{c}{e$\_$akari09} &
\multicolumn{1}{c}{$...$} & \multicolumn{1}{c}{JCMT850} & \multicolumn{1}{c}{eJCMT850} &
\multicolumn{1}{c}{$...$} & \multicolumn{1}{c}{IRAC1} & \multicolumn{1}{c}{eIRAC1} & \multicolumn{1}{c}{$...$}\\
\colhead{}&
\multicolumn{1}{c}{} & \multicolumn{1}{c}{[2.159]} & \multicolumn{1}{c}{} &  \multicolumn{1}{c}{} & \multicolumn{1}{c}{[3.4]} & \multicolumn{1}{c}{}&  \multicolumn{1}{c}{} &  \multicolumn{1}{c}{[70]} &   \multicolumn{1}{c}{} & \multicolumn{1}{c}{} & \multicolumn{1}{c}{[09]} & \multicolumn{1}{c}{} &
\multicolumn{1}{c}{} & \multicolumn{1}{c}{[850]} & \multicolumn{1}{c}{} &
\multicolumn{1}{c}{} & \multicolumn{1}{c}{[3.6]} & \multicolumn{1}{c}{} & \multicolumn{1}{c}{}
}
\startdata
G196.92-10.37$\_$C & $...$  &7.310e-03  &8.156e-07 & $...$  &1.845e-02 &4.606e-04 & $...$ &$--$ &$--$ & $...$  &1.972e-01  &3.200e-02 & $...$  &5.145e+00  &4.991e+00 & $...$  &$--$  &$--$ & $...$\\
 G200.34-10.97N & $...$  &8.589e-04  &8.834e-08 & $...$  &9.744e-04 &2.962e-05 & $...$ &$--$ &$--$ & $...$  &$--$  &$--$ & $...$  &6.507e-01  &6.349e-01 & $...$  &$--$  &$--$ & $...$\\
  G201.52-11.08 & $...$  &2.910e-03  &4.078e-07 & $...$  &4.047e-03 &1.011e-04 & $...$ &$--$ &$--$ & $...$  &$--$  &$--$ & $...$  &1.095e-01  &2.032e-02 & $...$  &$--$  &$--$ & $...$\\
G203.21-11.20W1 & $...$  &$--$  &$--$ & $...$  &1.630e-05 &6.532e-06 & $...$ &$--$ &$--$ & $...$  &$--$  &$--$ & $...$  &1.538e+00  &1.772e-01 & $...$  &$--$  &$--$ & $...$\\
G203.21-11.20W2 & $...$  &$--$  &$--$ & $...$  &1.463e-04 &1.112e-05 & $...$ &$--$ &$--$ & $...$  &$--$  &$--$ & $...$  &1.374e+00  &1.308e-01 & $...$  &$--$  &$--$ & $...$\\
G205.46-14.56M1$\_$B & $...$  &$--$  &$--$ & $...$  &$--$ &$--$ & $...$ &$--$ &$--$ & $...$  &$--$  &$--$ & $...$  &3.908e+00  &2.886e-01 & $...$  &$--$  &$--$ & $...$\\
G205.46-14.56M1$\_$A & $...$  &$--$  &$--$ & $...$  &2.344e-03 &6.616e-05 & $...$ &6.050e+00 &3.038e-01 & $...$  &$--$  &$--$ & $...$  &3.908e+00  &2.886e-01 & $...$  &3.339e-03  &1.680e-04 & $...$\\
G205.46-14.56M2$\_$A & $...$  &1.114e-02  &7.392e-06 & $...$  &5.769e-02 &2.819e-03 & $...$ &8.588e+00 &8.588e-01 & $...$  &$--$  &$--$ & $...$  &1.123e+00  &1.315e-01 & $...$  &3.007e-02  &1.507e-03 & $...$\\
G205.46-14.56M2$\_$B & $...$  &1.114e-02  &7.392e-06 & $...$  &5.769e-02 &2.819e-03 & $...$ &8.588e+00 &8.588e-01 & $...$  &$--$  &$--$ & $...$  &1.123e+00  &1.315e-01 & $...$  &3.007e-02  &1.507e-03 & $...$\\
G205.46-14.56M2$\_$C & $...$  &2.284e-02  &7.380e-06 & $...$  &1.350e-01 &3.225e-03 & $...$ &2.455e+01 &2.455e+00 & $...$  &1.432e+00  &7.350e-03 & $...$  &3.022e-01  &4.168e-02 & $...$  &1.620e-01  &8.115e-03 & $...$\\
G205.46-14.56M2$\_$D & $...$  &2.284e-02  &7.380e-06 & $...$  &1.350e-01 &3.225e-03 & $...$ &2.455e+01 &2.455e+00 & $...$  &1.432e+00  &7.350e-03 & $...$  &3.022e-01  &4.168e-02 & $...$  &1.620e-01  &8.115e-03 & $...$\\
G205.46-14.56M2$\_$E & $...$  &$--$  &$--$ & $...$  &$--$ &$--$ & $...$ &$--$ &$--$ & $...$  &$--$  &$--$ & $...$  &3.022e-01  &4.168e-02 & $...$  &$--$  &$--$ & $...$\\
G205.46-14.56M3 & $...$  &$--$  &$--$ & $...$  &$--$ &$--$ & $...$ &$--$ &$--$ & $...$  &$--$  &$--$ & $...$  &2.477e-01  &3.847e-02 & $...$  &$--$  &$--$ & $...$\\
G205.46-14.56N1 & $...$  &$--$  &$--$ & $...$  &$--$ &$--$ & $...$ &4.137e-01 &2.102e-02 & $...$  &$--$  &$--$ & $...$  &6.534e-01  &7.695e-02 & $...$  &5.672e-06  &$--$ & $...$\\
G205.46-14.56N2 & $...$  &$--$  &$--$ & $...$  &$--$ &$--$ & $...$ &6.514e-01 &3.293e-02 & $...$  &$--$  &$--$ & $...$  &3.022e-01  &4.168e-02 & $...$  &7.795e-06  &$--$ & $...$\\
G205.46-14.56S1$\_$A & $...$  &$--$  &$--$ & $...$  &2.591e-03 &9.003e-05 & $...$ &6.220e+01 &3.113e+00 & $...$  &$--$  &$--$ & $...$  &7.140e+00  &6.893e-01 & $...$  &$--$  &$--$ & $...$\\
G205.46-14.56S1$\_$B & $...$  &$--$  &$--$ & $...$  &$--$ &$--$ & $...$ &$--$ &$--$ & $...$  &$--$  &$--$ & $...$  &7.140e+00  &6.893e-01 & $...$  &$--$  &$--$ & $...$\\
\enddata

\tablenotetext{}{(This table contains all the cross-matching fluxes. T$_{bol}$ and L$_{bol}$ are estimated with the fluxes having good photometric accuracy (see text for details).}
\tablenotetext{}{(This table is available in its entirety in machine-readable form.}
\end{deluxetable*}
\end{longrotatetable}

\newpage
\appendix
   
\section{TM1+TM2+ ACA continuum images}


\setcounter{figure}{0} 
\renewcommand{\thefigure}{A\arabic{figure}} 
\begin{figure*}
\fig{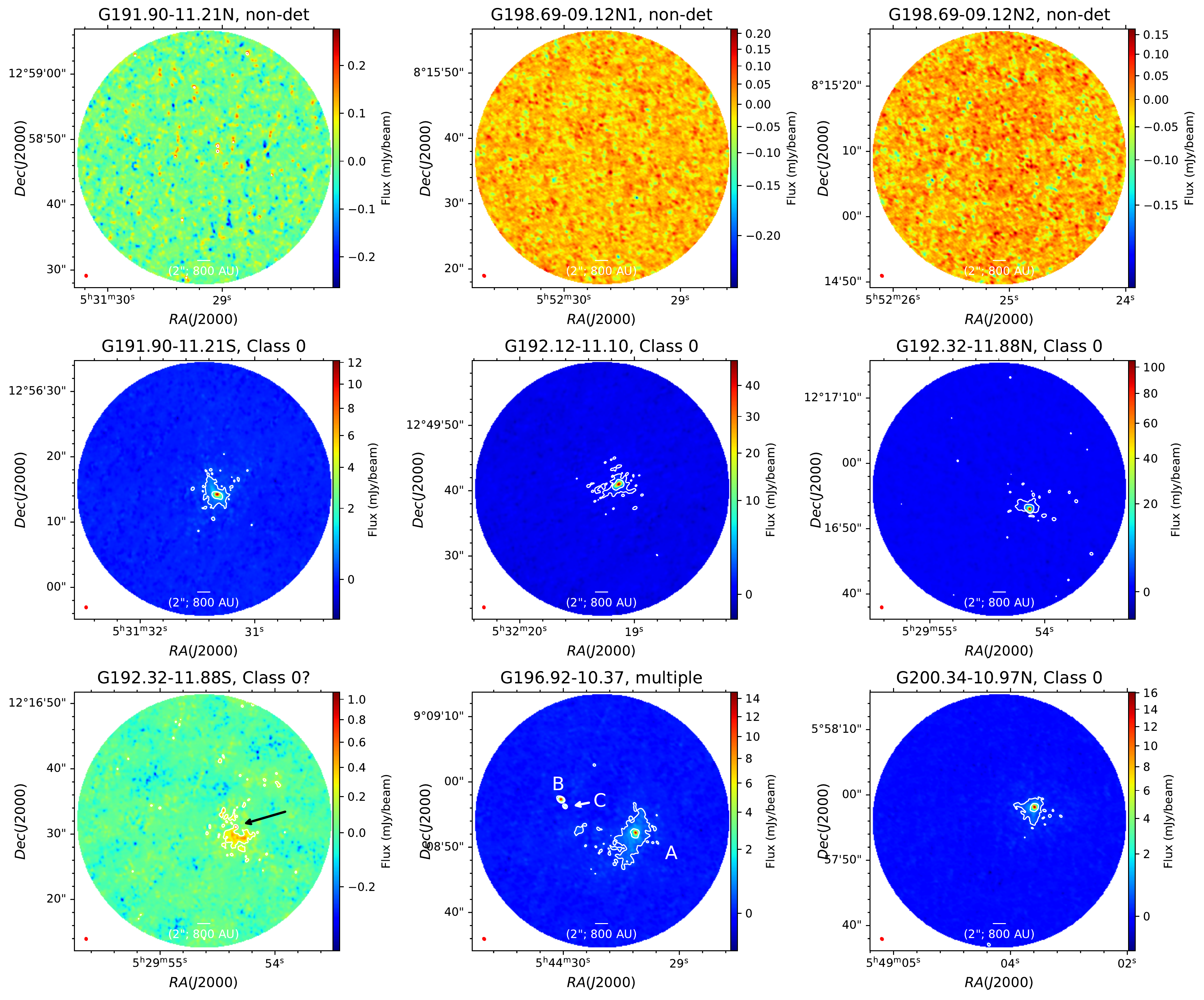}{0.98\textwidth}{}
\caption{$\lambda$-Orionis: The combined TM1, TM2 \& 7m ACA continuum images of non-detected dense cores and Class 0 systems (including multiples). The contours are are at 6 and 30 $\sigma$, where the corresponding $\sigma$s are tabulated in Table \ref{tab:targetobserved}.}
\label{cont1}
\end{figure*}

\begin{figure*}
\fig{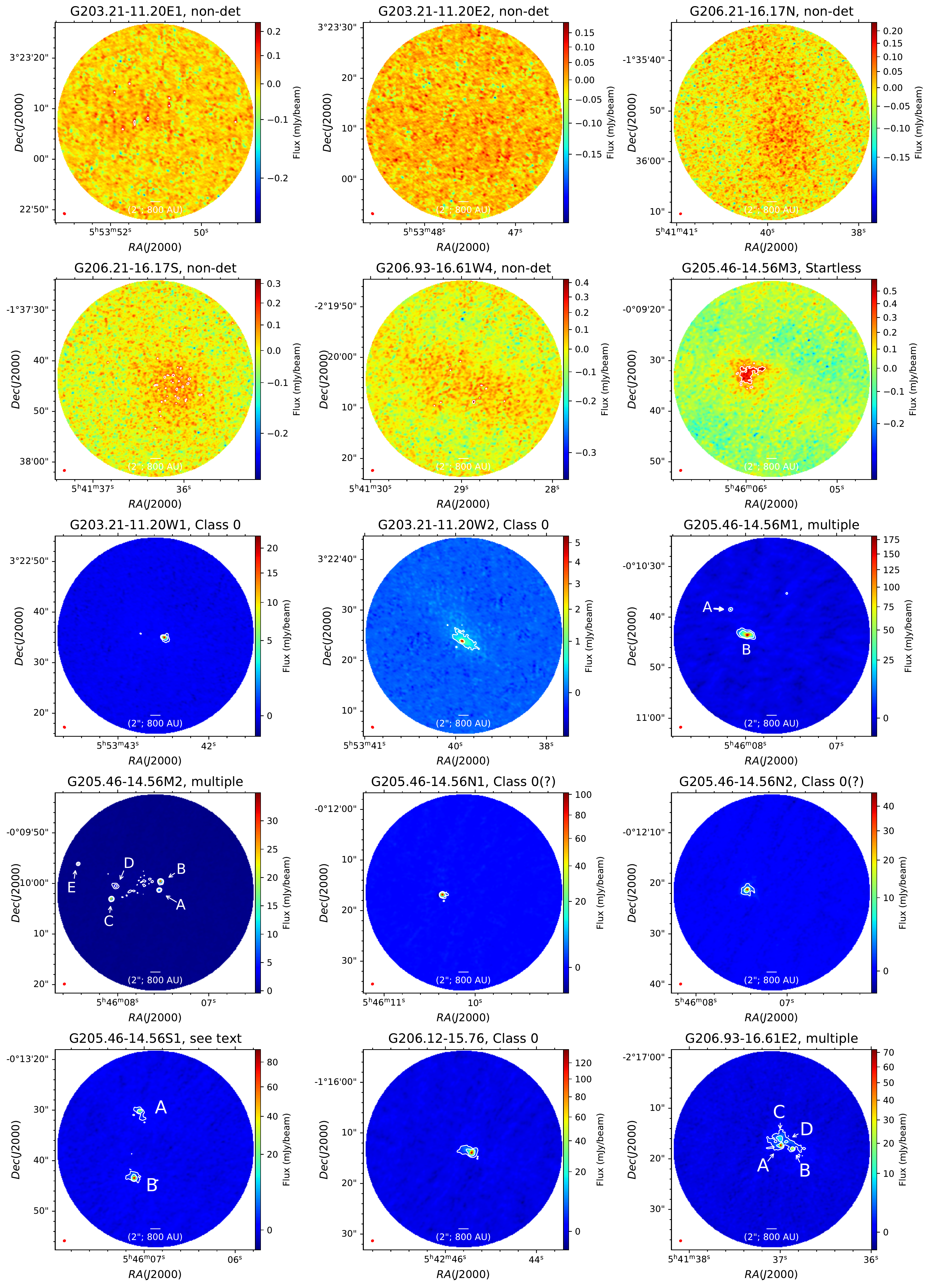}{0.98\textwidth}{}
\end{figure*}
\begin{figure*}
\setcounter{figure}{1} \renewcommand{\thefigure}{A\arabic{figure}} 
\fig{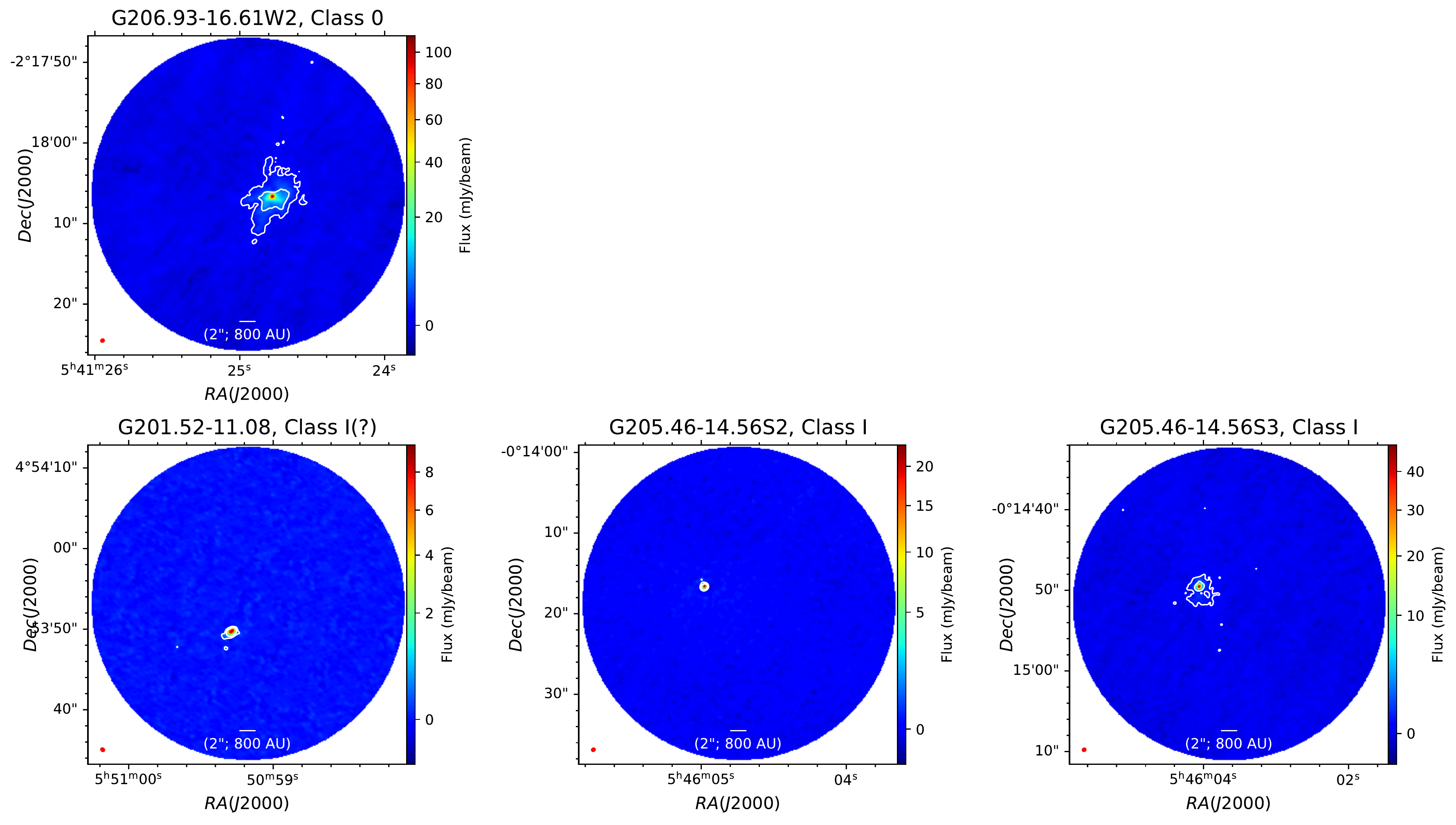}{0.98\textwidth}{}
\caption{Orion B: The combined TM1, TM2 \& 7m ACA continuum images of non-detected, starless dense cores, Class 0 and Class I systems (including multiples). The contours are are at 6 and 30 $\sigma$, where the corresponding $\sigma$s are tabulated in Table \ref{tab:targetobserved}.}
\label{cont2}
\end{figure*}

\begin{figure*}
\fig{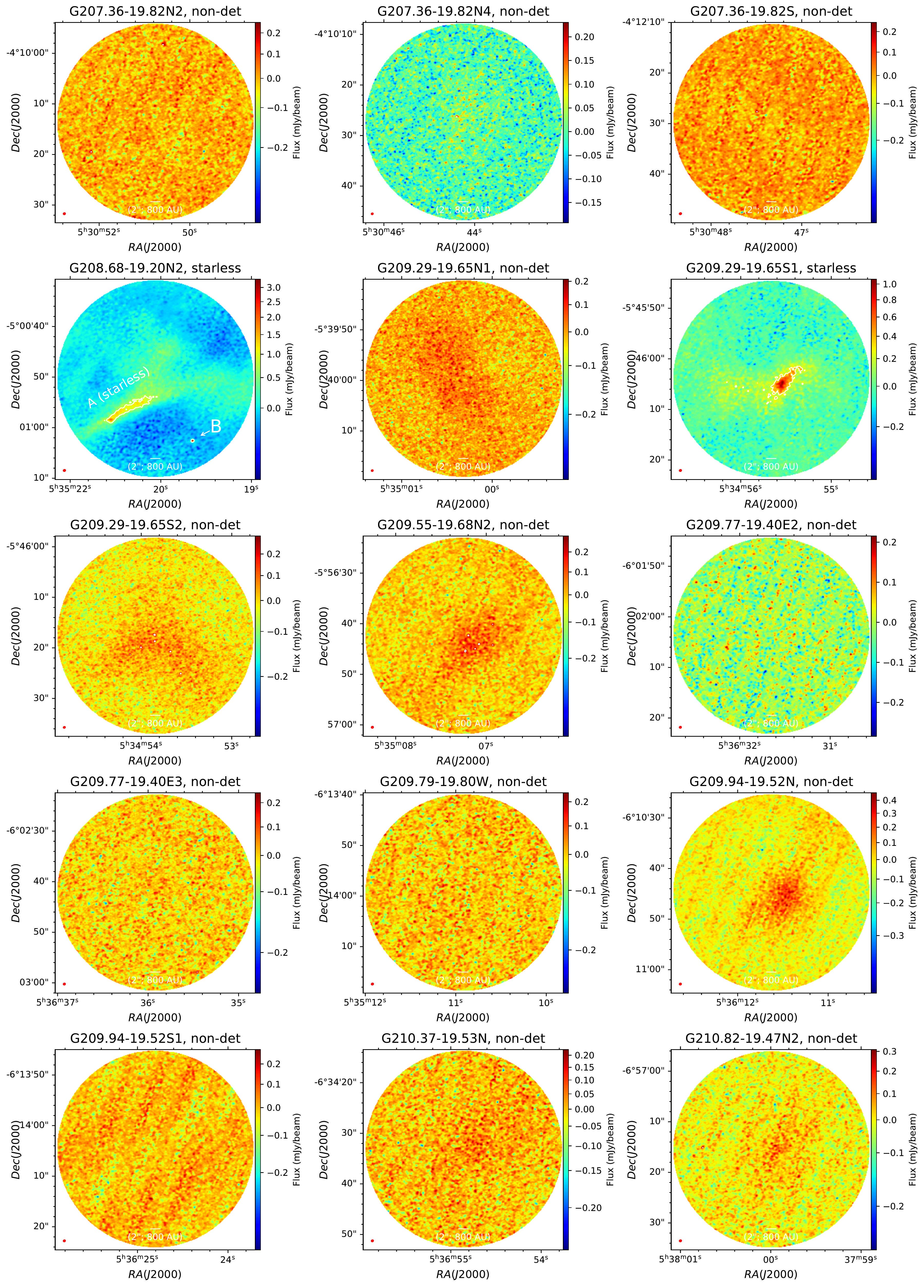}{0.78\textwidth}{}
\end{figure*}
\begin{figure*}
\setcounter{figure}{2} \renewcommand{\thefigure}{A\arabic{figure}} 
\fig{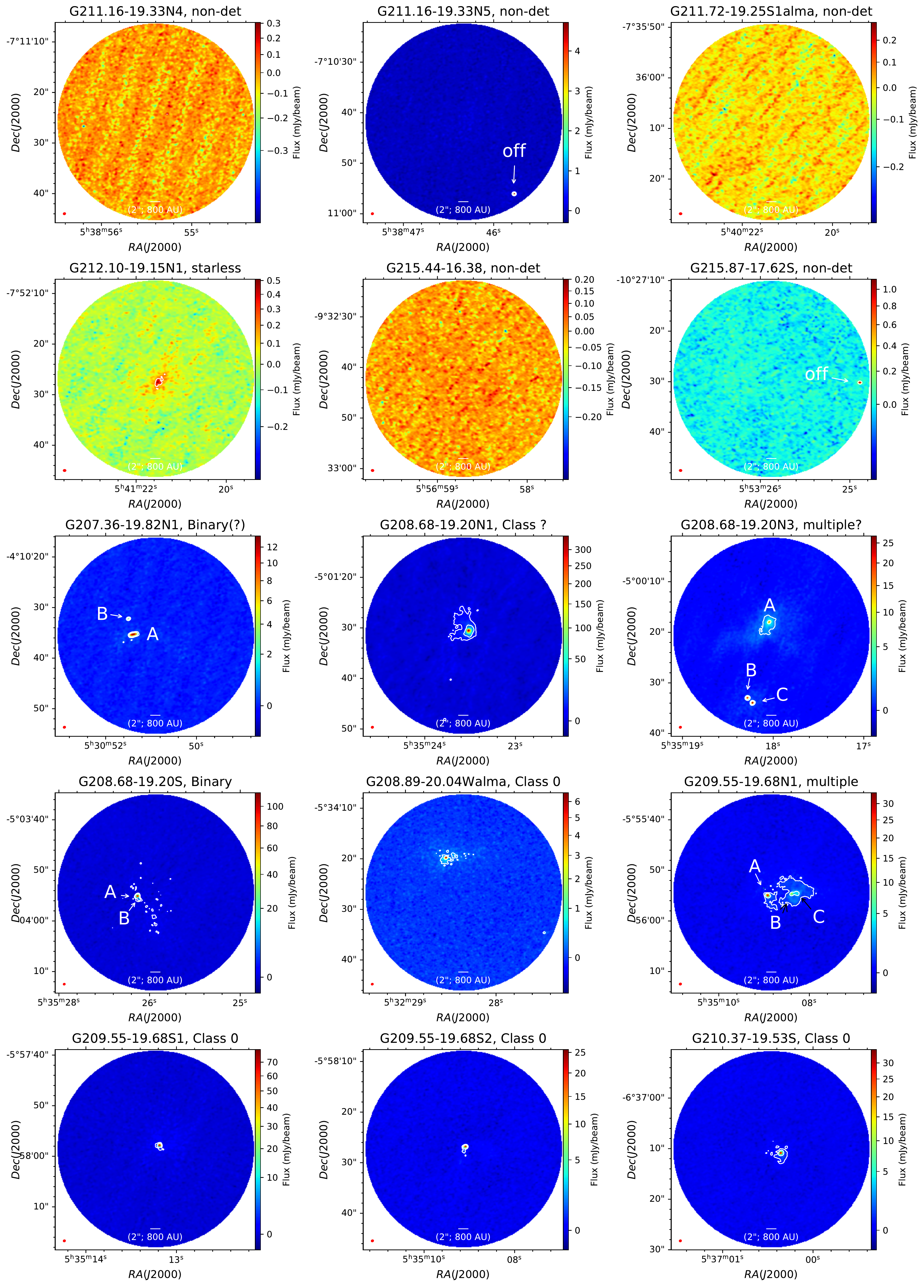}{0.78\textwidth}{}
\end{figure*}
\begin{figure*}
\setcounter{figure}{2} \renewcommand{\thefigure}{A\arabic{figure}} 
\fig{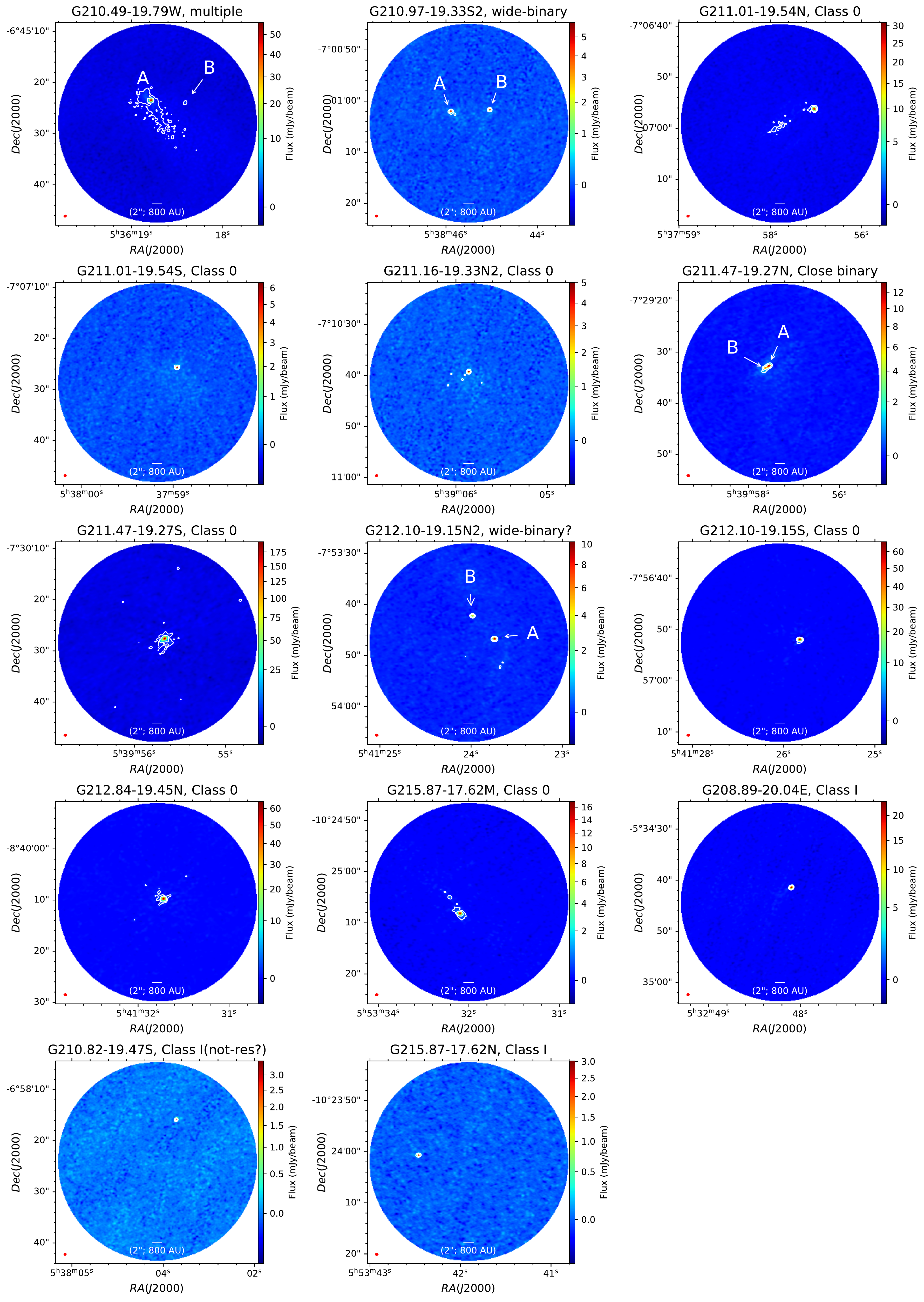}{0.78\textwidth}{}
\caption{Orion A: The combined TM1, TM2 \& 7m ACA continuum images of non-detected, starless dense cores, Class 0 and Class I systems (including multiples). The contours are are at 6 and 30 $\sigma$, where the corresponding $\sigma$s are tabulated in Table \ref{tab:targetobserved}.}
\label{cont3}
\end{figure*}

\begin{figure*}
\fig{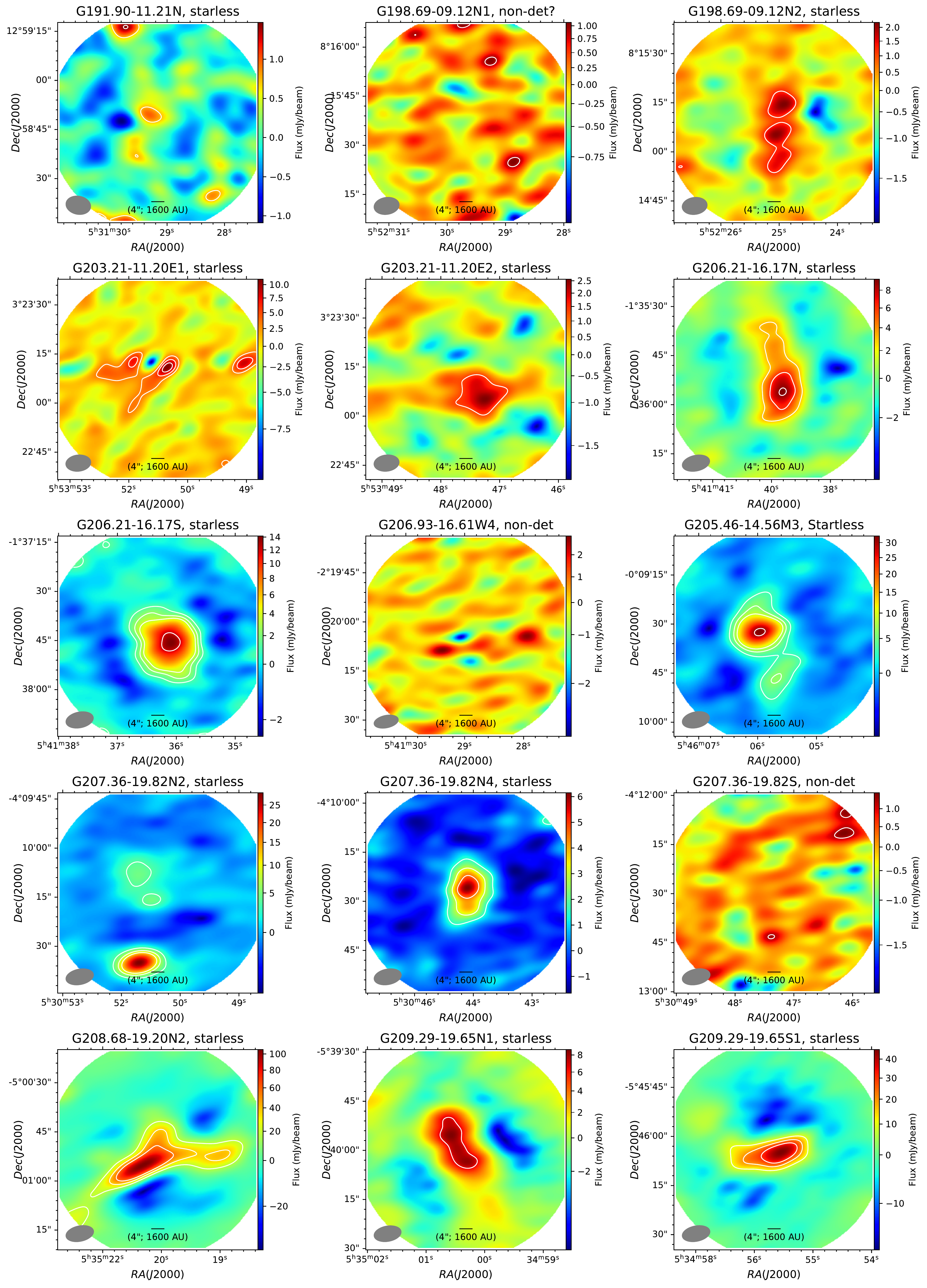}{0.95\textwidth}{}
\end{figure*}
\begin{figure*}
\setcounter{figure}{3} \renewcommand{\thefigure}{A\arabic{figure}} 
\fig{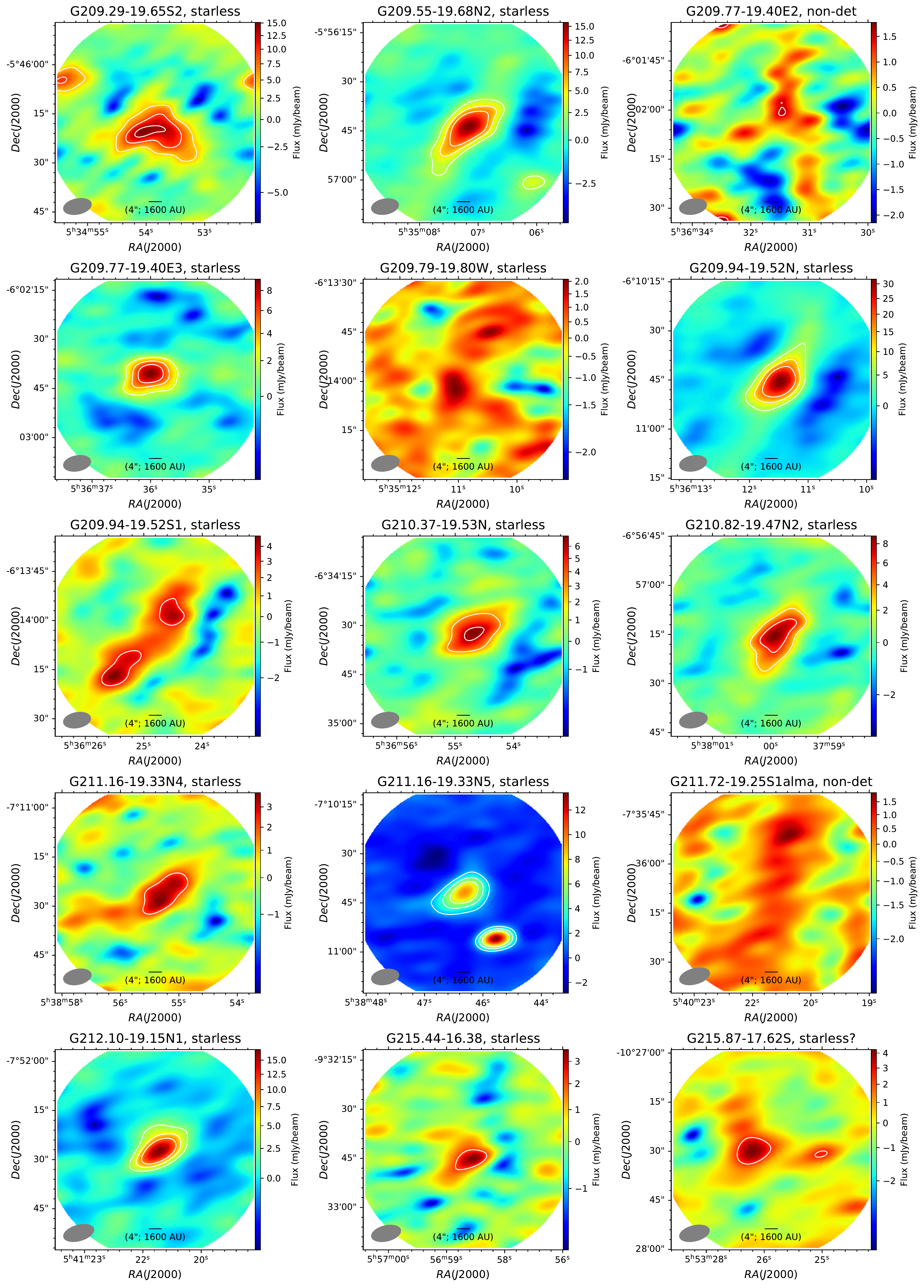}{0.95\textwidth}{}
\caption{All the 7m ACA continuum maps of the non-detected and starless dense cores in combined TM1, TM2 \& 7m ACA continuum images. The contours are are at 3, 6, 9  and 30 $\sigma$, where the corresponding $\sigma$s are tabulated in Table \ref{tab:targetobserved}}
\label{ACAcont}
\end{figure*}

\begin{figure*}
\fig{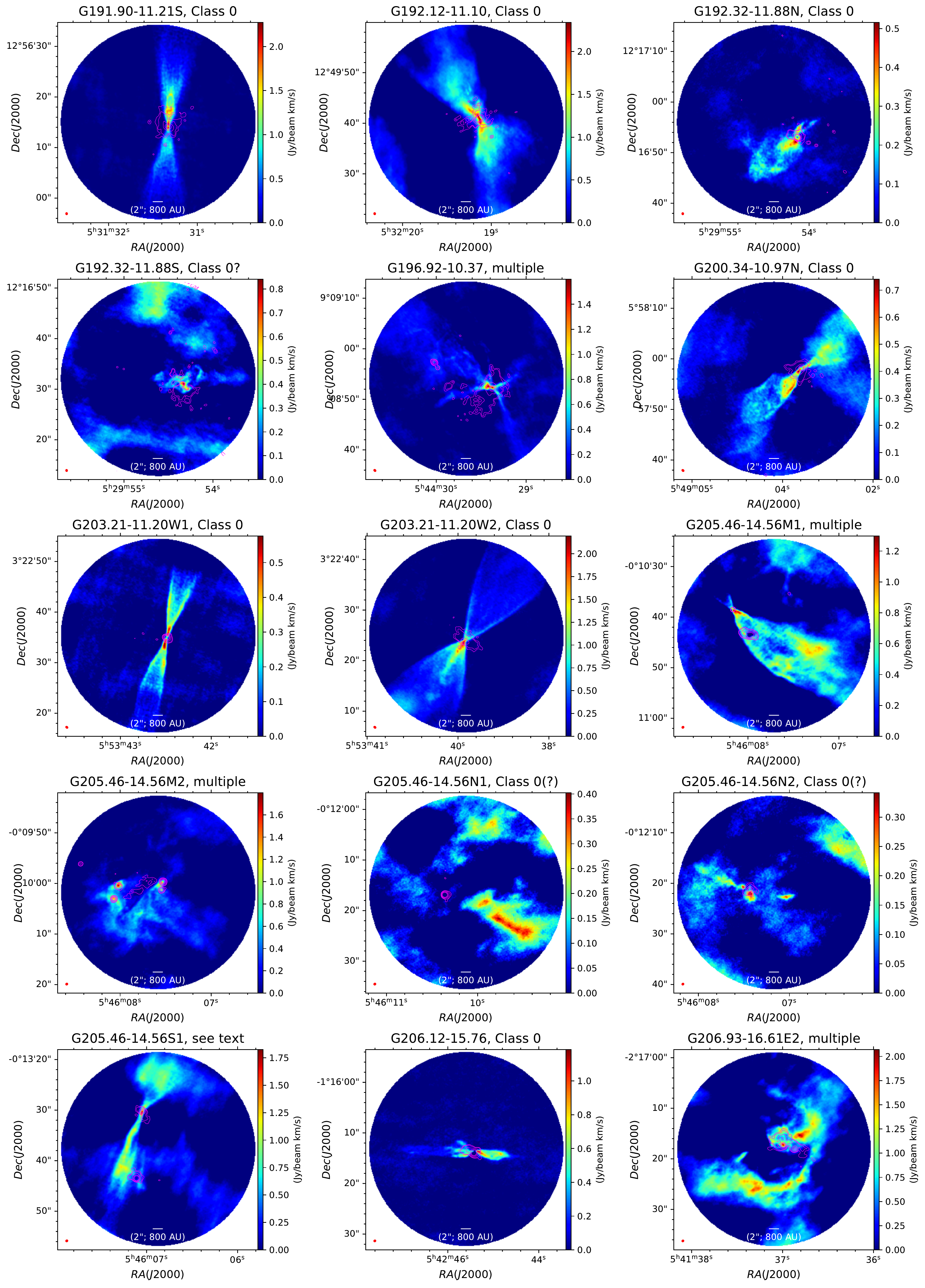}{0.95\textwidth}{}
\end{figure*}
\begin{figure*}
\setcounter{figure}{4} \renewcommand{\thefigure}{A\arabic{figure}} 
\fig{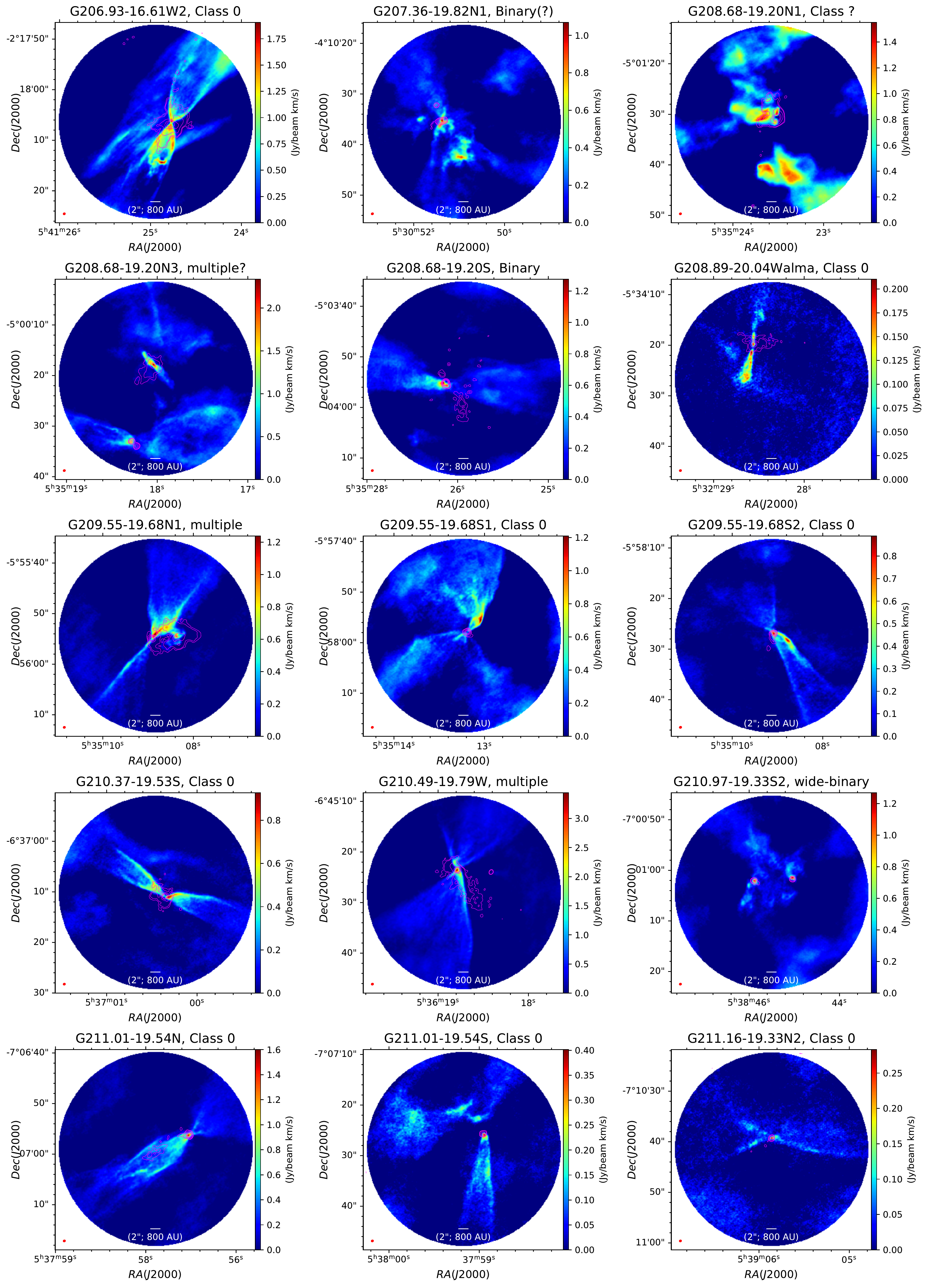}{0.95\textwidth}{}
\end{figure*}
\begin{figure*}
\setcounter{figure}{4} \renewcommand{\thefigure}{A\arabic{figure}} 
\fig{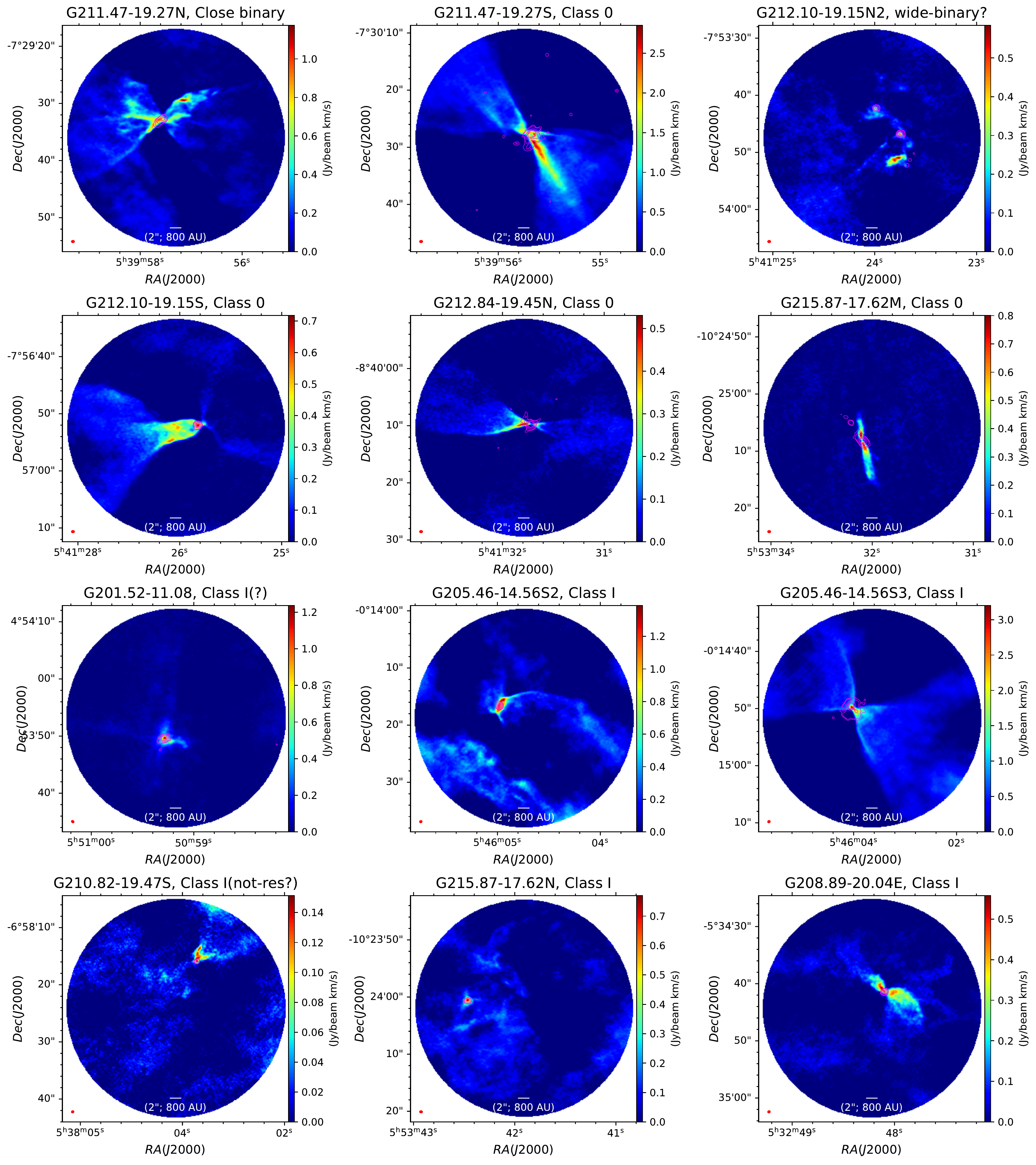}{0.95\textwidth}{}
\caption{Velocity-integrated CO maps showing the outflow structures of the protostellar sources. The magenta contours are 4, 6, 18, 50, 100 $\sigma$ of combined TM1, TM2 \& 7m ACA continuum emission, where the the corresponding $\sigma$s are tabulated in Table \ref{tab:targetobserved}.}
\label{alloutflows}
\end{figure*}


\bibliography{sample63}{}
\bibliographystyle{aasjournal}

\end{document}